\documentclass[longauth]{aa} 
\usepackage{graphicx}
\usepackage{txfonts}
\usepackage{xcolor}
\usepackage{hyperref}

\newcommand{\orcid}[1]{\href{https://orcid.org/#1}{\includegraphics[width=10pt]{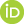}}}
\newcommand{\gaia}{{\em Gaia}}
\newcommand{\lc}{lightcurve} 
\newcommand{\lcs}{lightcurves}
\newcommand{\newsource}{NewSource}
\newcommand{\newsources}{NewSources}
\newcommand{\oldsource}{OldSource}
\newcommand{\dmag}{OldSourceDeltaMag}
\newcommand{\meanrms}{OldSourceMeanRms}
\newcommand{\skewVonN}{OldSourceSkewVonN}
\newcommand{\sourceid}{sourceId}

\newcommand{\gsa}{{\em GSA}}

\authorrunning{S. T. Hodgkin et al.}

\graphicspath{{}{../}} 

\begin{document} 


    \title{\gaia\ Photometric Science Alerts}

%

\author{
    S.~T.~Hodgkin\inst{\ref{inst1}}\orcid{0000-0002-5470-3962}
    \and D.~L.~Harrison\inst{\ref{inst1},\ref{inst2}}\orcid{0000-0001-8687-6588}
    \and E.~Breedt\inst{\ref{inst1}}\orcid{0000-0001-6180-3438}
    \and T.~Wevers\inst{\ref{inst1},\ref{inst3}}\orcid{0000-0002-4043-9400}
    \and G.~Rixon\inst{\ref{inst1}}\orcid{0000-0003-4399-6568}
    \and A.~Delgado\inst{\ref{inst1},\ref{inst4}}
    \and A.~Yoldas\inst{\ref{inst1}}
    \and Z.~Kostrzewa-Rutkowska\inst{\ref{inst5},\ref{inst6}}
    \and {\L}.~Wyrzykowski\inst{\ref{inst7}}\orcid{0000-0002-9658-6151}
    \and M.~van~Leeuwen\inst{\ref{inst1}}
    \and N.~Blagorodnova\inst{\ref{inst8}}
    \and H.~Campbell\inst{\ref{inst9}}
    \and D.~Eappachen\inst{\ref{inst6},\ref{inst8}}
    \and M.~Fraser\inst{\ref{inst10}}\orcid{0000-0003-2191-1674}
    \and N.~Ihanec\inst{\ref{inst7},\ref{inst11}}
    \and S.~E.~Koposov\inst{\ref{inst12},\ref{inst1}}
    \and K.~Kruszy\'{n}ska\inst{\ref{inst7}}\orcid{0000-0002-2729-5369}
    \and G.~Marton\inst{\ref{inst13}}\orcid{0000-0002-1326-1686}
    \and K.~A.~Rybicki\inst{\ref{inst7}}\orcid{0000-0002-9326-9329}
    \and A.~G.~A.~Brown\inst{\ref{inst5}}\orcid{0000-0002-7419-9679}
    \and P.~W.~Burgess\inst{\ref{inst1}}
    \and G.~Busso\inst{\ref{inst1}}
    \and S.~Cowell\inst{\ref{inst1}}
    \and F.~De~Angeli\inst{\ref{inst1}}\orcid{0000-0003-1879-0488}
    \and C.~Diener\inst{\ref{inst1}}
    \and D.~W.~Evans\inst{\ref{inst1}}\orcid{0000-0002-6685-5998}
    \and G.~Gilmore\inst{\ref{inst1}}\orcid{0000-0003-4632-0213}
    \and G.~Holland\inst{\ref{inst1}}
    \and P.~G.~Jonker\inst{\ref{inst8},\ref{inst6}}\orcid{0000-0001-5679-0695}
    \and F.~van~Leeuwen\inst{\ref{inst1}}
    \and F.~Mignard\inst{\ref{inst14}}
    \and P.~J.~Osborne\inst{\ref{inst1}}
    \and J.~Portell\inst{\ref{inst15}}\orcid{0000-0002-8886-8925}
    \and T.~Prusti\inst{\ref{inst16}}\orcid{0000-0003-3120-7867}
    \and P.~J.~Richards\inst{\ref{inst17}}
    \and M.~Riello\inst{\ref{inst1}}\orcid{0000-0002-3134-0935}
    \and G.~M.~Seabroke\inst{\ref{inst18}}\orcid{0000-0003-4072-9536}
    \and N.\ A.\ Walton\inst{\ref{inst1}}\orcid{0000-0003-3983-8778}
    \and P\'eter~\'Abrah\'am\inst{\ref{inst13},\ref{inst42}}\orcid{0000-0001-6015-646X}
    \and G.~Altavilla\inst{\ref{inst19},\ref{inst20}}\orcid{0000-0002-9934-1352}
    \and S.~G.~Baker\inst{\ref{inst18}}\orcid{0000-0002-6436-1257}
    \and U.~Bastian\inst{\ref{inst21}}\orcid{0000-0002-8667-1715}
    \and P.~O'Brien\inst{\ref{inst22}}
    \and J.~de~Bruijne\inst{\ref{inst16}}\orcid{0000-0001-6459-8599}
    \and T.~Butterley\inst{\ref{inst23}}\orcid{0000-0002-2853-0834}
    \and J.~M.~Carrasco\inst{\ref{inst15}}\orcid{0000-0002-3029-5853}
    \and J.~Casta{\~n}eda\inst{\ref{inst24}}\orcid{0000-0001-7820-946X}
    \and J.~S.~Clark\inst{\ref{inst25}}
    \and G.~Clementini\inst{\ref{inst26}}\orcid{0000-0001-9206-9723}
    \and C.~M.~Copperwheat\inst{\ref{inst27}}\orcid{0000-0001-7983-8698}
    \and M.~Cropper\inst{\ref{inst18}}\orcid{0000-0003-4571-9468}
    \and G.~Damljanovic\inst{\ref{inst28}}\orcid{0000-0002-6710-6868}
    \and M.~Davidson\inst{\ref{inst12}}
    \and C.~J.~Davis\inst{\ref{inst29}}
    \and M.~Dennefeld\inst{\ref{inst30}}
    \and V.S.~Dhillon\inst{\ref{inst31},\ref{inst32}}\orcid{0000-0003-4236-9642}
    \and C.~Dolding\inst{\ref{inst18}}
    \and M.~Dominik\inst{\ref{inst33}}\orcid{0000-0002-3202-0343}
    \and P.~Esquej\inst{\ref{inst4}}\orcid{0000-0001-8195-628X}
    \and L.~Eyer\inst{\ref{inst34}}\orcid{0000-0002-0182-8040}
    \and C.~Fabricius\inst{\ref{inst15}}\orcid{0000-0003-2639-1372}
    \and M.~Fridman\inst{\ref{inst35},\ref{inst36}}\orcid{0000-0002-6946-5506}
    \and D.~Froebrich\inst{\ref{inst37}}\orcid{0000-0003-4734-3345}
    \and N.~Garralda\inst{\ref{inst15}}
    \and A.~Gomboc\inst{\ref{inst35}}\orcid{0000-0002-0908-914X}
    \and J.~J.~Gonz\'alez-Vidal\inst{\ref{inst24}}
    \and R.~Guerra\inst{\ref{inst53}}\orcid{0000-0002-9850-8982}
    \and N.~C.~Hambly\inst{\ref{inst12}}\orcid{0000-0002-9901-9064}
    \and L.~K.~Hardy\inst{\ref{inst31}}
    \and B.~Holl\inst{\ref{inst34}}\orcid{0000-0001-6220-3266}
    \and A.~Hourihane\inst{\ref{inst1}}
    \and J.~Japelj\inst{\ref{inst38}}\orcid{0000-0001-7484-6859}
    \and D.~A.~Kann\inst{\ref{inst39}}\orcid{0000-0003-2902-3583}
    \and C.~Kiss\inst{\ref{inst13}}\orcid{0000-0002-8722-6875}
    \and C.~Knigge\inst{\ref{inst60}}
    \and U.~Kolb\inst{\ref{inst25}}\orcid{0000-0001-8670-8365}
    \and S.~Komossa\inst{\ref{inst40}}
    \and \'A. K\'osp\'al\inst{\ref{inst13},\ref{inst41},\ref{inst42}}\orcid{0000-0001-7157-6275}
    \and G.~Kov\'acs\inst{\ref{inst43}}
    \and M.~Kun\inst{\ref{inst13}}
    \and G.~Leto\inst{\ref{inst44}}\orcid{0000-0002-0040-5011}
    \and F.~Lewis\inst{\ref{inst27},\ref{inst45}}\orcid{0000-0003-3352-2334}
    \and S.~P.~Littlefair\inst{\ref{inst31}}\orcid{0000-0001-7221-855X}
    \and A.~A.~Mahabal\inst{\ref{inst46},\ref{inst47}}\orcid{0000-0003-2242-0244}
    \and C.~G.~Mundell\inst{\ref{inst48}}\orcid{0000-0003-2809-8743}
    \and Z.~Nagy\inst{\ref{inst13}}\orcid{0000-0002-3632-1194}
    \and D.~Padeletti\inst{\ref{inst49}}
    \and L.~Palaversa\inst{\ref{inst34},\ref{inst50}}
    \and A.~Pigulski\inst{\ref{inst51}}\orcid{0000-0003-2488-6726}
    \and M.~L.~Pretorius\inst{\ref{inst52}}
    \and W.~van~Reeven\inst{\ref{inst53}}
    \and V.~A.~R.~M.~Ribeiro\inst{\ref{inst54},\ref{inst55},\ref{inst8}}\orcid{0000-0003-3617-4400}
    \and M.~Roelens\inst{\ref{inst34}}
    \and N.~Rowell\inst{\ref{inst12}}
    \and N.~Schartel\inst{\ref{inst53}}
    \and A.~Scholz\inst{\ref{inst33}}\orcid{0000-0001-8993-5053}
    \and A.~Schwope\inst{\ref{inst56}}\orcid{0000-0003-3441-9355}
    \and B.~M.~Sip\H{o}cz\inst{\ref{inst43}}\orcid{0000-0002-3713-6337}
    \and S.~J.~Smartt\inst{\ref{inst58}}
    \and M.~D.~Smith\inst{\ref{inst37}}
    \and I.~Serraller\inst{\ref{inst15}}
    \and D.~Steeghs\inst{\ref{inst59}}\orcid{0000-0003-0771-4746}
    \and M.~Sullivan\inst{\ref{inst60}}\orcid{0000-0001-9053-4820}
    \and L.~Szabados\inst{\ref{inst13}}\orcid{0000-0002-2046-4131}
    \and E.~Szegedi-Elek\inst{\ref{inst13}}\orcid{0000-0001-7807-6644}
    \and P.~Tisserand\inst{\ref{inst30}}\orcid{0000-0003-4237-0520}
    \and L.~Tomasella\inst{\ref{inst61}}\orcid{0000-0002-3697-2616}
    \and S.~van~Velzen\inst{\ref{inst5}}\orcid{0000-0002-3859-8074}
    \and P.~A~Whitelock\inst{\ref{inst52},\ref{inst62}}\orcid{0000-0002-4678-4432}
    \and R.~W.~Wilson\inst{\ref{inst23}}\orcid{0000-0002-6122-7052}
    \and D.~R.~Young\inst{\ref{inst58}}\orcid{0000-0002-1229-2499}
    }
    
    \institute{Institute of Astronomy, Madingley Road, Cambridge, CB3 0HA, UK. \email{sth@ast.cam.ac.uk, dlh@ast.cam.ac.uk}\label{inst1}
    \and 
        Kavli Institute for Cosmology, Institute of Astronomy, Madingley Road, Cambridge, CB3 0HA, UK\label{inst2}
    \and 
        European Southern Observatory, Alonso de Córdova 3107, Vitacura, Casilla 19001, Santiago, Chile\label{inst3}
    \and 
        RHEA Group for ESA/ESAC, Camino bajo del Castillo s/n, Urbanización Villafranca del Castillo, Villanueva de la Ca\~{n}ada, 28692 Madrid, Spain\label{inst4}
    \and 
        Leiden Observatory, Leiden University, PO Box 9513, NL-2300 RA Leiden, the Netherlands\label{inst5}
    \and 
         SRON, Netherlands Institute for Space Research, Sorbonnelaan 2, 3584 CA, Utrecht, The Netherlands\label{inst6}
    \and 
        Warsaw University, Astronomical Observatory, Department of Physics, Al. Ujazdowskie 4, 00-478, Warszawa, Poland\label{inst7}
    \and 
        Department of Astrophysics/IMAPP, Radboud University, P.O. Box 9010, 6500 GL Nijmegen, The Netherlands\label{inst8}
    \and 
        Department of Physics, Faculty of Engineering and Physical Sciences, University of Surrey, Guildford, Surrey, GU2 7XH, UK\label{inst9}
    \and 
        School of Physics, O'Brien Centre for Science North, University College Dublin, Belfield, Dublin 4, Ireland\label{inst10}
    \and 
        Isaac Newton Group of Telescopes, Apdo. 321, E-38700 Santa Cruz de la Palma, Canary Islands, Spain\label{inst11}
    \and 
        Institute for Astronomy, University of Edinburgh, Royal Observatory, Blackford Hill, Edinburgh EH9 3HJ, UK\label{inst12}
    \and 
        Konkoly Observatory, Research Centre for Astronomy and Earth Sciences, E\"{o}tv\"{o}s Lor\'{a}nd Research Network, Konkoly Thege 15-17, H-1121 Budapest, Hungary\label{inst13}
    \and 
        Université C{\^o}te d'Azur, Observatoire de la C{\^o}te d'Azur, CNRS, Laboratoire Lagrange, France\label{inst14}
    \and 
        Departament de F\'isica Qu\`antica i Astrof\'isica, Institut de Ci\`encies del Cosmos (ICCUB), Universitat de Barcelona (IEEC-UB), Mart\'i Franqu\`es 1, 08028 Barcelona, Spain\label{inst15}
    \and 
        European Space Agency (ESA), European Space Research and Technology Centre (ESTEC), Keplerlaan 1, 2201AZ, Noordwijk, The Netherlands\label{inst16}
    \and 
        STFC, Rutherford Appleton Laboratory, Harwell, Didcot, OX11 0QX, UK\label{inst17}
    \and 
        Mullard Space Science Laboratory, University College London, Holmbury St Mary, Dorking, Surrey RH5 6NT, UK\label{inst18}
    \and 
        ELTE E\"otv\"os Lor\'and University, Institute of Physics, P\'azm\'any P\'eter s\'et\'any 1/A, 1117 Budapest, Hungary\label{inst42}
    \and 
        INAF -- Osservatorio Astronomico di Roma, Via Frascati 33, I-00078, Monte Porzio Catone (Roma), Italy\label{inst19}
    \and 
        Space Science Data Center -- ASI, Via del Politecnico SNC, I-00133 Roma, Italy\label{inst20}
    \and 
        ARI/ZAH, Heidelberg University, Heidlelberg, Germany\label{inst21}
    \and 
        School of Physics and Astronomy, University of Leicester, University Road, Leicester, LE1 7RH, UK\label{inst22}
    \and 
        Centre for Advanced Instrumentation, Department of Physics, University of Durham, South Road, Durham DH1 3LE, UK\label{inst23}
    \and 
        DAPCOM for Institut de Ci{\`e}ncies del Cosmos, Universitat de Barcelona (IEEC-UB), Mart{\'i} i Franqu{\`e}s 1, E-08028 Barcelona, Spain\label{inst24}
    \and 
        School of Physical Sciences, The Open University, Walton Hall, Milton Keynes MK7 67AA, UK\label{inst25}
    \and 
        INAF - Osservatorio di Astrofisica e Scienza dello Spazio di Bologna, via Piero Gobetti 93/3, 40129 Bologna, Italy\label{inst26}
    \and 
        Astrophysics Research Institute, Liverpool John Moores University, 146 Brownlow Hill, Liverpool L3 5RF, UK\label{inst27}
    \and 
        Astronomical Observatory (street Volgina 7, 11060 Belgrade, Serbia)\label{inst28}
    \and 
        National Science Foundation\label{inst29}
    \and 
        Sorbonne Universit\'es, UPMC Univ Paris 6 et CNRS, UMR 7095, Institut d'Astrophysique de Paris, IAP, F-75014 Paris, France\label{inst30}
    \and 
        Department of Physics and Astronomy, University of Sheffield, Sheffield S3 7RH, UK\label{inst31}
    \and 
        Instituto de Astrof\'{i}sica de Canarias, E-38205 La Laguna, Tenerife, Spain\label{inst32}
        \newpage
    \and 
        SUPA, School of Physics \& Astronomy, University of St Andrews, North Haugh, St Andrews, KY16 9SS, UK\label{inst33}
    \and 
        Department of Astronomy, University of Geneva, Chemin Pegasi 51, 1290 Versoix, Switzerland\label{inst34}
    \and 
        Center for Astrophysics and Cosmology, University of Nova Gorica, Vipavska 13, 5000 Nova Gorica, Slovenia\label{inst35}
    \and 
        University of Lethbridge\label{inst36}
    \and 
        Centre for Astrophysics and Planetary Science, School of Physical Sciences, University of Kent, Canterbury CT2 7NH, UK\label{inst37}
    \and 
        ESA , ESAC, Camino Bajo del Castillo s/n, Urb. Villafranca del Castillo, 28692 Villanueva de la Ca{\~n}ada, Spain\label{inst53}
    \and 
        Anton Pannekoek Institute for Astronomy, University of Amsterdam, Science Park 904, 1098 XH Amsterdam, The Netherlands\label{inst38}
    \and 
        Instituto de Astrof\'isica de Andaluc\'ia (IAA-CSIC), Glorieta de la Astronom\'ia s/n, 18008 Granada, Spain\label{inst39}
    \and 
        School of Physics and Astronomy, University of Southampton, Southampton, SO17 1BJ, UK\label{inst60}
    \and 
        Max-Planck-Institut f{\"u}r Radioastronomie, Auf dem H{\"u}gel 69, 53111 Bonn, Germany\label{inst40}
    \and 
        Max Planck Institute for Astronomy, K\"onigstuhl 17, 69117 Heidelberg, Germany\label{inst41}
    \and 
        DiRAC Institute, Department of Astronomy, University of Washington, Seattle, WA 98195, USA\label{inst43}
    \and 
        INAF-Osservatorio Astrofisico di Catania, via S. Sofia 78, 95123 Catania, Italia\label{inst44}
    \and 
	    Faulkes Telescope Project, School of Physics and Astronomy, Cardiff University, The Parade, Cardiff, CF24 3AA, Wales\label{inst45}
    \and 
        Division of Physics, Mathematics and Astronomy, California Institute of Technology, Pasadena, CA 91125, USA\label{inst46}
    \and 
        Center for Data Driven Discovery, California Institute of Technology, Pasadena, CA 91125, USA\label{inst47}
    \and 
        Department of Physics, University of Bath, Claverton Down, Bath, BA2 7AY\label{inst48}
    \and 
        ZARM, Center of Applied Space Technology and Microgravity, Bremen University, Germany\label{inst49}
    \and 
        Ru{\dj}er Bo\v{s}kovi\'c Institute, Bijeni\v{c}ka cesta 54, Zagreb, Croatia\label{inst50}
    \and 
        Instytut Astronomiczny, Uniwersytet Wroc{\l}awski, Kopernika 11, 51-622 Wroc{\l}aw, Poland\label{inst51}
    \and 
        South African Astronomical Observatory, PO Box 9, Observatory 7935, South Africa\label{inst52}
    \and 
        Instituto de Telecomunica\c{c}\~oes, Campus Universit\'ario de Santiago, 3810-193 Aveiro, Portugal\label{inst54}
    \and 
        Departamento de F\'isica, Universidade de Aveiro, Campus Universit\'ario de Santiago, 3810-193 Aveiro, Portugal\label{inst55}
    \and 
        Leibniz-Institut f\"ur Astrophysik Potsdam (AIP), An der Sternwarte 16, 14482 Potsdam, Germany\label{inst56}
    \and 
        Astrophysics Research Centre, School of Mathematics and Physics, Queen's University Belfast, Belfast BT7 1NN, UK\label{inst58}
    \and 
        Department of Physics, University of Warwick, Gibbet Hill Road, Coventry CV4 7AL, UK\label{inst59}
    \and 
        INAF Osservatorio Astronomico di Padova, Vicolo dell’Osservatorio 5, 35122 Padova, Italy\label{inst61}
    \and 
        Department of Astronomy, University of Cape Town, 7701 Rondebosch, South Africa\label{inst62}
    }

\date{Received March 12, 2021; accepted May 20, 2021}


\abstract
{Since July 2014, the \gaia\ mission has been engaged in a high-spatial-resolution, time-resolved, precise, accurate astrometric, and photometric survey of the entire sky.}
{We present the \gaia\ Science Alerts project, which has been in operation since 1 June 2016. We describe the system which has been developed to enable the discovery and publication of transient photometric events as seen by \gaia.}
{We outline the data handling, timings, and performances, and we describe the transient detection algorithms and filtering procedures needed to manage the high false alarm rate. We identify two classes of events: (1) sources which are new to \gaia and (2) \gaia{} sources which have undergone a significant brightening or fading. Validation of the \gaia\ transit astrometry and photometry was performed, followed by testing of the source environment to minimise contamination from Solar System objects, bright stars, and fainter near-neighbours.}
{We show that the \gaia\ Science Alerts project suffers from very low contamination, that is there are very few false-positives. We find that the external completeness for supernovae, $C_E=0.46$, is dominated by the \gaia\ scanning law and the requirement of detections from both fields-of-view. Where we have two or more scans the internal completeness is $C_I=0.79$ at 3 arcsec or larger from the centres of galaxies, but it drops closer in, especially within 1 arcsec.}
{The per-transit photometry for \gaia\ transients is precise to 1 per cent at $G=13$, and 3 per cent at $G=19$. The per-transit astrometry is accurate to 55 milliarcseconds when compared to \gaia\ DR2. The \gaia\ Science Alerts project is one of the most homogeneous and productive transient surveys in operation, and it is the only survey which covers the whole sky at high spatial resolution (subarcsecond), including the Galactic plane and bulge.}



\maketitle



\section{Introduction}
\label{sec:introduction}

On 19 December 2013, the European Space Agency (ESA) launched its \gaia{} satellite, which was the start of an ambitious project to measure the parallaxes of a billion stars in the Milky Way. \gaia{} started scientific operations in July 2014 and completed the 5-year nominal mission on 16 July 2019, but the spacecraft is in good health and the data collection and processing is still ongoing as an extended mission phase. Although the final data release of the nominal mission is still to come (DR4, the extended mission will be released as DR5), the survey has already had a transformational impact on a broad range of fields, including white dwarfs \citep{GentileFusillo19}, hypervelocity stars \citep{Boubert18}, cosmological gravitational lensing \citep{Lemon19}, and the merger history of the Galaxy \citep{Belokurov18}.

In order to make the astrometric measurements, \gaia{} scans the full sky repeatedly. The exact number of observations and observing cadence of a given source depends on its location on sky, but each source will be observed $\sim$140 times over the lifetime of the survey \citep[see e.g.][]{boubert20}. Typically a pair of observations, separated by 106.5 minutes, is followed by another pair of observations two to four weeks later. Each observation consists of a 50-second long white-light ($G$-band) lightcurve, sampled every 5\,s, which can also be used for variability detection on very short timescales \citep{wevers2018, roelens2017, roelens2018}. Hence \gaia{} samples the sky on a range of timescales, allowing us to search these time series observations for transient variables. The detected transients are published as a public alerts stream, known as \gaia{} Science Alerts (hereafter {\em GSA}). Throughout the lifetime of the survey so far, \gsa{} has undergone several changes, in particular as more data became available, making it possible to introduce more reliable and efficient detection algorithms. This paper focuses on the current operational state, but it includes the details of important changes throughout the development. 

\gsa{} has been designed to produce notifications for transient phenomena, that is to say any event which would benefit from a timely reaction, and thus to avoid a potential science loss. \gsa{} is an added-value science product to the main astrometric goals of the satellite mission; the survey is not optimised for transient detection or completeness of transient populations. Nevertheless, \gaia{} has numerous advantages compared to ground-based transient surveys. Its space-based location means that biases due to weather or variable seeing are eliminated. It also benefits from a high dynamic range, high spatial resolution ($\sim0.1$\arcsec), high photometric precision (1\% at $G = 13$; 3\% at $G = 19$), and all-sky coverage, including the Galactic plane which most ground-based surveys avoid because of crowding. Each observation also includes a low resolution ($R\sim100$) `blue photometer/red photometer' (BP/RP) slitless spectrum, which provides colour information at every epoch. A comparison between \gsa{} and a number of other existing and planned transient surveys is presented in Table~\ref{tab:surveys}.
 
\begin{figure*}
\includegraphics[width=\textwidth]{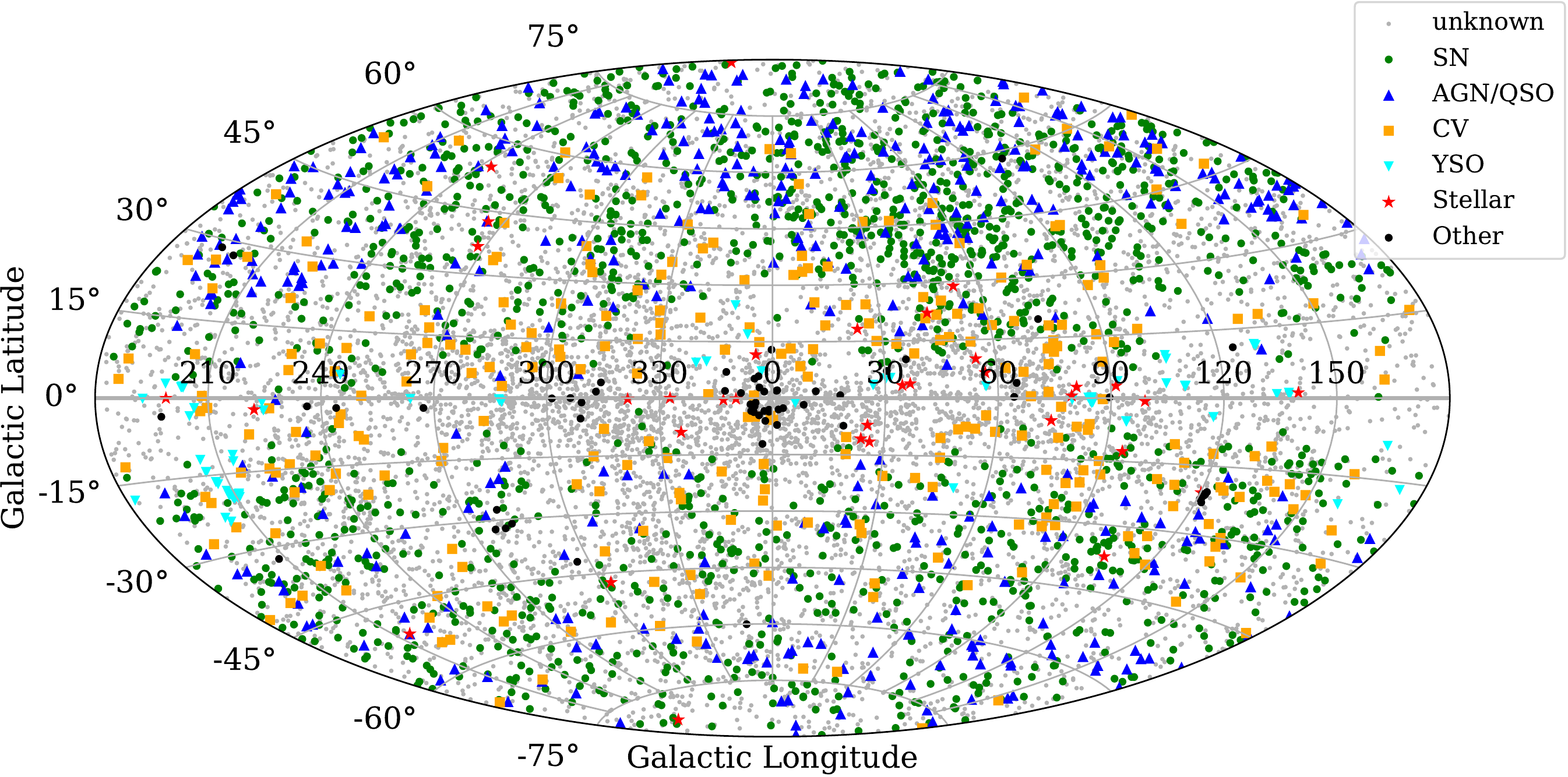}
\caption{Alerts detected by \gaia{} up to the end of 2019, plotted in Galactic coordinates. Alerts with unknown identifications are shown in grey, and spectroscopically confirmed alerts are highlighted in colour. The category `other' includes microlensing events, galactic novae and X-ray binaries.
\label{fig:alertsky}}
\end{figure*}

Up to 31 December 2019, 10765 alerts have been published, covering the full sky (Figure~\ref{fig:alertsky}). The alert detection is ongoing and currently alerts are published at a rate of 12 per day (see Section~\ref{sec:results}). We note that pulsating stars, regular variables, and eclipsing binary stars are excluded from this alert stream, as far as possible, as such variables are processed and published separately by the \gaia{} collaboration \citep[e.g.][]{Eyer19}.

\begin{table}[]
    \centering
    \begin{tabular}{lcccc}
        \hline
        Survey      & $\Omega_{\rm fov}$ & Platescale & m$_{\rm lim}$ & $\dot{\Omega}$ \\
                    &      (deg$^2$) & (arcsec)   &           & (deg$^2$ hr$^{-1}$) \\
        \hline
        Gaia (2 FOVs) &         0.9  & 0.06$\times$0.18  &      20.7 & 81.4 \\
        ASAS-SN     &          73.0  &             7.8   &      17   & 1294 \\
        ATLAS       &          60.0  &             1.9   &      20.0 & 5684 \\ 
        CRTS-2      &          19.0  &             1.5   &      19.5 & 1628 \\
        PS1         &           7.0  &             0.3   &      21.8 &  630 \\
        ZTF         &           47   &             1.0   &      20.4 & 3760 \\
        LSST        &           9.6  &             0.2   &      24.7 &  842 \\
        \hline\\
    \end{tabular}
    \caption{Comparison between \gaia{} and other existing or planned transient surveys (\citealt{bellm16}). For each survey we list the instantaneous field-of-view ($\Omega_{fov}$; \gaia{} has two fields-of-view), the size of the pixels, the limiting magnitude, and the areal survey rate ($\dot{\Omega}$). We note that \gaia{} and ASAS-SN are the only surveys which cover the whole sky. We further note that all the other transient surveys employ difference-imaging techniques to identify transients, while \gsa{} is a purely catalogue driven survey (see discussion in Section~\ref{sec:spurious}).}
    \label{tab:surveys}
\end{table}

Approximately 25 per cent of the \gaia\, alerts have been classified (Section~\ref{sec:results}), including previously known objects. The majority of classifications are from ground-based spectroscopic observations, with a small sample classified mainly through photometry (e.g. microlensing event classification includes a model fit to the lightcurves). The alert stream is currently available in its entirety to the public (\url{http://gsaweb.ast.cam.ac.uk/alerts}), so that alerts can be followed up on by anyone interested. Currently, the majority of spectroscopically identified alerts are supernovae due to large-scale supernova follow-up by, for example, PESSTO \citep[Public ESO Spectroscopic Survey of Transient Objects;][]{Smartt15}, NUTS (Nordic Optical Telescope Unbiased Transient Survey), and the Zwicky Transient Facility \citep[ZTF;][]{Bellm19}. Among the large number of Type-I and Type-II supernovae observed so far, \gaia\, also discovered a number of unusual supernovae, such as the extremely UV-bright super-luminous supernova (SLSN) Gaia16apd \citep{Kangas16,Nicholl17}, Gaia17biu, which is a hydrogen-poor SLSN and by a factor of almost 3, the nearest SLSN known to date \citep{Xiang17,Dong17,Bose18}, and Gaia16bvd, the first example of a pair-instability supernova \citep{Gomez2019}. \gsa{} is currently the second-largest contributor of transients to the IAU Transient Name Server\footnote{TNS; the official IAU mechanism for reporting new astronomical transients, \url{https://www.wis-tns.org/}}.

Other highlights so far include the discovery of the first fully-eclipsing AM\,CVn binary Gaia14aae \citep{Rixon14,Campbell15}, the fifth alert that was published by \gsa. The outburst that led to this discovery is the only outburst of this object that has been observed so far. Subsequent follow-up observations have resulted in high-precision measurements of the binary parameters that had not been possible for this class of object before \citep{Green18}. 

The high photometric and astrometric precision ($\sim$50 milli-arcseconds per transit) also makes \gaia\, sensitive to gravitational microlensing events, and several microlensing candidates have already been alerted on. Microlensing events occur when a star crosses our line of sight towards a distant background star and is observed as a temporary magnification of the background starlight. In 2016 \gaia\, detected the first binary microlensing event in the Galactic disc, Gaia16aye. The \gaia\, data, along with subsequent time series follow-up observations, afforded a full solution of the binary parameters, showing that this is a K giant doubly lensed by a main sequence binary \citep{Wyrzykowski2020}. The observations illustrate the potential for measuring the mass function of dark objects through microlensing.

\begin{figure}
\includegraphics[width=\columnwidth]{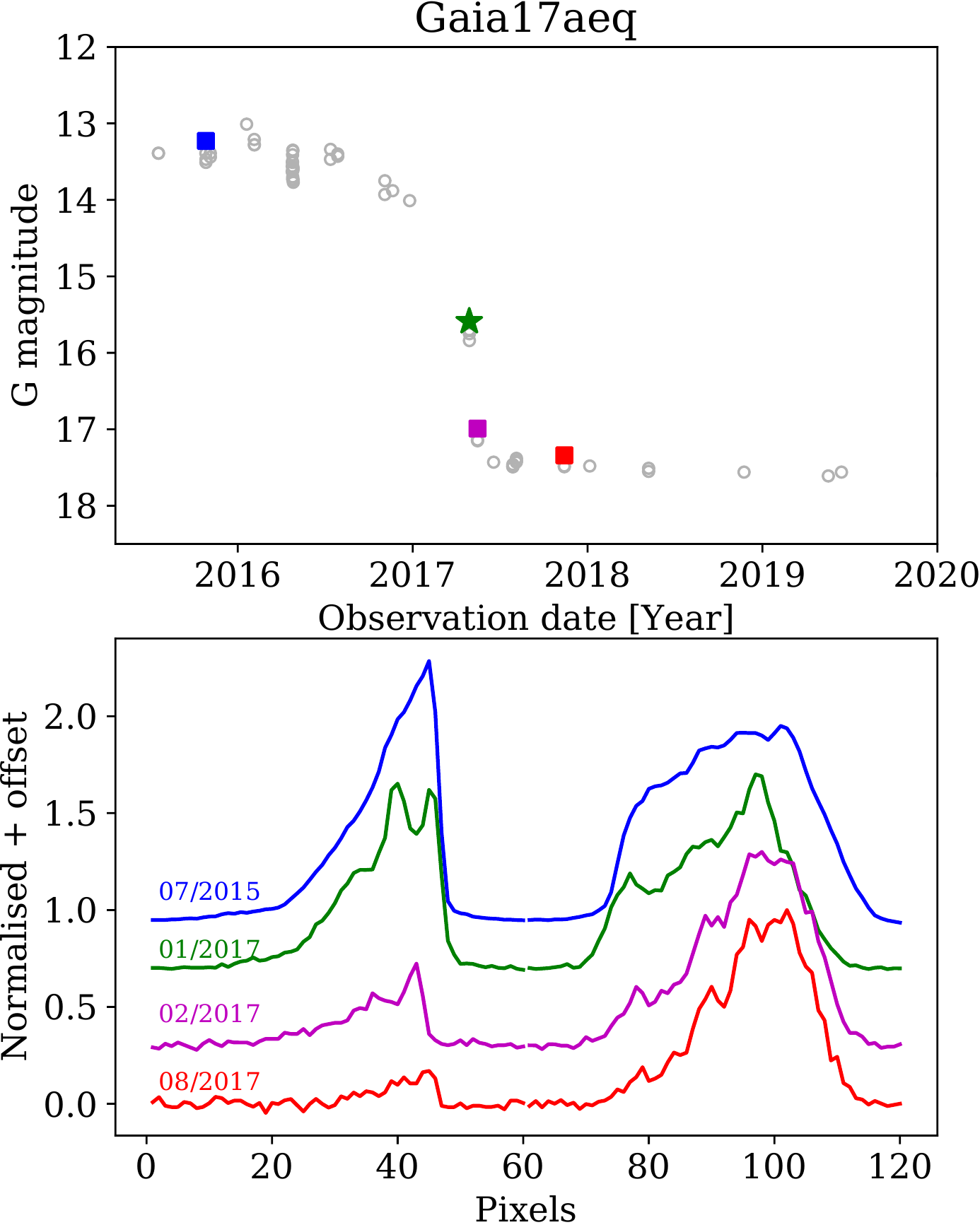}
\caption{Lightcurve (upper panel) and spectral variation (lower panel) of ASASSN-13db/Gaia17aeq. The points in the lightcurve for which the BP (lower left) and RP (lower right) spectra are shown are indicated with filled symbols in the same colour. The \gaia\, alert was issued when the target faded, at the point indicated by the star symbol.
\label{fig:17aeq} }
\end{figure}

A unique feature of \gsa{} is that it is also able to alert on sources that fade significantly. In this way, many new Young Stellar Objects (YSOs) and other `dipping' sources, such as VY\,Scl stars, have been discovered or alerted on. Gaia17aeq is shown as an example in Figure~\ref{fig:17aeq}. This is an EXor variable --- a YSO with a large proto-stellar accretion disc, characterised by large amplitude eruptive variability. It was originally discovered in outburst by the ASAS-SN survey as ASASSN-13db. A second, long-lasting outburst was underway when \gaia's nominal observations started
\citep[see][]{Sicilia-Aguilar17} and \gsa{} detected the accretion state change when it started to fade again towards quiescence (star symbol in Fig.\,\ref{fig:17aeq}). The time-series BP/RP spectra clearly illustrate the dramatic colour and spectral changes that accompany the flux variation in accretion events like these. ASASSN-13db/Gaia17aeq is the lowest mass star known to show outbursts like these \citep{Holoien14}. \citet{Kashi19} suggested that ASASSN-13db/Gaia17aeq may also be a luminous red nova, with the long-lasting outburst resulting from the disruption of the inner accretion disc or the accretion of a planet, but \citet{cieza18} confirmed its nature as an EXor variable, using ALMA observations of its dust disc. Several other YSO outbursts have been discovered as a result of the flaring activity observed by \gaia\, (e.g. Gaia18dvy -- \citealt{szegedielek20}, Gaia18dvz -- \citealt{Hodapp2019} and Gaia19ajj -- \citealt{Hillenbrand2019}). A detailed study of Gaia17bpi \citep{Hillenbrand2018} showed that this FU\,Ori-type outburst started in the mid-infrared, appearing at optical wavelengths approximately 1.5 year later. This is the first of these outbursts to be detected at both parts of the spectrum and it serves as direct tests of accretion disc models in these large discs.

Finally, \gaia\, is making contributions to the growing field of transients occurring in the very centres of galaxies (in spite of incompleteness in these regions, see \citet{Kostrzewa2018} and Section~\ref{sec:tnscomp}). One such event -- Gaia16aax -- has been detected in a galaxy hosting a known QSO where the centre brightened by about 1 mag over 1 year, before fading back to its pre-outburst state over more than two years. Both the photometric and spectroscopic variability show a dramatic change. The outburst of Gaia16aax can be explained by a change in the accretion flow onto the central black hole or could have been caused by a tidal disruption event \citep{Cannizzaro2020}.

In this paper we describe the operational state of the \gaia\, Science Alerts survey. Section~\ref{sec:dataflow} gives a full technical description of the data flow including the ingestion of the main data, the alert detectors, filtering methods, eyeballing and publication. The main results are described in Section~\ref{sec:results}, which includes a summary of the \gsa{} event rate, the photometric and astrometric precision of the candidates, and their main properties. The purity and completeness of the survey is discussed in Section~\ref{sec:discussion} and we summarise in Section~\ref{sec:summary}. We also include appendices with additional information on the cyclic processing of \gaia{} data and subsequent catalogue changes (Appendix~\ref{appx:idu}), the photometric calibration of \gsa{} (Appendix~\ref{appx:podc}), details of the computing cluster (Appendix~\ref{appx:comp}), and a complete list of abbreviations used in this paper (Appendix~\ref{appx:abbreviations}).

Throughout the paper, where we have performed analysis of the \gsa{} detection rates (and contamination rates), or considered the performance of the photometry or astrometry, we have set a fixed range of observational dates, encompassed in a fixed set of Initial Data Treatment (IDT, \citealt{Fabricius-1}) runs. These runs and dates are: run 1046 (earliest data point: 2016-07-11 04:45:53) to run 4724 (latest data point: 2019-12-30 09:35:49). The starting point was set as the point in time when the largest part of our system had stabilised. 

%
%
%
%

\section{Data flow: From observation to alert}
\label{sec:dataflow}

\begin{figure}
\includegraphics[width=\columnwidth]{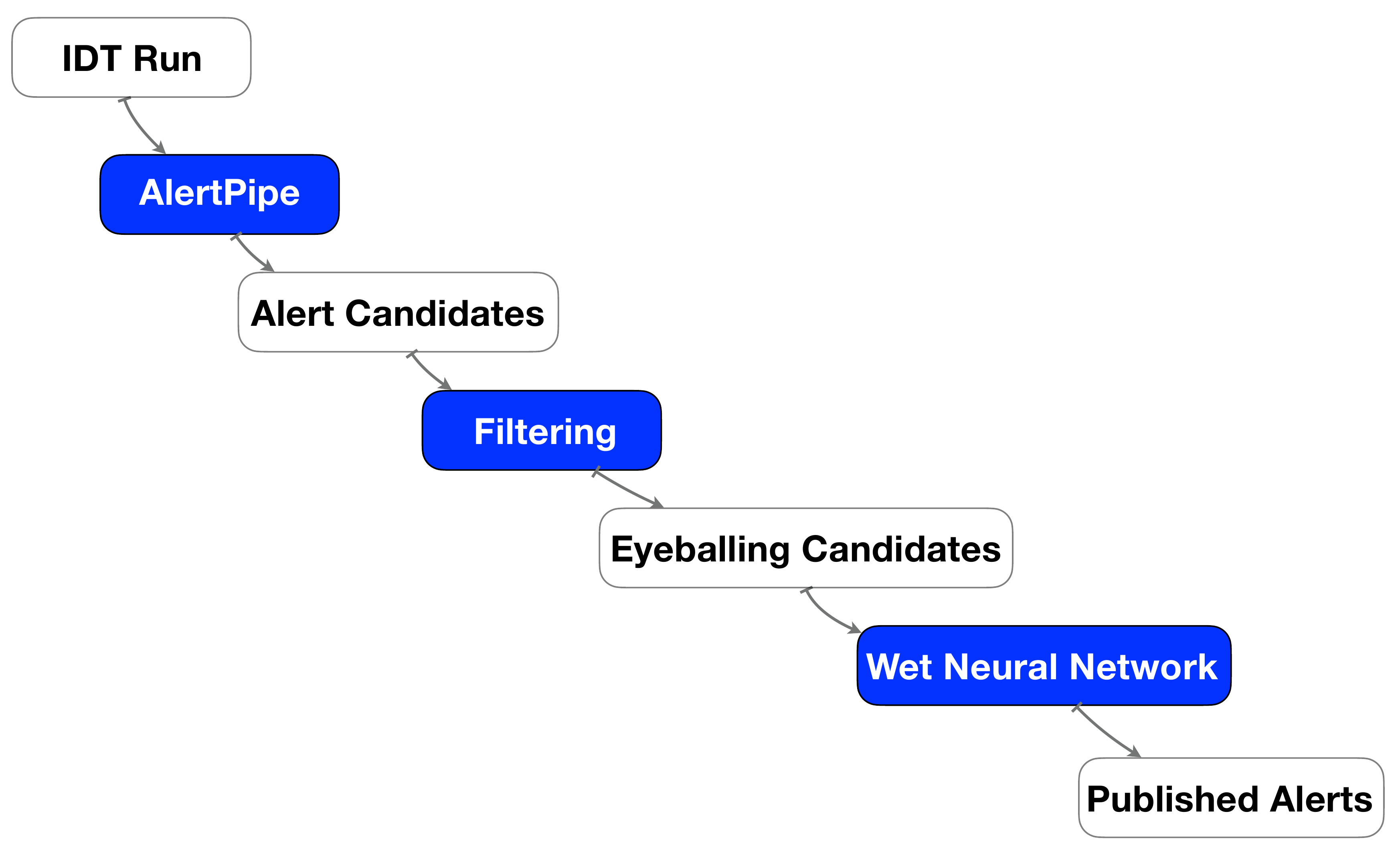}
\caption{Schematic of the data flow and processing performed by the \gsa\ project. Unfilled boxes indicate data, blue boxes (dark grey) show processes. {\em Wet Neural Network} refers to the eyeballing, voting and commenting process performed by humans.}
\label{fig:dataflow} 
\end{figure}

\gaia{} is at heart a time-domain experiment, measuring exquisitely precise astrometry and photometry with a well-defined observational depth and cadence. However, the daily processing of \gsa\ cannot accumulate, and iteratively calibrate, data in the same way that is used for the main \gaia\ data releases. In this section we discuss how \gsa\ proceeds from the on-board measurements taken by the \gaia\ spacecraft to the eventual publication of transient astronomical phenomena. We pay attention to how we curate the large data flow, apply simple calibrations, and filter out spurious detections, resulting in a viable and scientifically useful stream of transient events. An overview of the principal steps is described here (see also Figure~\ref{fig:dataflow}). Firstly, sources are detected and observed by \gaia\ as the spacecraft rotation and precession brings them through the fields of view (FOV, Section~\ref{sec:observation}). Next, observations are downlinked and forwarded via the Mission Operations Centre (MOC) to ESA's Science Operations Centre (SOC) for processing (Section~\ref{sec:downlink}). SOC collates the telemetry from \gaia\ and performs the Initial Data Treatment (IDT), extracting positions and fluxes of the sources from the pixel data. The results are copied to the various data-processing centres of the Data Processing and Analysis Consortium (DPAC, \citealt{mignard08}), including the one at Institute of Astronomy Cambridge (known as DPCI) where alerts processing takes place (Section~\ref{sec:idt}). \gsa\ processes the data of the current IDT run\footnote{Processing requires having all of the IDT output relating to a given source, but IDT's output is not organised cleanly by source, because to do so would be expensive and inefficient. In practice, the alerts pipeline runs once per IDT run (typically one day of observation) and cannot begin until the last output for that run arrives at DPCI.}, filtering the observations by quality, applying an on-the-fly photometric calibration, detecting transient features in the lightcurves, and flagging events suspected to arise from specific instrumental effects, as well as transients of astrophysical sources that are not worthy of alerts (e.g. known periodic variable stars and Solar System objects). This stage produces a list of candidate alerts (see Section~\ref{sec:lightcurves} and after for details). All data are handled by the \gsa\ PostgreSQL database which make use of the Quad Tree Cube (Q3C) software \citep{koposov06}. Further filtering removes the alert candidates that are probably due to interference effects from neighbouring sources (Section~\ref{sec:spurious}). Human inspection (i.e. eyeballing) identifies those candidates suitable for publication (Section~\ref{sec:eyeball}). Finally, the chosen alerts are published immediately to the World Wide Web via the Alerts Website, TNS entries and VOEvents (see Section~\ref{sec:publish}).

Each alert is published with a timestamp corresponding to the observation time by \gaia\ (in barycentric coordinate time, TCB) as well as the time of publication of the alert (in Coordinated Universal Time, UTC). The latency between the two timestamps is the sum of: (1) the time from observation until downlink of the data to MOC (commonly less than 12 hours, but significantly more in exceptional cases), (2) processing time at MOC and SOC, mainly in IDT (typically around 10 hours), (3) time for automatic processing at DPCI (typically from 3--6 hours, but rising to $\sim$24 hours for scans that run tangentially along the Galactic plane), and (4) time for human evaluation at DPCI (see Section~\ref{sec:eyeball}).

\begin{figure}
\includegraphics[width=\columnwidth]{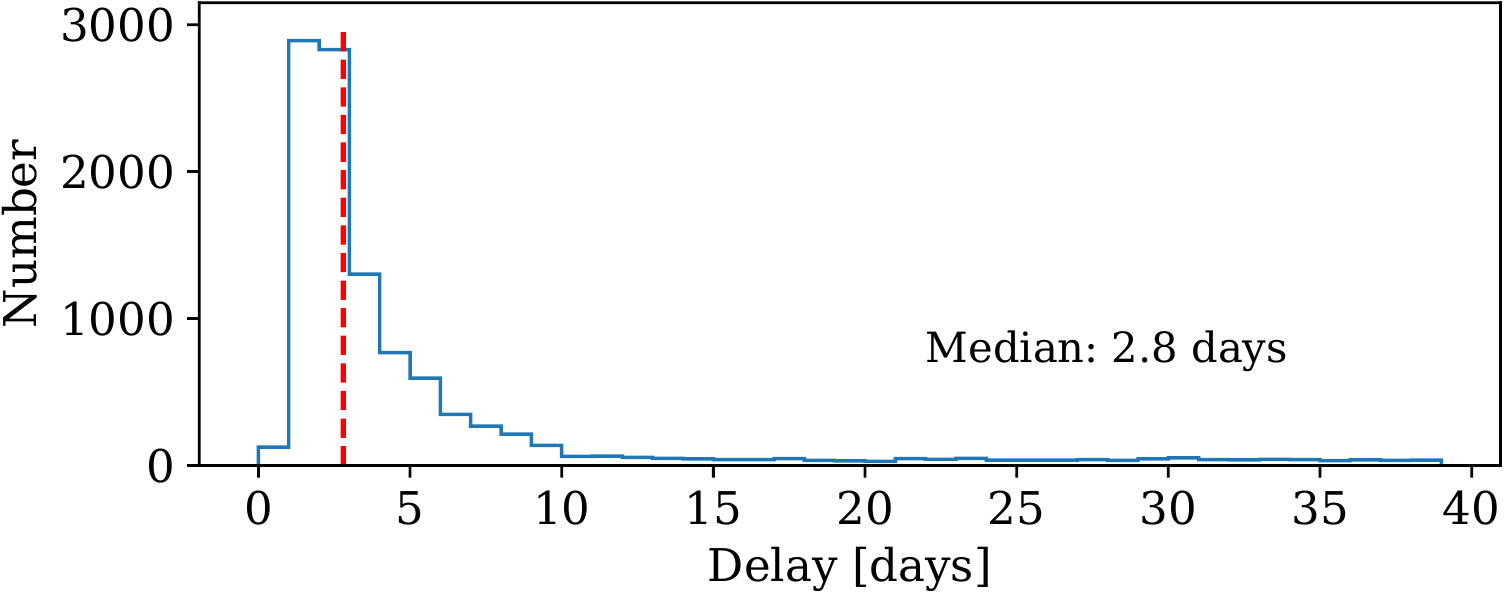}
\caption{Histogram of the delay (in days) between the \gaia\ observation and the publication of an alert. The long tail is the result of allowing the two FOVs that we require the alert to be seen in, to be separated by up to 40 days. 12 per cent of the alerts have a publication delay longer than 10~days. The median delay is 2.8~days and is indicated by a dashed vertical line. 
\label{fig:delay}}
\end{figure}

Alerts, therefore, typically appear between 24 and 96 hours after the triggering observations (median delay is 2.8 days, see Figure~\ref{fig:delay}). There is also a long tail, which corresponds to the delay between detection in two different FOVs (up to 40 days), discussed in Section~\ref{sec:newsource}.

\subsection{Observations and data types used in alerts processing}
\label{sec:observation}

\gaia\ is a drift-scan survey with two telescopes whose FOVs are separated by 106.5 degrees. The closely controlled rotation of the spacecraft scans the two FOVs, which are both pointed perpendicular to the spin axis, across the sky once every six hours. Precession of the spin axis, and the satellite's orbit around the Sun, varies the part of the sky observed on each rotation.

Each pass of a source across a FOV is termed a `transit', and this is the fundamental unit of observation. In a transit, a source crosses first the sky mapper (SM) CCD, then nine CCDs (except row 4 which has eight CCDs) of the astrometric field (AF), then the CCDs of the blue and red photometers (BP and RP) where the light is dispersed by prisms to obtain low-resolution spectra, then finally the grid of the Radial Velocity Spectrometer (RVS). The SM and AF measurements are in white light \citep[covering 330-1050nm, ][]{evans2018}.

The on-board algorithms responsible for the detection, selection and confirmation of sources are described in \cite{debruijne2015} and \cite{prusti2015}. The magnitude limit for retaining an observation is $G=20.7$.

In alerts processing, use is made of the following IDT data: (1) fluxes measured on each AF CCD; (2) positions of the source on each AF CCD along and across the scan direction, extracted by means of a PSF/LSF fitting (Point/Line Spread Function, \citealt{Fabricius-1}); (3) the calculated RA, Dec; (4) integrated fluxes for the transit in BP and RP, plus the colour derived from their combination; (5) individual pixels of the BP and RP measurements; (6) matching of transits to sources in the working catalogue; (7) status flags describing the reliability of the IDT results.

Alerts processing does not use the SM data, the raw pixel values from the AF CCDs, or the RVS data (although RVS data were reported for a small number of alerts, for a limited time, see Section~\ref{sec:publish}).

\subsection{Downlinking of data}
\label{sec:downlink}

Typically, data from the \gaia\ spacecraft can be transmitted to three ground-stations (operated by ESA) at Malarg\"ue (Argentina), Cebreros (Spain), and New Norcia (Australia). More recently, NASA Deep Space Network stations have also been used during some of the recent Galactic plane scans. The actual contact time is adjusted to match the predicted downlink data volume for the day, typically ${\sim}$8--10\:h, covered by one of the three antennae (two are used if the data rate is very high).

It is worth noting that the typical amount of (compressed) science data downlinked to the ground is some 40 gigabytes per day. Small onboard data losses (photometry and astrometry) can be caused by shortages of ground-station contact periods (e.g. in times when \gaia\ scans along the Galactic plane), amounting to zero for bright objects ($G< 16$ mag), a few per cent for $G=16-20$~mag, around 10 per cent for $G=20-20.5$~mag, and $\sim$25 per cent for fainter objects (see \citealt{prusti2015} for details). 

The MOC, located at the European Space Operations Centre (ESOC) in Darmstadt buffers the data packets and forwards them to the SOC near Madrid. SOC marshals the data into the standard formats of DPAC, and runs IDT.

\subsection{Initial data treatment}
\label{sec:idt}

The main role of IDT is to generate self-contained raw data records, extract the fluxes and centroids for SM, AF and BP/RP CCDs, and to match transits to catalogue sources \citep{Fabricius-1}. These processes are done in a time-constrained computer system where fully consistent processing is foregone in favour of prompt delivery to other data processing centres; both are subject to data artefacts that can cause false alerts. IDT breaks its operations into runs, where a typical run contains roughly one mission day of data. 

IDT reconstructs the spacecraft attitude to enable generation of the first on-ground attitude (OGA1), and thus the computation of source positions in sky coordinates (RA, Dec), to a required accuracy of $\leq$100 milliarcseconds (\citealt{Fabricius-1}, but see Section~\ref{sec:astrometry} for a discussion of the \gsa\ astrometric precision which we find to be ${\sim}55$ milliarcseconds). These reasonably accurate coordinates are used in the cross-match between the transits in the current IDT run and the \gaia\ working catalogue of the current data reduction cycle\footnote{\gaia\ Data Release 3 (DR3) is based on Cycle~03 processing, while the \gaia\ alerts included in DR3 are based on the Cycle~01 and Cycle~02 IDT working catalogues}. 

\subsubsection{IDT new sources}

A transit which can be associated with an IDT working catalogue source is assigned the appropriate \sourceid\ (defined in \citealt{BAS020}), while one that cannot, triggers the generation of a new \sourceid\, (which is added into this catalogue).

The magnitude limit for detection of a source by \gaia\ is $G=20.7$. Some 15--20 per cent of all \gaia\ detections are spurious detections on board (\citealt{Fabricius-1}), and $\sim$80 per cent of these cases are flagged in IDT. The most common causes of spurious detection include: diffraction spikes, bright sources from the other FOV, major planets (especially Venus), diffuse objects, duplicated detections, cosmic rays and hot CCD columns (see \citealt{Fabricius-1} for a detailed description of the causes and mitigation strategies). Occasionally, large numbers of new sources can be generated when the OGA1 attitude solution for the spacecraft suffers an excursion. This can arise when the spacecraft suffers disturbances from external micro-meteoroid hits. Later processing, and in particular the Astrometric Global Iterative Solution (AGIS, \citealt{lindegren16}) do a much better job of modelling these excursions, but these are beyond the timescale constraints of the IDT and \gsa\ systems. Rarely, \gaia\ also detects very large numbers of prompt particle events, associated with Solar coronal mass ejections. The largest such event occurred on 10 September 2017\footnote{https://blogs.esa.int/rocketscience/2017/11/03/unexpected-view-from-gaia-the-galactic-surveyor/}, which resulted in our system flagging spacecraft revolutions 5651.4 to 5659 as bad.

At the beginning of operations, the working catalogue was the Input \gaia\ Source List (IGSL, \citealt{smart14}), and the maximum radius for cross-match was set to 2.0~arcseconds. During this first phase, large numbers (up to millions) of new sources were generated in each IDT run, due to incompletenesses and inaccuracies in the IGSL, as well as from spurious detections. Over time, as the working catalogue has been improved, this cross-match radius has been reduced to 1.0~arcsecond. The number of new sources arising via this channel has been significantly reduced.

\subsubsection{Cyclical processing and catalogue update}

IDT is a daily process using a working catalogue which is updated as new detections arise. The primary data-releases of \gaia\ are derived from cyclical reprocessing of the whole data set, in which a new catalogue is formed from consideration of all observations. When IDT replaces its working catalogue with the new, cyclical catalogue, \gsa\ experiences disruption. This can lead to gaps in the alert lightcurves, or to \lcs\ being the union of observations of several sources in the new catalogue. See Appendix~\ref{appx:idu} for details.

\subsection{\gsa\ \lc{} processing}
\label{sec:lightcurves}

\begin{figure}
    \centering
    \includegraphics[width=\columnwidth]{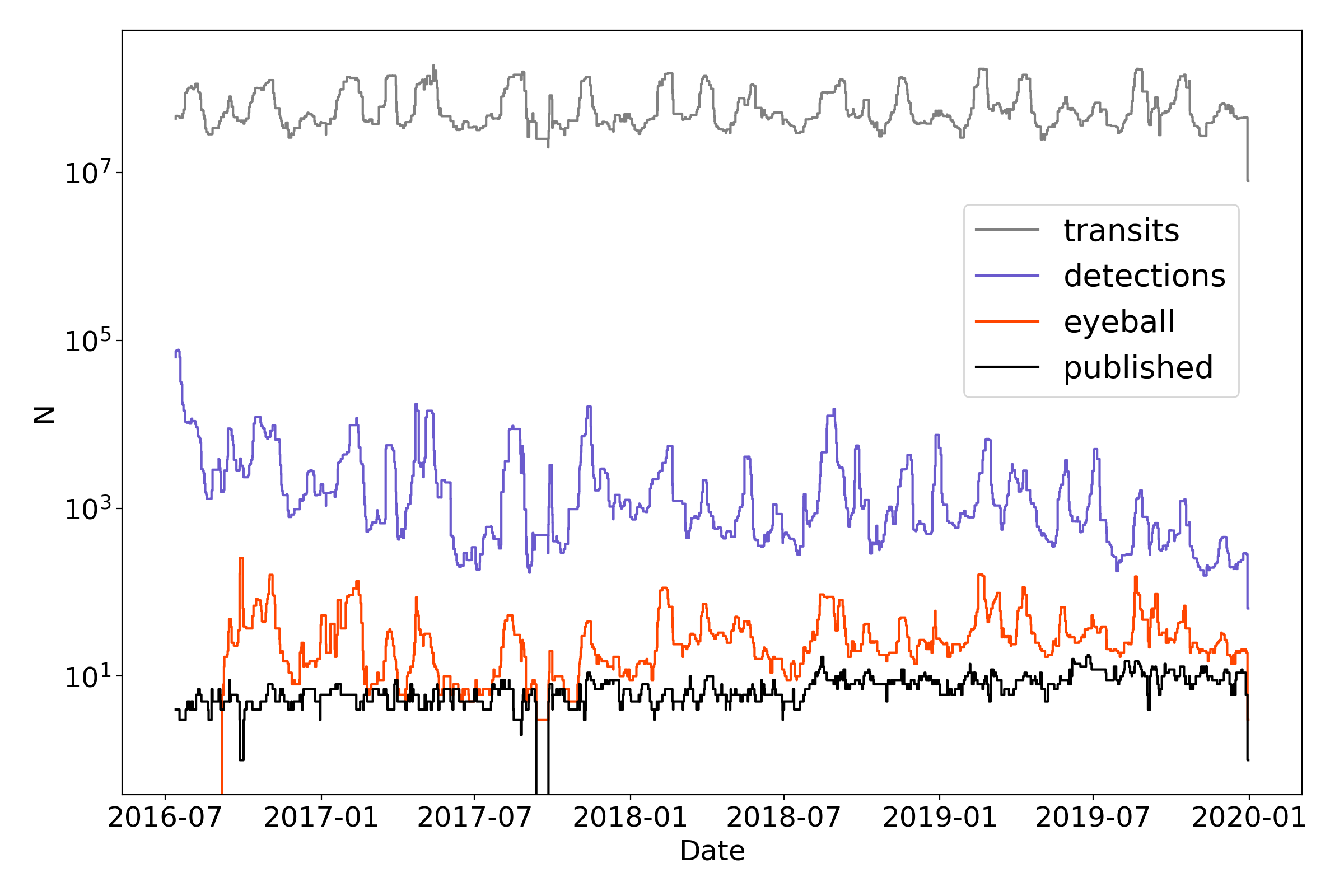}
    \caption{Histograms showing the time evolution of four quantities (all are totals per-IDT run): (1) numbers of transits processed by AlertPipe (in grey), (2) numbers of automated alert detections from AlertPipe (in blue), (3) numbers of alerts presented to eyeballers after additional (mostly environmental) filtering (in red), (4) numbers of alerts published from each run (in black). A 7-day running median filter has been applied to all totals. Note that no records of eyeballing statistics were preserved for the first months of 2016.}
    \label{fig:howmany}
\end{figure}

The data in an IDT run represent new observations of sources (identified by their \sourceid) in the \gaia\ working catalogue, and observations of new sources. \gsa\ processing starts with the building of \lcs\ for all these sources. A full \lc\ is the union of all observations assigned to that \sourceid\ by IDT over all runs, with photometric calibration applied on-the-fly, and precise to 3 per cent at $G=19$ (see Appendix~\ref{appx:podc}). A typical run contains on average 60 million observations (transits) arising from some 37 million sources, but the amount of data in each run varies widely according to the current direction of \gaia's scanning across the Galactic disc (see Figure~\ref{fig:howmany}).

With the lightcurves formed, \gsa\ processing proceeds to the evaluation of each source that has received new data in the current run. The principal steps for each lightcurve are: (1) Filtering to remove untrusted transits from the lightcurve (Section~\ref{sec:transitfilter}); (2) Detection of transients, using four different algorithms (see Section~\ref{sec:detection}); (3) Automated classification of transients to flag possible artefacts from instrumental effects and astrophysical transients not suitable for an alert. These can include  excursions in the spacecraft attitude from nominal pointing, proximity to minor planets, or already classified long period variable stars (from DR2 \citealt{Eyer19}, see also Section~\ref{sec:spurious}).

This gives a list of alert candidates for the run. A typical run produces a few thousand candidates (see Figure \ref{fig:howmany}). Over the duration of the mission, the number of candidates detected by \gsa{} processing has decreased, while the number of published candidates has increased, demonstrating a trend of increased efficiency.

\subsection{Filtering of bad transits}
\label{sec:transitfilter}
An observed transit may be eliminated from a lightcurve for a number of reasons. For example, the details (flux, position) of the transit may be flagged by IDT as improperly extracted from the pixel data. Alternatively, the transit may have been observed when \gaia\ was not in a stable state, as when the mirrors were being heated to remove condensates. Sometimes, the readout parameters may be inconsistent with the magnitude of the source\footnote{\gaia\ may observe multiple sources simultaneously on the same TDI (time-delayed integration) line (\citealt{debruijne2015}). The readout parameters are set to suit the brightest of such sources, and this may compromise observations of the fainter sources.}. Another example occurs when the scatter in the distribution of fluxes obtained from the individual CCDs is significantly higher than expected from photon statistics. This is evidence of interference from sources in the opposite FOV.

Where a transit is filtered, its flux is not used in detecting transients. If an alert is published for that source, the transit appears in the published lightcurve with no stated magnitude.

\subsection{Alert detection algorithms}
\label{sec:detection}

Transients are detected in the lightcurves formed from the white light $G$-band fluxes measured by the AF CCDs. Four detection algorithms are applied to detect different kinds of events.

\subsubsection{New source detector}
\label{sec:newsource}

This reacts to sources that brighten from below \gaia's detection threshold. A source not previously seen and rising to $G<19$ is considered as an alert candidate.

To defeat the many sources of systematic noise, some other criteria must be met for the detector to report a candidate. The source must be seen in both FOVs; many effects result in a spurious detection in only one FOV (see e.g. \citealt{wevers2018} and \citealt{Kostrzewa2020} for more detailed discussions). The location of the source must have passed through \gaia's FOV at least 10 times previously without detection (calculated using HEALpix with a resolution of $\sim$40 arcseconds). Due to instrumental and resource limits, not all transits of all sources are recorded, with fainter sources (in crowded regions in particular) more likely to be lost before transmission to Earth.

Observations of a newly-visible source may be split between IDT runs, and this would cause the detector to miss them if no single run contains detections in both FOVs. To avoid this, the detector aggregates all observations of the source in the current IDT-run and in all previous runs. Any transits in the current run and in the preceding 40 days are potentially from a new source. Any older transits are taken to indicate a previously known source and the \newsource{} detector is not triggered.

Requiring detection in both FOVs improves the cleanliness of the alert stream at the expense of completeness. An alternative approach~\citep{Kostrzewa2020} would be to alert on each detection of a previously unknown source: one alert per FOV transit. This would be suited to finding brief, faint transients such as possible optical counterparts of gravitational wave events, but at the expense of increased contamination. Work on implementing this detector is ongoing.

\subsubsection{Old source delta-magnitude detector}

This detects gross changes in the brightness of sources already in the IDT working catalogue. It reacts to the more extreme events (e.g. cataclysmic variables) but can also detect supernovae that are not resolved spatially from the nuclei of their host galaxies (where the galaxy is in the \gaia\ catalogue, otherwise this would be a \newsource{} alert). 

Measurements in the lightcurve obtained within 40 days of the most recent measurement are analysed for transient behaviour, while the mean and standard deviation of older measurements are taken as a historic baseline for comparison. To become an alert candidate, the lightcurve must have at least two transits that differ from the historic mean by at least one magnitude and by three times the standard deviation of the baseline.

The scatter of measured positions on the sky is used to rule out cases where transits of two separate (barely-resolved) sources have been mixed. To survive as an alert candidate, the source must have a standard deviation in position of less than 0.1 arcseconds. This may have a negative impact on transients arising in marginally resolved sources such as galaxies.

\subsubsection{Old source mean-rms detector}

This detector is similar to the old source delta-magnitude detector, above, but detects smaller changes in the quieter lightcurves. The minimum change in brightness is reduced to 0.15 magnitudes, but the deviating transits must change by at least six times the standard deviation of the baseline flux.

\subsubsection{Skewness/Von Neumann detector}
This detector, hereafter called \skewVonN{}, exploits the available source history to search for slower photometric variability. It was designed to cover a parameter space that is complementary to the other detectors. It is based on slicing a parameter space consisting of the third moment of the distribution of magnitudes (the skewness) and the von Neumann statistic $\eta$. The latter is defined as the ratio of the mean square successive difference to the variance \citep{Neumann1941} :

\begin{equation}
\eta = \frac{\delta^2}{s^2} = \frac{\frac{1}{n-1} \sum\limits_{j=1}^{n-1}{(m_{j+1} - m_{j})^2}}{s^2},
\end{equation}

where $n$ is the number of datapoints in each lightcurve, $s$ is the standard deviation of the lightcurve, and $m_j$ are measured magnitudes in the $G$-band.
A strong positive serial correlation between datapoints leads to a low von Neumann statistic, which signifies smooth variability, as opposed to single outliers or non-variable lightcurves which result in large $\eta$ values (see e.g. \citealt{wevers2018}, \citealt{Kostrzewa2018} for an application to \gaia\ data). The skewness metric can be used to remove stochastic / periodic variability.

One advantage of the \skewVonN{} detector is that it is well suited to finding relatively low amplitude events with high fidelity, such as microlensing events, variable AGN, and YSOs. The need for a sustained upward/downward trend in the lightcurve makes this detector robust against artefacts and outliers. The downsides are that (i) it requires sufficient history -- it was only brought into operation in May 2019 -- and (ii) several outlying data points are required before detection can be triggered, thus there is a delay between the start of the event and its detection. 

\subsection{Spurious alerts}
\label{sec:spurious}

To all intents and purposes, \gsa\ is a catalogue-driven transient survey, because two-dimensional AF pixel data are not available for the vast majority of sources. The strength of many of the extant ground-based transient surveys, including ZTF \citep{Bellm19}, ASAS-SN \citep{Shappee14} and PanSTARRS \citep{chambers16}, is that they employ difference-imaging techniques, thus the operators and users can ultimately inspect the images, and decide on the veracity of each event. For \gsa\ this lack of an image, and constraints on the release of \gaia\ data ahead of formal data releases, pushes us to deliver a high-purity alert stream, whereby a high degree of candidate vetting and rejection is performed in house.

Some statistics for the processing of \gsa{} are shown in Figure~\ref{fig:howmany}. Each run of AlertPipe handles on average 60 million transits for 37 million sources (maximum values can reach in excess of 300 million transits for 200 million sources). The vast majority of these measurements are not unusual, or are easily identifiable as spurious (e.g. big dippers, attitude excursions etc, see below), leading to a median raw detection rate of ${\sim}1000$ alerts per run, thus about 30 per million sources show anomalous flux behaviour. More detailed filtering, particularly exploring the environment of the candidates (see Section~\ref{sec:environment}) leads to a reduction in the median number of candidate alerts by a further factor of ${\sim}50$. Thus about 20 candidates per run survive to the phase of human eyeballing, and about half of these are published.

Not everything which is found by the detectors is something we wish to alert on and publish. There are many types of false positives, some of which are the real behaviour of real sources (such as periodic variables and asteroids), some of which are spurious behaviour of real sources (such as an increase in flux due to a bright star or planet lying nearby in the along-scan (AL) or across-scan (AC) direction from the source), and some are completely spurious sources (such as apparent new sources reported during attitude excursions, which are in effect the misplaced detections of old sources). Here we describe the mitigations we have put in place for some of the leading causes of false positive alerts. It is worth noting that there has been significant evolution in the rates of the differing types of false positives throughout the \gaia\ mission. These have arisen from (i) changes to the on-board \gaia\ detection parameters, (ii) improved mitigations in IDT, combined with updates to the working catalogue, (iii) evolution in our own understanding of the data and identification of spurious events. As an example, in the first half of operations during 2016, we employed a source-density map of the sky to reject all transients found in the most crowded regions (the density map was constructed from the \gsa\ database). Once more thorough environmental filters were developed and tested, the use of the density map was discontinued (i.e. by June 2016).

\begin{figure}
    \centering
    \includegraphics[width=\columnwidth]{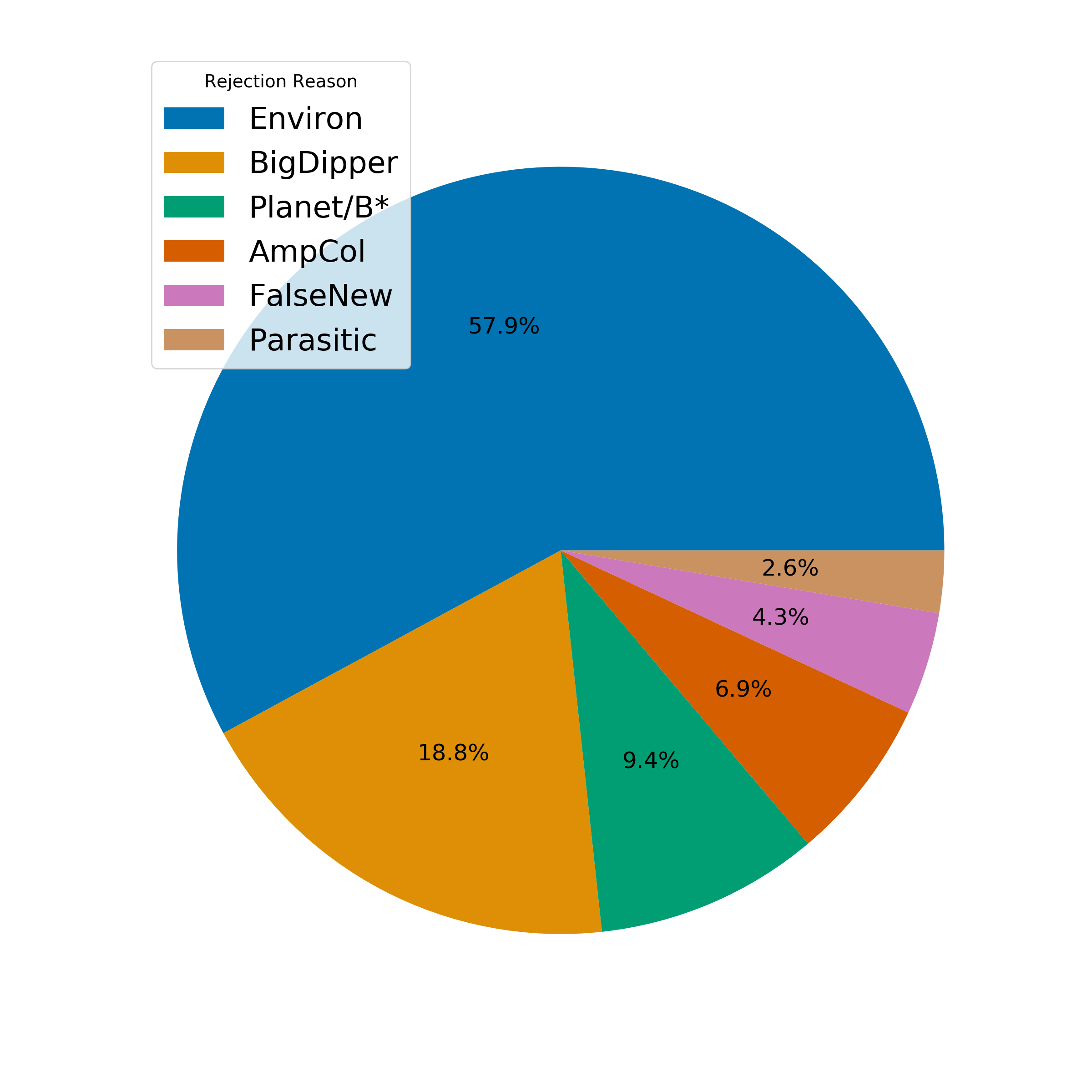}
    \caption{Most common classes of spurious alert that are rejected automatically by the \gsa\ system (Section~\ref{sec:spurious}). {\bf Environ:} alert rejected after assessment of near neighbour(s) within 10 arcsec, {\bf BigDipper:} alerting transit is in wing of bright star, which leads to a fainter measurement being associated with the star (i.e. the window is effectively misplaced, Section~\ref{sec:bigdipper}), {\bf Planet/B*: } alert likely caused by influence of bright star or planet in vicinity, {\bf AmpCol:} likely Mira-like variable on the basis of historic flux-scatter and extreme red colour, {\bf FalseNew:} source was new to the \gsa\ database, but not new to IDT. This could be due to missing or late arriving data for an IDT run which was not ingested into the database, {\bf Parasitic:} second FOV source affecting flux within transit for alerting source.}
    \label{fig:APpie}
\end{figure}

In the following sections we detail the most common types of false positive, which are either trapped and rejected in AlertPipe, or in two cases, flagged for inspection by the eyeball team. The most common types of automatically rejected alerts are summarised in Figure~\ref{fig:APpie}.  The two classes of candidate false alert that require a human decision are (i) Solar System objects (SSO), and (ii) variable star in \gaia\ DR2 (Section \ref{sec:varstar}). The first case is very rare, and almost always is unrelated to the alert (e.g. a faint SSO is reasonably close to a bright CV candidate). For the second case, the human almost always follows the flag, however due to occasional misclassifications in the variable catalogue, we do not automatically reject candidate alerts that are cross-matched to classified variable stars in Gaia DR2.

\subsubsection{Environment: Alignments in AL/AC directions, Planets and Bright stars}
\label{sec:environment}

While planets and the very brightest of stars can induce spurious alert candidates over a large area (${\sim}2$ degree radius, \citealt{Fabricius-1}), less bright stars still have an impact, albeit over a smaller area around their locations. 
A bright star can cause a spurious alert candidate as a result of flux from a diffraction spike entering into the window of the alert candidate, and producing an apparent increase in magnitude. The amount of additional flux depends on the magnitude of the bright star, the separation, and the alignment with respect to the orientation of the scanning direction (the spikes are asymmetric and aligned in AL and AC).

\begin{figure}
    \centering
    \includegraphics[width=\columnwidth]{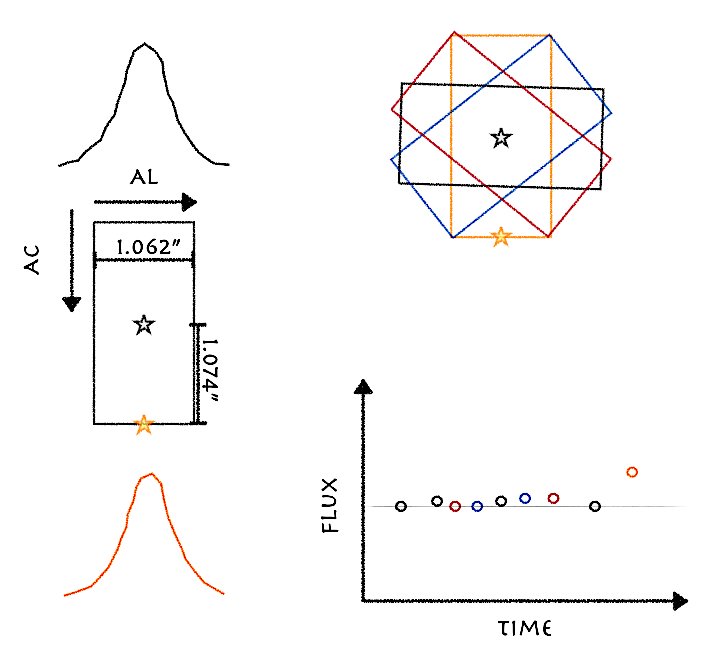}
    \caption{Transits of sources fainter than $G=13$ are one-dimensional with the assigned window divided into different samples in the AL direction with each sample spanning the full length of the window in AC. We illustrate, above and below the acquisition window (left panel), the approximate shape of the LSF provided through these one-dimensional samples. If a near neighbour aligns with the source in the AC direction this can cause an enhancement in the flux recorded for the source without disturbing the goodness-of-fit statistics of the transit in question. This sketch shows the impact of this arrangement on the lightcurve of the source; how the rectangularly shaped windows can capture flux from neighbouring sources in preferential directions, and hence how this may generate a spurious alert.}
    \label{fig:nearneighbourInAC}
\end{figure}

Alignments in the AC direction between the bright star and the candidate are particularly difficult to deal with as there is no noticeable impact on the goodness-of-fit statistics of the candidate. 
A common arrangement which produces a large population of spurious \meanrms\ candidates is a neighbour between 1 to 2 arcseconds away in the AC direction. This alignment can result in a significant amount of flux from the neighbouring source entering into the window of the alerting candidate source and producing artificial brightness variations.
This arrangement and its effects is illustrated in Figure~\ref{fig:nearneighbourInAC}.

The environment of every alert candidate, therefore, must be examined to reject such artificial flux variations. This assessment is performed in sky coordinates, rather than \gaia\ detector coordinates. 
Although sources from both FOVs can be adjacent in pixel coordinates, their differential motion will vary their separation, and thus lead to a variation in flux across the AFs (i.e. within a transit). These alerts are weeded out.
The amount of additional flux required to induce an alert candidate depends on the historical magnitude of the candidate and the detector type (faint \meanrms\ alert candidates are the most vulnerable to this effect). Mitigation of this effect may then be expected to depend on the detector, the historical magnitude of the alert candidate and the magnitudes and angular separations to neighbouring stars which are as bright or brighter than the alert candidate. 

The exact implementation of the environmental assessment is based on empirically derived magnitudes and angular separation distances as well as computational considerations. Extending the environmental search out beyond 10 arcseconds for every alert candidate becomes infeasible in terms of CPU time. 
Instead for the brightest of sources (planets and the top 30 brightest stars) the environmental search is done in reverse by finding all the alert candidates near them.
There is a subset of alert candidates caused by environmental effects, therefore, which could survive to the eyeballing stage (see Section~\ref{sec:eyeball}) should there be a bright enough star beyond 10 arcseconds. 
However, these are sufficiently few in number to be dealt with at that stage.

While not an every day occurrence, when the location of Jupiter, Saturn or Venus is near the scanning path of \gaia\ they can cause many spurious alert candidates by increasing the apparent fluxes of sources. The same is true for some of the brightest stars in the sky. Hence all alert candidates within 2 degrees of a planet or one of the top 30 brightest stars in the sky are assessed.

The local (within 10 arcsecond) environmental assessment is performed as follows:
\begin{itemize}
    \item{All alert candidates are rejected if they are not the brightest source by at least 1 magnitude in a 1.5 arcsecond radius about their median position. For \dmag{} and \meanrms{} alert candidates this radius is extended to 2 arcseconds.}
    \item{\newsource, \dmag{} and \meanrms{} alert candidates are rejected if there is a neighbouring source within 10 arcseconds which is brighter than $G=12$ mag.}
    \item{\newsource{} alert candidates with a source in the AL or AC direction within 10 arcseconds and $G<17$ may be rejected depending on the relative magnitudes of the two sources, as spurious non-blacklisted detections may still occur due to the AL/AC PSF spikes of these sources.}
    \item{The rejection criteria for \dmag{} and \meanrms{} alert candidates are stricter, as sources fainter than $G=17$ in the AL/AC directions may still cause a brightening in an existing source even if they cannot cause an entirely spurious detection. For these alert candidates, if the alert is due to the brightening of the candidate, any source within 10 arcseconds in the AL/AC direction may lead to a rejection, again depending on the relative magnitudes of the two sources.}
    \item{Note that \skewVonN{} candidates undergo less filtering as the detector is sensitive to long-term changes rather than short term ones produced by unfavourable alignments.}
\end{itemize}

All \dmag{} and \meanrms{} alert candidates with an historic magnitude fainter than 19, and within 2 degrees of a planet or one of the top 30 brightest stars in the sky, are immediately discarded.
For the remaining alert candidates, their positions in AL and AC with respect to the planet or bright star are evaluated, and any candidates in a predetermined box around the planet are rejected. The size of the box is determined by the area in which there is a clear excess of alerts.
The size of the box is larger in area for \meanrms{} candidates, than for \dmag{} candidates, whereas \skewVonN{} alert candidates use the same exclusion region as the \dmag{} candidates if the change in magnitude from the historic magnitude is less than 0.5 magnitudes. If the change in magnitude is greater than this, they are not automatically discarded.
The box size is always at least a degree wide in AL and always more than 1.5 degrees in AC.

\subsubsection{Big Dippers}
\label{sec:bigdipper}

Early on it was noticed that there were large numbers of alert candidates which had alerted due to their associated source having dimmed by several magnitudes. These sources were predominately in the magnitude range $13 \leq G \leq 17$. Further investigations revealed the position of the alerting transit to be offset from the median source position. It is thought that these observations are due to bright-star artefacts, where the on-board algorithm detects the spikes of the point-spread function, resulting in a fainter measurement for the same sourceId.

For brighter stars ($G<13$) these spurious offset transits are successfully blacklisted by IDT, at least in the vicinity of the star itself, and hence removed from the data-stream for AlertPipe. However, this proved not to be the case for fainter stars and additional processing was required to remove the resultant alerts from the list of candidates. This was done by evaluating the median position of the source and rejecting any alert caused by a drop in magnitude and a transit located more than 0.3 arcseconds from this position. Note a transit which brightens and is offset is not rejected, to allow the discovery of supernovae whose host galaxies are detected by \gaia. 
This source of spurious alerts was significantly reduced once IDT updated their algorithm to include the region around fainter sources when blacklisting transits which are due to this effect. 

\subsubsection{Attitude excursion: hits, clanks}

Large scale attitude excursions are rare events, but when they occur they can render the data unusable. In \gsa, an indicative measure of the reconstructed attitude is achieved per IDT run by accumulating the offsets in AL and AC of each transit of a \newsource{} alert candidate to the median position of the source in question (we recall that a \newsource{} alert must have at least two transits in order to be an alert candidate). The width of the distribution of the AL and AC offset may then be compared against that expected. These offsets may also be displayed as a function of time, highlighting periods of excess error. Diagnostic plots are created for each IDT run, and form part of the final verification process described in Section~\ref{sec:eyeball}.
Additionally, as large scale attitude excursions generate many spurious \newsource{} alert candidates, any \newsource{} alert candidate which does not have at least two transits located within 0.3 arcseconds from the median position of the source, is rejected automatically.
Smaller shorter-term attitude excursions rely on the final inspection step prior to publication for their rejection, where the location of the alerting transits are compared to those of the other transits belonging to the source.

\subsubsection{Prompt particle events and parasitics}

Prompt particle events (PPEs) are high-energy particles, such as cosmic rays or trapped protons from the Solar wind, which may cause noise in the signal read out from \gaia's CCDs. 
Parasitics are instances where a source from the other FOV happens to be projected onto the same location on the AF CCDs \citep{wevers2018}. As the AC rate is different for the other FOV, and thus the star-path is not parallel, this projection only contributes to a few of the AF CCDs along the transit rather than all of them.
It is for this reason that we require eight reliable (as defined using IDT's flags) AF flux measurements per transit and take the median value (and its error computed by median absolute deviation statistics) for the value of the transit's flux and its error.

In addition, for the \meanrms{} alert candidates, the goodness-of-fit (GoF) measures of the PSF/LSF to the transits are used as an additional means to reject suspicious candidates. The GoFs belonging to the alerting transits are compared against the expected GoF from the historical transits, and if there are too many significant outliers in the alerting transits the candidate is rejected.  
Note that the GoF has a magnitude dependence so this method is not applied to the \dmag{} alert candidates.

\subsubsection{Solar System objects}
\label{sec:sso}

As part of the DPAC processing system, the predicted \gaia\ transits of SSOs are calculated roughly every year and shared with \gsa\ (\citealt{mignard16}). The transit times are accurate to $<0.02$ seconds and account for planned changes to the \gaia\ Nominal Scanning Law.

If an alert candidate is found to be within 2 arcminutes of the expected location of a known SSO as seen by \gaia\ the candidate is flagged as a tentative match, if it is within 2 arcseconds then it is flagged as a probable match. An associated match probability is calculated, which depends not only on angular separation, but also on the magnitude difference between SSO and alert candidate.
The flagging does not remove the candidate automatically, but this information is retained for the final verification step prior to publications, see Section~\ref{sec:eyeball}, where the likelihood that the alert candidate is due to the observation of the SSO may be assessed.

\subsubsection{High amplitude variables: known and unknown}
\label{sec:varstar}
\gaia\ DR2 included classifications for more than 550\,000 variable stars, many of which are periodic (\citealt{holl18}). From 2018, we began to compare the \gsa\ candidates to these DR2 tables, and flag them if already classified (see Figure~\ref{fig:APpie}). But the DR2 candidates, drawn from 22 months of data, are not a complete sample, so additional strategies were devised to automatically identify the large numbers of high amplitude variables (such as Miras) which were still being seen in the \dmag{} alert candidates.

Searching for periodicity proved problematic given the poor and non-uniform sampling of the \lcs, however cuts in the colours and in statistics which are indicative of a high scatter in the \lc\ have proved useful in removing many Miras prior to the final verification procedure (see Section~\ref{sec:eyeball}).
These cuts were empirically derived using the data itself, selecting cuts on parameters which would remove as many candidates previously rejected by the final verification step as possible without resulting in the loss of any published alerts.
If the median colour (BP--RP) of the source is $>4.0$ and the median absolute deviation (MAD) of the magnitude is $>0.3$, the alert candidate is rejected.
Additionally, if the median colour is $>4.4$ and the kurtosis of the magnitude $>0.4$, the alert candidate is rejected.
This results in a reduction of the order of 40~per~cent in the number of high amplitude long period variables surviving to the verification (eyeballing) stage.

\subsubsection{Salvaged Alerts}
\label{sec:salvage}

Our filtering approach errs on the side of caution, to avoid placing excessive burden on the eyeballing process (see Section~\ref{sec:eyeball}). This would suggest that we generate a pure sample of events, but with reduced completeness (see Sections~\ref{sec:purity} and~\ref{sec:tnscomp} for more discussion). We know \citep{Kostrzewa2018} that an independent search for transients in galactic nuclei can find bona-fide events missed by \gsa, however the extra eyeballing required prohibits daily operation.

To mitigate against some of these lost events, we introduced (27 June 2017) a method for salvaging alerts discovered by our detectors, but rejected by the filters. There are four scenarios which we include, and which are passed to the eyeballers:

\begin{enumerate}
    \item Transients which are near a known galaxy in the LEDA catalogue (\citealt{makarov14}).
    \item Transients which are spatially-coincident with externally reported events. We maintain a comprehensive list of events discovered in other surveys within the \gsa\ database (see Section~\ref{sec:eyeball} for more details).
    \item An additional and independent filter (LWfilter) was brought into operation in July 2018. LWfilter uses auxiliary data from other surveys to classify the source (star, galaxy, AGN). Additionally, each alerting lightcurve is fit with a microlensing model (\citealt{paczynski96}) in order to identify potential microlensing events. BP--RP colour is also used to identify blue flares (e.g. CVs and Be stars) and very red variables (e.g. long-period variables, such as Miras). Alerts are then inspected visually (by the Warsaw team) and added into the list for eyeballing. Until the end of 2019 (IDT run 4724, i.e. spanning 18 months) this filter added 323 alerts for Eyeballing, primarily (85 per cent) from the \skewVonN{} detector.
    \item We also salvage candidates that are spatially coincident with a bespoke set of catalogues of YSOs (compiled by some of the authors). These include: (1) a catalogue of optically selected YSOs, (2) a catalogue of YSOs based on Spitzer observations (compiled from published articles), (3) a catalogue of confirmed YSOs from the Spitzer c2d survey \citep{Young15c2d}, (4) a catalogue of candidate YSOs from the Spitzer c2d survey, and (5) a list of candidate YSOs published by \cite{marton19}.
\end{enumerate}

Salvaging does not make a large difference to the numbers of alerts we publish. In a 2 year period from 1-Jul-2018 to 30-Jun-2020 (IDT runs 4026--4956), we published a total of 7568 alerts, of which 945 (around 12 per cent) came through the salvaging route. The breakdown for the 4 channels listed above are: {\bf 1.} 187 events, {\bf 2.} 365 events, {\bf 3.} 340 events, {\bf 4.} 53 events. Over half of the candidates were rejected by the filters because there were nearby neighbour sources, or because the IDT cross-match split the event across multiple sourceIds (sometimes in error).

\subsection{Eyeballing}
\label{sec:eyeball}

After detection and filtering, surviving alert candidates are subject to human evaluation using a web application, the Eyeballing App. This presents team-members with a series of figures and charts displaying \gaia\ and ancillary data. These data are used by the eyeballer to rank the candidate with a score between -1 and +1. A comment box is provided for the eyeballer to describe the event for the community\footnote{The comment is limited to 100 characters, and draws on the eyeballer's experience to try to describe the event as succinctly as possible. The eyeballer may sometimes make an estimate of a possible classification.}, and a dialogue box enables internal discussion between eyeballers. Votes from a minimum of two eyeballers, with a net score of +2, are required for an alert to be deemed publishable. A total of 15 people have contributed to the eyeballing of \gaia\ Alerts over the duration.

The \gaia\ data made available to the eyeballer include: The calibrated lightcurve, including the photometric scatter within a transit; The line spread function goodness-of-fit vs. time of the alerting source, derived from the image parameter determination in IDT \citep{Fabricius-1}; All near-neighbour \gaia\ transits within 10 arcsecs of the alert, projected in RA-Dec and AL-AC directions; Radial distribution of all neighbour transits out to 10 arcseconds (magnitude versus separation); Uncalibrated BP/RP spectra showing the evolution of the source before and after alert (if available); The probability of a known Solar System object crossing the FOV; A flag if the source is already classified as a long-period variable star in \gaia\ DR2 (\citealt{holl18}); \gaia\ DR2 parameters (including parallax, proper motion, BP/RP colour); HR diagram with the candidate superimposed (when possible).

Between 2014 and August 2018 we also applied a classifier (GS-TEC, \citealt{blagorodnova14}) to the raw BP/RP spectra, and shared the results with the eyeballers. GS-TEC takes a Bayesian approach to model observed spectra, using a constructed reference spectral library and literature-driven priors. GS-TEC can classify SN, AGN and stars down to $G$=19, however the classifier was disabled due to its significant execution time.

Auxiliary data are parsed from a variety of sources, and presented to the eyeballers, to help understand the context of a \gaia\ transient detection. 

\begin{figure}
    \centering
    \includegraphics[width=\columnwidth]{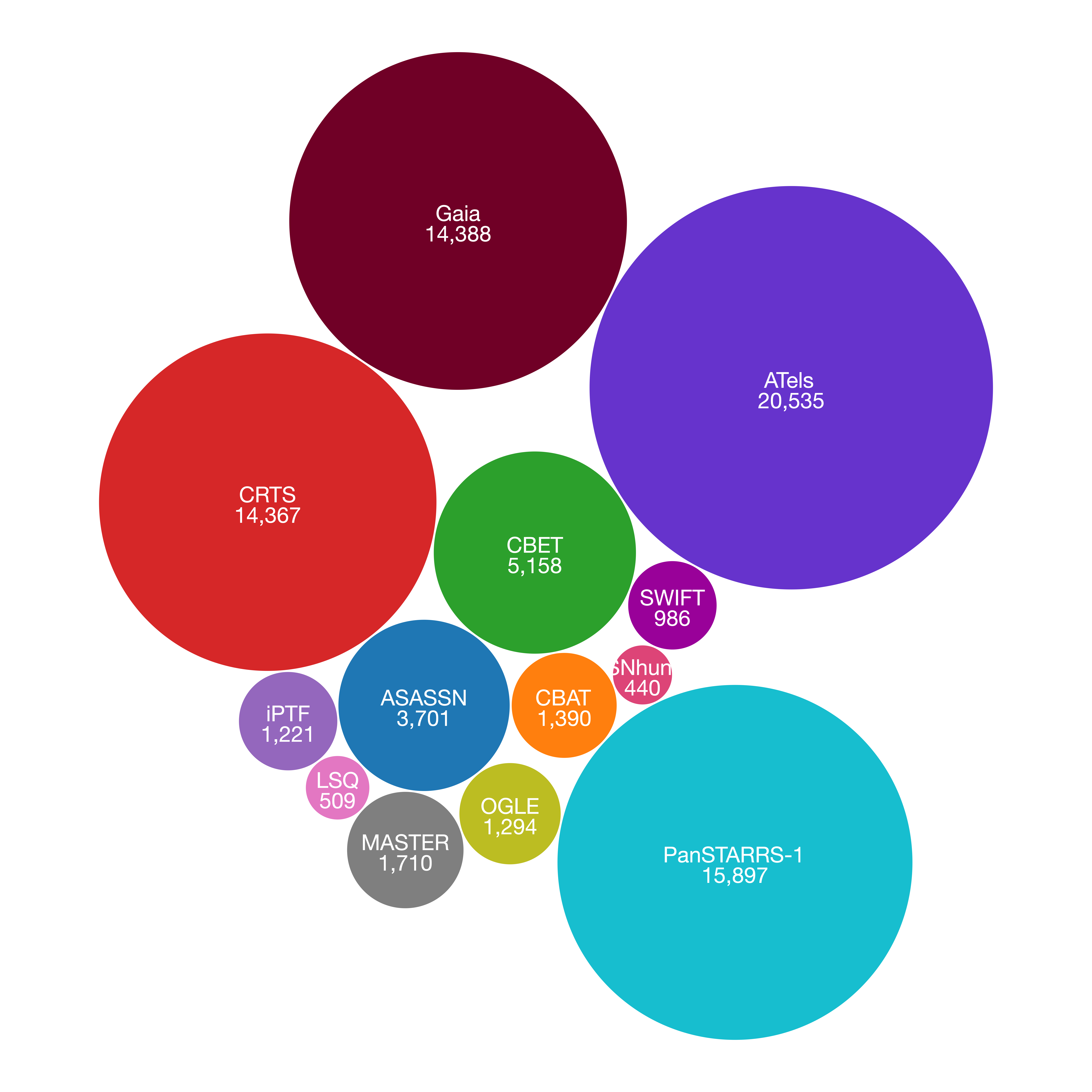}
    \caption{Circles are proportional in area to the unique numbers of objects/events (also shown in text), compiled for the \gsa\ database, and colour-coded by the data source. The circle for \gaia\ is shown for comparison. References for the surveys are given in the text. The data are taken from a snapshot of our archive on 24 November 2020.}
    \label{fig:bubble}
\end{figure}

\begin{itemize}
    \item To allow a visual inspection of the alert’s location, the Eyeballing App shows the Aladin Lite \citep{Boch14} and SDSS finding charts.
    \item Results of positional queries to the Simbad \citep{simbad}, NED and VSX databases\footnote{The NASA/IPAC Extragalactic Database (NED) is funded by the National Aeronautics and Space Administration and operated by the California Institute of Technology. VSX is the International Variable Star Index database, operated at AAVSO, Cambridge, Massachusetts, USA}, to determine whether it is an already-known transient or variable object.
    \item The list of YSOs described in Section~\ref{sec:salvage}.
    \item To aid the rejection of spurious transients arising from contamination by Solar System objects, we also display data on nearby planets, their satellites, and minor planets. In the early phases of \gsa\ we used SkyBot (Berthier et al. 2006), but we now exploit ephemerides shared within DPAC (see also Section~\ref{sec:sso}).
    \item Results of positional cross-match against our own tables of transient events, assembled from the hourly parsing of a significant collection of other publicly available transient surveys. An ETL (Extract Transform Load) system gathers discoveries reported by the major transient survey websites: Transient Name Server (TNS), Catalina Real-Time Transients (\citealt{drake2009}), ASAS-SN (\citealt{Shappee14}, Pan-STARRS1 (\citealt{kaiser10}), OGLE IV (\citealt{kozlowski13}, \citealt{wyrzykowski14}), MASTER (\citealt{lipunov10}), iPTF (\citealt{law09}), La Silla Quest (\citealt{baltay13}) and IAU Central Bureau for Astronomical Telegrams (CBAT\footnote{\url{http://www.cbat.eps.harvard.edu/index.html}}). Every hour, a total of 27 websites are scraped for data that are transformed, cleaned, homogenised and stored in the \gsa\ database. In a similar manner, Astronomer’s Telegrams are automatically parsed and stored in the database, accounting for the very diverse formats in the content of these HTML pages. The data stored in the \gsa\ database for the external transient surveys is shown in Figure~\ref{fig:bubble}.
    \item These data also contain classification information for large numbers of transient events which are shared with the eyeballer, and used at the point of publication. Classifications often arrive to the database after publication of an alert. As part of the publisher app, these can be viewed and the alert record updated (at the discretion of the operator). The bulk of classifications are reported via TNS (supernovae for the most part), but we also receive classifications on microlensing events from the Warsaw group through the publisher app (see \citealt{Wyrzykowski2020}).
\end{itemize}

\subsection{Publication}
\label{sec:publish}

Once eyeballing is complete, successful alerts are made publicly available to the astronomical community in several formats: via a dedicated website in CSV, HTML and RSS formats with permanent URLs for every published alert\footnote{\url{http://gsaweb.ast.cam.ac.uk/alerts}}; via the IAU-Transient Name Server\footnote{\url{https://www.wis-tns.org/}}; as VOEvents using the 4 Pi Sky broker\footnote{\url{https://4pisky.org/voevents/}}. The \gsa\ web application (\citealt{delgado17}, \citealt{delgado19}) has a public facing side where a set of the information is published, and a restricted area for administration and bookkeeping. The \gaia\ alerts catalogue can also be visualised on an All-Sky interface developed using Aladin Lite \citep{Boch14} enabling the display of alerts by time or individually.

\begin{figure}
    \centering
    \includegraphics[width=\columnwidth]{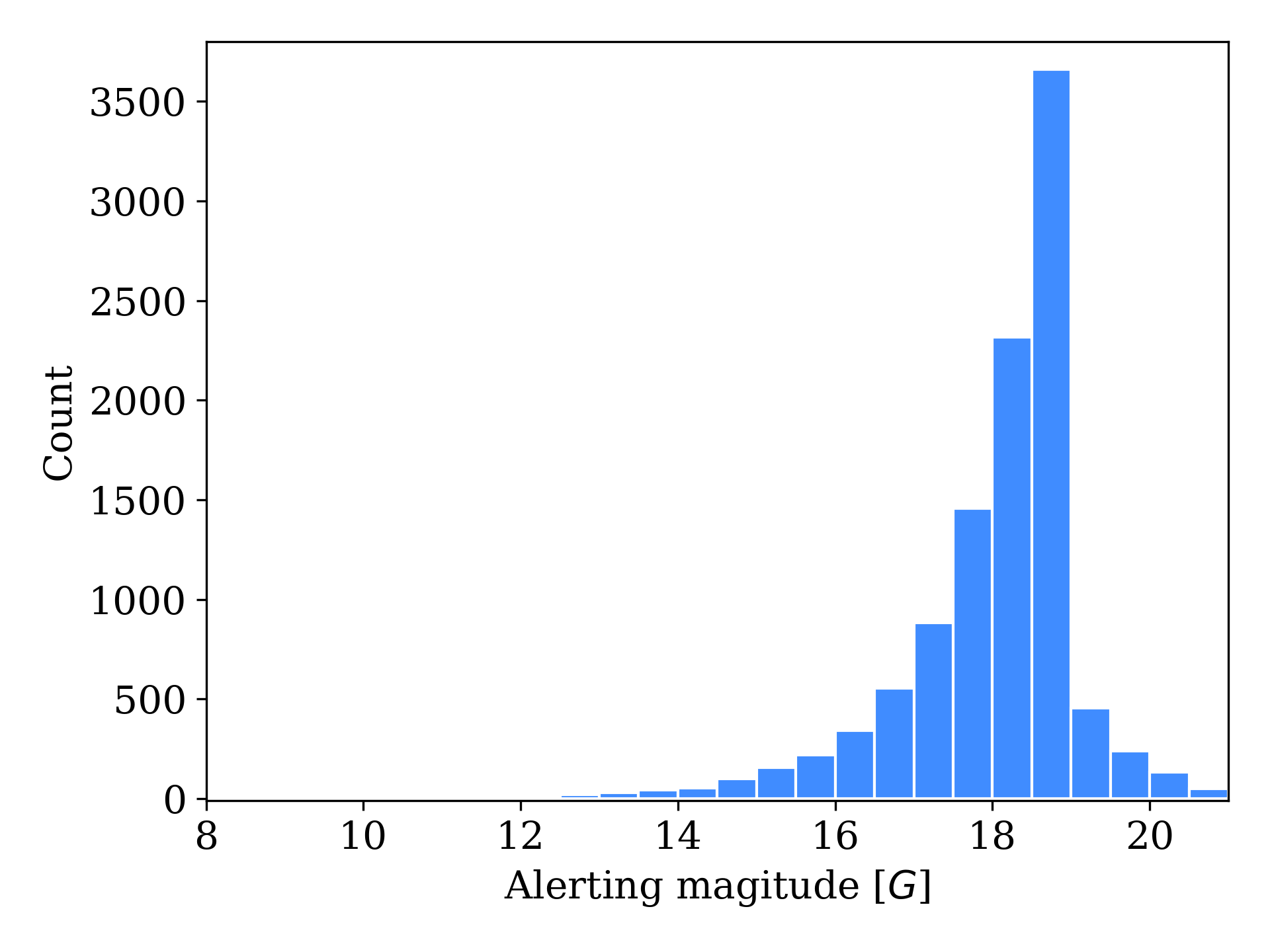}
    \caption{Histogram showing the number of published alerts as a function of the alerting magnitude, covering the start of operations (September 2014) until the end of 2019.}
    \label{fig:maghist}
\end{figure}

For a limited time, the set of information published also included a small number of RVS transit spectra (27 spectra for 12 alerts: see \citealt{seabroke20}). This number is small for several reasons: (1) most detected alerts are much fainter than the limiting magnitude of RVS ($G_{\mbox{RVS}} = 16.2$ mag, while $G\sim17$ mag for the alerts) see Figure~\ref{fig:maghist}; (2) RVS covers only four of the seven Gaia CCD rows; (3) RVS spectra have much lower signal-to-noise than the other Gaia measurements at the same magnitude; (4) the pipeline used to produce the RVS spectra of alerts did not process blended windows or take into account flux from sources without windows (an issue because the majority of alerts with RVS spectra are close to the Galactic plane). The RVS pipeline now treats these issues and all RVS transit spectra will be published in Gaia’s fourth data release. This should provide additional useful diagnostic information for the brightest alerts.

Once an alert is published for a source, the alert page is permanent. New data from \gaia\ concerning that source are added to its lightcurve as they become available. Hence, the published description of the source is mutable and represents the most recent information available. The state of the source at the time of the alert is preserved in the VOEvent document released to the 4 PI SKY event-broker \citep{staley16} at the time that the alert is first raised. Once an alert is raised on a source, no second alert can be raised on the same source, even when subsequent events occur, for example in the case of repeated outbursts. There are a handful of exceptions (e.g. Gaia16acr$\equiv$Gaia16adx and Gaia16ade$\equiv$Gaia16aey) where new events in the same source are attached to a new \sourceid{} arising from the IDT cross-match algorithm (see Section~\ref{sec:idt}). Note that these duplicated alerts will also be included in the DR3 data release.

If an alert candidate does not pass the aforementioned filtering/eyeballing steps, future observations can raise another alert for the same source, which will then be re-evaluated, possibly leading to publication. Between the IDT runs 1046 and 4724 inclusive there were 556 published alerts which had previously alerted but were not published at that time (out of a total of 9969 alerts, i.e. 5.6 per cent).

%
%
%
%

\section{Results}
\label{sec:results}

\subsection{Alert rate}

\begin{figure}
\includegraphics[width=\columnwidth]{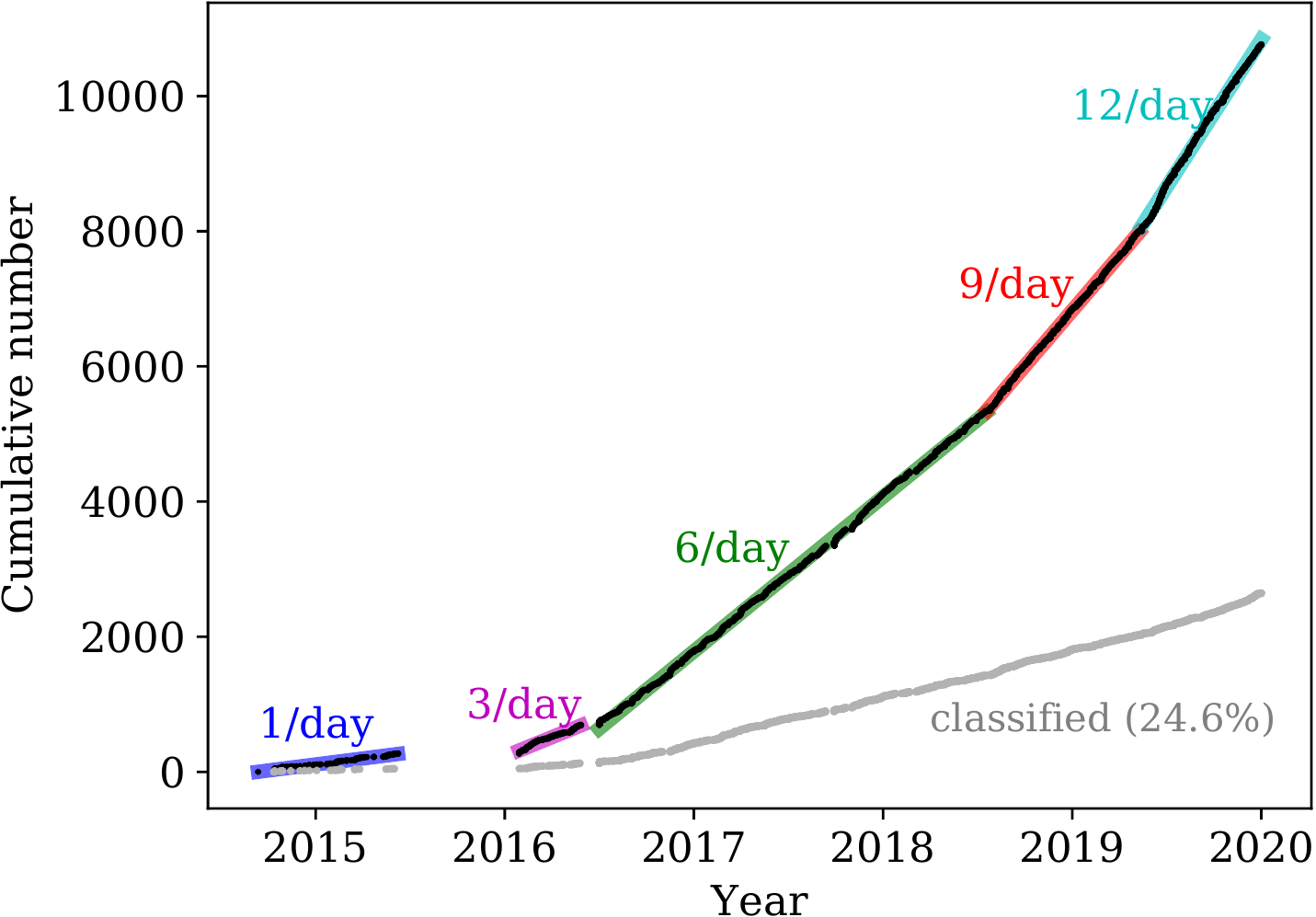}
\caption{Cumulative number of alerts detected as a function of publication date. Changes in the alert rate correspond to changes and improvements to AlertPipe, as described in this paper, which allowed us to identify alerts more reliably. Grey line shows the cumulative number of classified alerts, reaching almost 25 per cent of all alerts by the end of 2019 (see Section~\ref{sec:class}). Figure includes alerts published from the start of operations until the end of December 2019.}
\label{fig:rate}
\end{figure}

The cumulative number of published alerts is shown in Figure~\ref{fig:rate} and shows a number of distinct phases (indicated with different colours in the figure):

\begin{itemize}
    \item {\bf From September 2014 -- June 2015:} An initial commissioning  phase, spanning almost 300 days, where we published alerts at a rate of about one event per day.
    \item {\bf From June 2015 -- January 2016:} A pause in publication, where we developed most of the filters discussed in Section~\ref{sec:environment}, to minimise the rate of spurious detections.
    \item {\bf January 2016:} Restarting of operations for AlertPipe and Alert publication. A density map was implemented to ignore events arising in the most crowded regions of sky.
    \item {\bf June 2016:} Following further improvements to the filters, and removal of the density map restriction.
    \item {\bf April 2018:} Eyeballing App introduced.
    \item {\bf May 2019:} \skewVonN{} detector introduced.
\end{itemize}

Figure~\ref{fig:maghist} shows the magnitude distribution of alerts detected and published by \gsa. Some transients have alerting magnitudes fainter than G=19, our nominal threshold for detection. This can happen for a number of reasons: (1) the first detection of a brightening source is fainter than G=19, but subsequent measurements are brighter, (2) a source which was previously brighter than G=19, fades, (3) in earlier phases of operations we allowed publications of fainter transients.

\subsection{Alerts as a function of class}
\label{sec:class}

A long-standing problem with transient surveys is the rate of classification compared to the (usually much higher) rate of discovery. For \gsa{} we see that almost 25 per cent of alerts discovered up to the end of 2019 were classified. These classifications were obtained from numerous sources (including Simbad, ADS, TNS, and Astronomical Telegrams as described in Section~\ref{sec:eyeball}) and may well be incomplete. The majority of classifications were obtained spectroscopically (and are dominated by SNe). However in the case of microlensing events, a classification could be derived from modelling of the lightcurve alone. An analysis of all events reported to the Transient Name Server in 2019 \citep{kulkarni20}, showed that only around 10 per cent of events were classified. This is typically limited by access to ground-based facilities, where the problem scales with magnitude as shown in Figure~4 of \citet{kulkarni20}, that is fainter objects are less frequently classified (although they also note that there is a bright tail of events, dominated by \gsa, which remain unclassified and are likely stellar in origin). This is supported by Figure~\ref{fig:map_classified}, but see also Section~\ref{sec:purity}.

\begin{figure}
    \centering
    \includegraphics[width=\columnwidth]{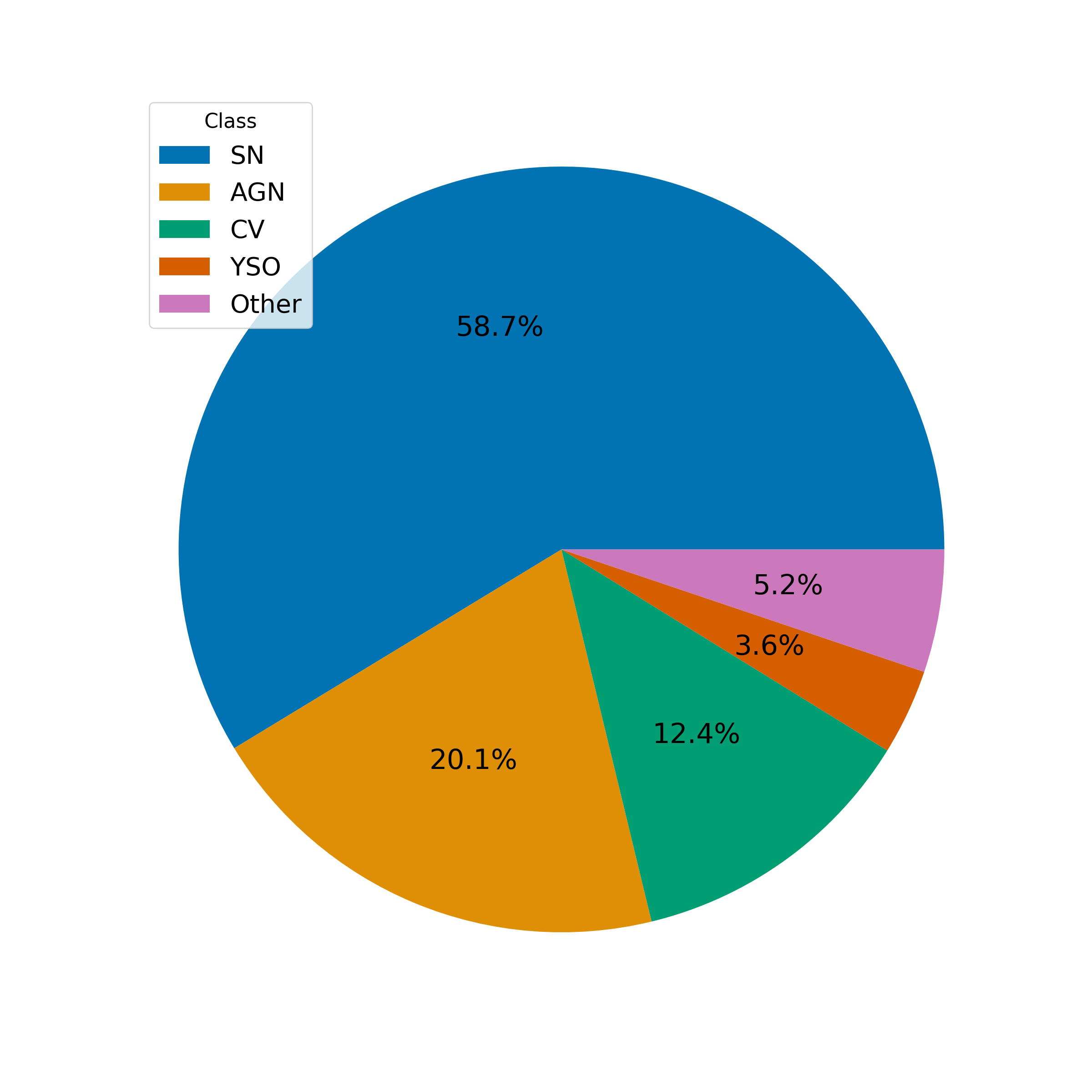}
    \caption{Distribution of the most common classifications for alerts from \gsa. As discussed in the text, there is strong bias in the rate of follow-up and classification in favour of events which look like supernovae or other extragalactic transients. }
    \label{fig:class}
\end{figure}

\begin{figure*}
    \centering
    \includegraphics[width=\textwidth]{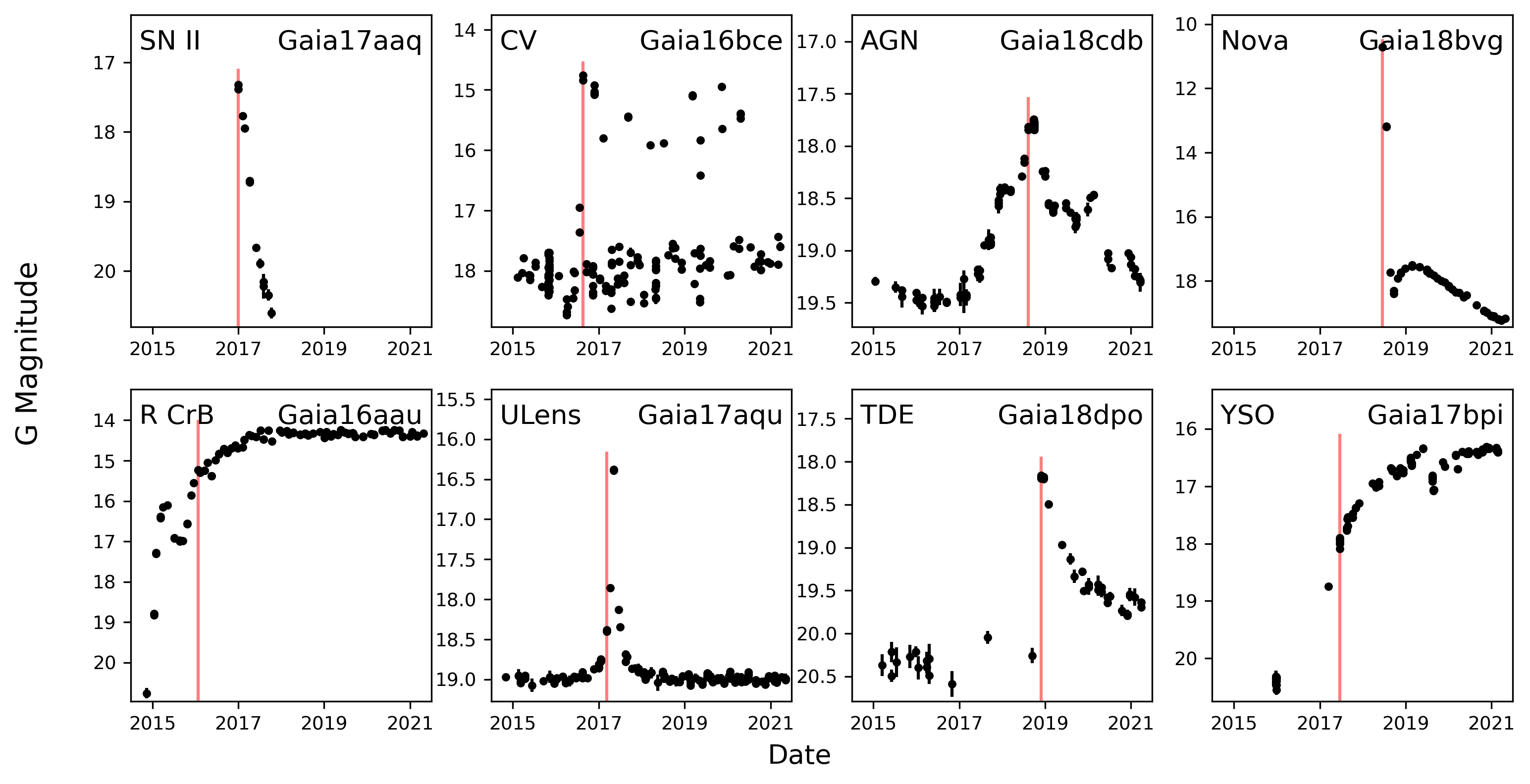}
    \caption{\gaia{} lightcurves for 8 alerts spanning a range of classes. {\bf Top Row:} type II supernova, cataclysmic variable, active galactic nucleus, nova. {\bf Bottom row:} R CrB star, microlensing event, tidal disruption event, young stellar object. The red vertical line illustrates the alerting transit for the event. The y-axis gives the \gaia{} magnitude, and the shared x-axis spans mid-2015 to mid-2021.}
    \label{fig:lightcurves}
\end{figure*}

For the classified \gsa{} alerts, Figure~\ref{fig:class} shows the most common transient classes. We see that supernovae are dominant amongst classified alerts, followed by AGN (this includes QSOs and BL~Lac objects) and then CVs. A full list of these broad classifications is given in Table~\ref{tab:class}, and some illustrative example lightcurves for 8 different classified alerts are shown in Figure~\ref{fig:lightcurves}.

\begin{table}
    \caption{\gsa{} classifications by number and percentage (of classified alerts)}
    \label{tab:class}
    \centering
    \begin{tabular}{l|r|r}
    {\bf Class} & {\bf Number} & {\bf \% of Classified} \\
    \hline
    SN & 2019 & 59.0\\
    AGN & 717 & 20.1\\
    CV & 442 & 12.4\\
    YSO & 125 & 3.5\\
    Star & 55 & 1.5\\
    Microlensing & 37 & 1.0\\
    Nova & 29 & 0.8\\
    SLSN & 22 & 0.6\\
    TDE & 9 & 0.2\\
    R CrB & 9 & 0.2\\
    XRB & 8 & 0.2\\
    Symbiotic & 7 & 0.2\\
    SN Impostor & 3 & 0.1 \\
    SSO & 2 & 0.1\\
    \hline
    \end{tabular}
\end{table}

A large fraction of the classifications come from dedicated spectroscopic programmes such as PESSTO (\citealt{Smartt15}) and Spectral Energy Distribution Machine (SEDM, \citealt{blagorodnova18}), and therefore they are heavily biased towards supernova discovery by design. Consequently, the class fractions derived from these classifications are not representative of the entire sample of \gsa{} events.

\subsection{Spatial distribution}

\begin{figure}
\includegraphics[width=\columnwidth]{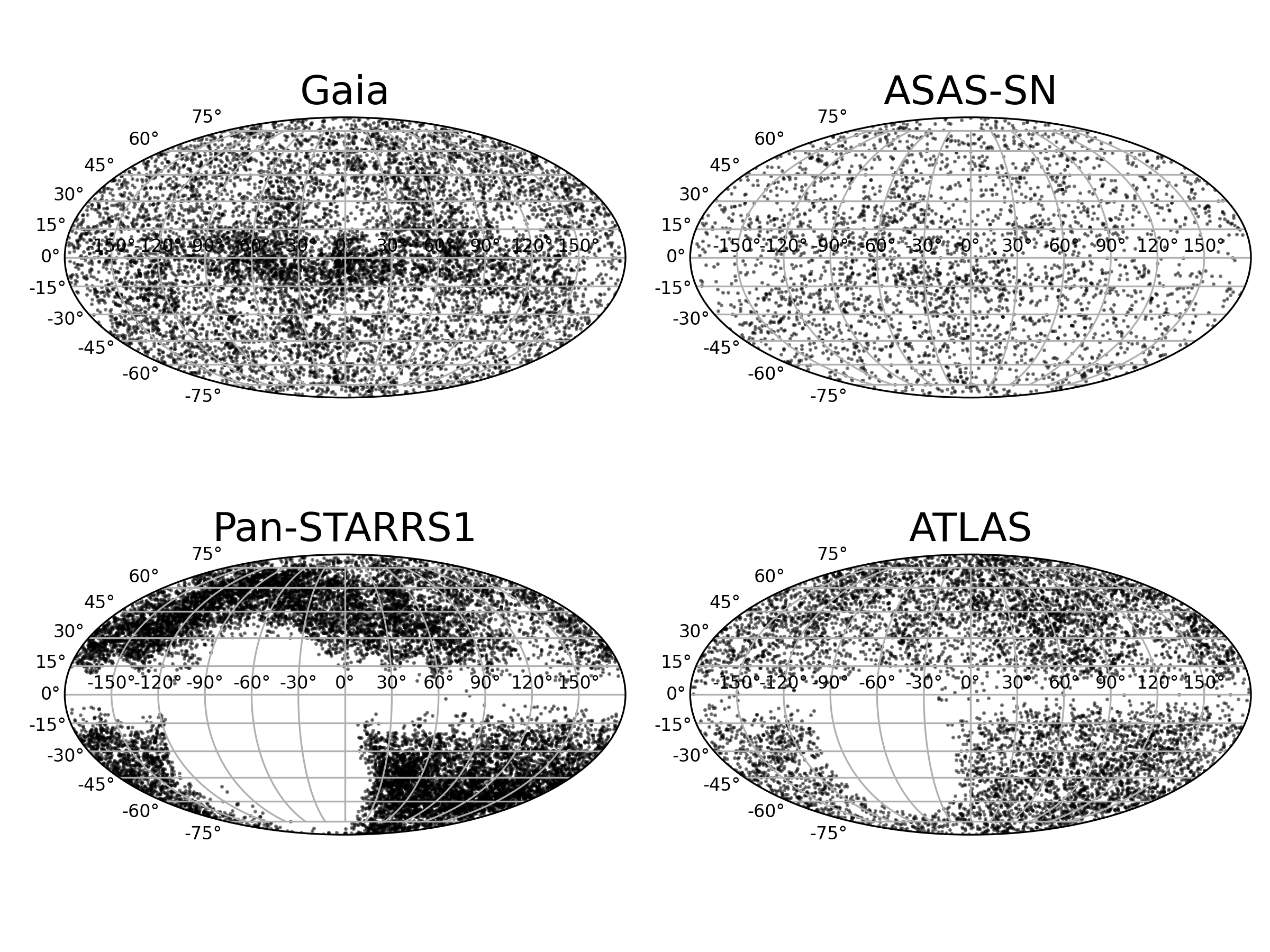}
\caption{Survey coverage for \gsa{} (top-left), compared to three other transient surveys: ASAS-SN (top-right), ATLAS (bottom-right) and PanSTARRS (bottom-left). Data are presented in Galactic coordinates, with the centre of the Galaxy at the centre of each figure. Transients were all compiled using our local database as described in Section~\ref{sec:eyeball}, and reported during the date range June 11 2016 to December 31 2019 inclusive.}
\label{fig:survey_coverage}
\end{figure}

In Figure~\ref{fig:survey_coverage} we show 9969 \gaia{} alerts published from observations made between 11 June 2016 and 31 December 2019 inclusive. We compare the distribution on the sky (in Galactic coordinates) with three other ongoing transient surveys for the same time range: ASAS-SN (3120 events), Pan-STARRS1 (15\,086 events), and ATLAS (7804 events). We note a number of interesting features in the distributions. Firstly, only the \gaia{} and ASAS-SN surveys sample the Galactic plane, while the majority of surveys, including Pan-STARRS1 and ATLAS avoid this crowded region. Secondly, only ASAS-SN and \gsa{} are all-sky surveys, the others are based on data taken from a single ground-based observatory, and thus miss a significant fraction of the southern hemisphere. Finally we note that \gaia{} sees an excess of events in the Galactic plane which is not seen by ASAS-SN, perhaps because the latter uses bluer filters (which will be more affected by extinction), is shallower, and has lower spatial resolution.

\begin{figure}
    \centering
    \includegraphics[width=\columnwidth]{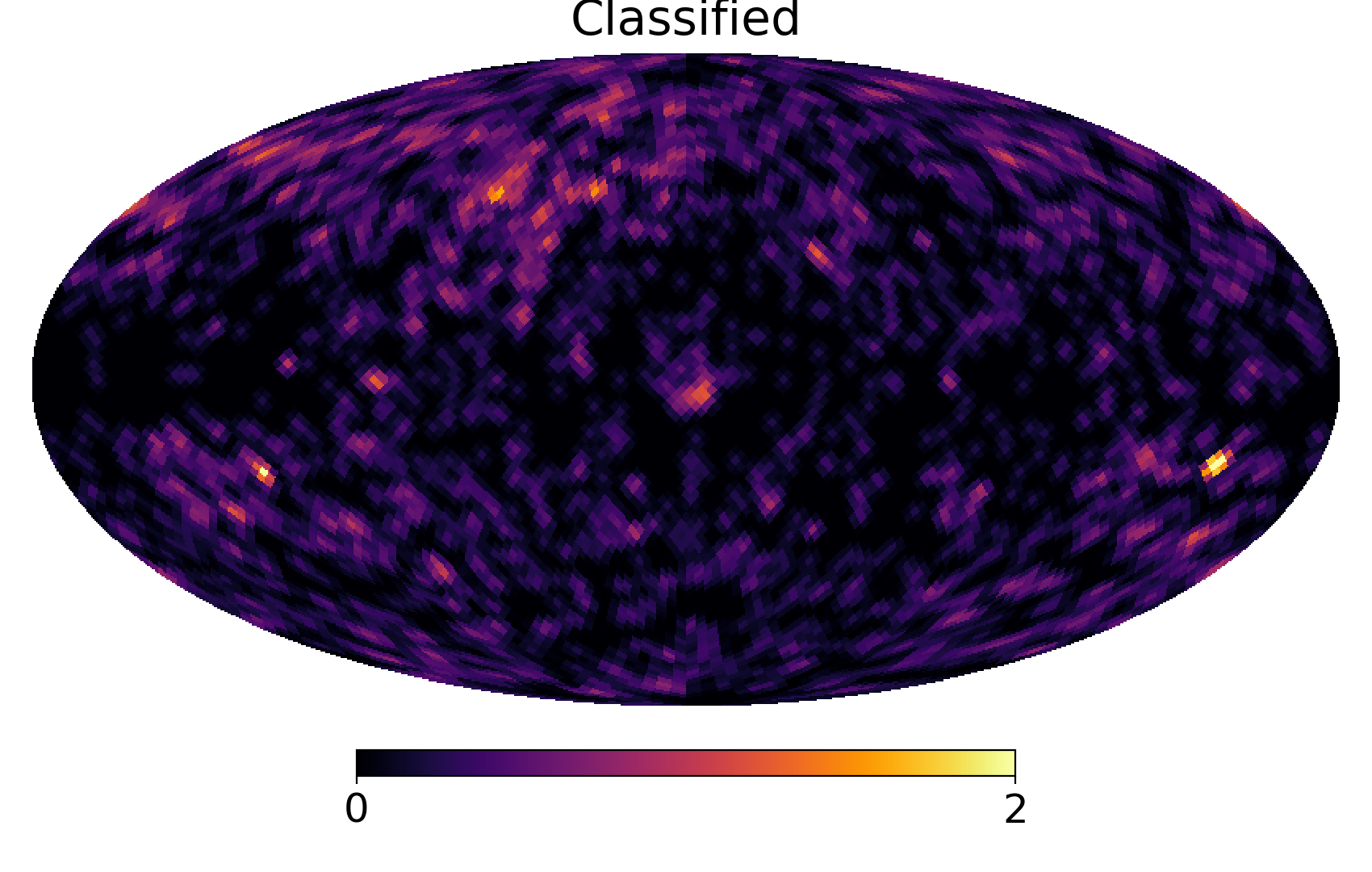}
    \includegraphics[width=\columnwidth]{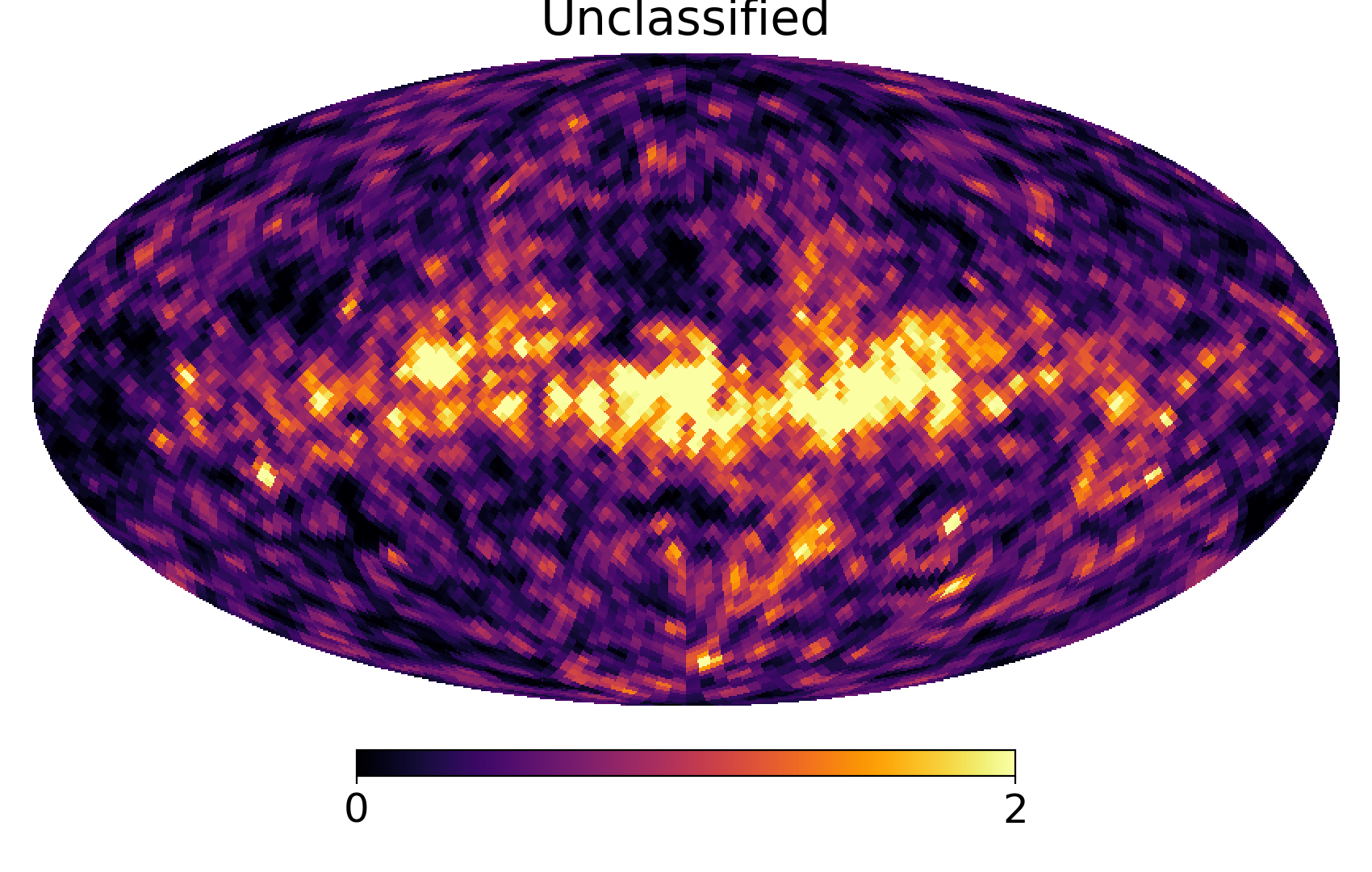}
    \caption{Distribution on the sky for classified (upper panel) and unclassified (lower panel) alerts. The map is in Galactic coordinates, using a HEALpix grid with a resolution of about 1.8 degrees (level 5, NSIDE=32, 12288 pixels), and has been lightly smoothed.
}
    \label{fig:map_classified}
\end{figure}

The overdensity in the plane is further explored in Figure~\ref{fig:map_classified} which shows in two panels the distribution of classified and unclassified \gaia{} alerts. The classified and unclassified alerts are anti-correlated with Galactic latitude. This is perhaps unsurprising, as the main follow-up campaigns (e.g. PESSTO, NUTS) are focused on extragalactic events (such as supernovae and tidal disruption flares), and so avoid the plane by design. We can infer from this that the majority of unclassified alerts are Galactic in origin, and thus the statistics presented in Table~\ref{tab:class} are not reflective of the true breakdown of the \gsa{} transient classes.

\subsection{Astrometry}
\label{sec:astrometry}
As discussed in Section~\ref{sec:dataflow} \gsa{} uses astrometry derived by IDT. Previously, \cite{wevers2019} has shown that the median separation between the \gaia{} alert coordinates and \gaia\, DR2 is 62 milliarcseconds (with a standard deviation of 40 milliarcseconds). We independently compared the per-transit RA and Dec positions for 10\,461 distinct alerts comprising just over 240\,000 individual transits, measured between 15 January 2016 and 31 December 2019, with the positions published in \gaia\, DR2. We find the separations between the two coordinate systems are reasonably well described by a Rayleigh Distribution, albeit with a slight excess in the tail to larger separations, presumably arising from systematic differences between the actual and predicted \gaia\, spacecraft attitude (i.e. consistent with spacecraft hits, \citealt{Fabricius-1}). The best fit model results in an average offset of 55 milliarcseconds, with no dependency on magnitude.

We also note that \cite{Yaron19} compared the transient positions between published \gaia{} alerts and a number of surveys, finding the following median separations (amongst others): \gaia--ZTF: 0.12 arcsec; \gaia--Pan-STARRS1: 0.12 arcsec; \gaia--ASAS-SN: 1.17 arcsec.

They conclude that \gsa{} astrometric measurements will be treated as {\it ground truth}, which is to say that the TNS coordinates will be updated to the \gaia{} positions if and when published.

\subsection{Photometry}
\label{sec:photometry}


All transit photometry is calibrated on-the-fly via a database function. The parameters used by the function are derived from the Photometric One Day Calibration (PODC, see Appendix~\ref{appx:podc}), and are generated with a one-day cadence. The operation to build the calibration is run roughly monthly, thus at the time of alert, the calibration can be a month out of date (and sometimes more).

\begin{figure}
    \centering
    \includegraphics[width=\columnwidth]{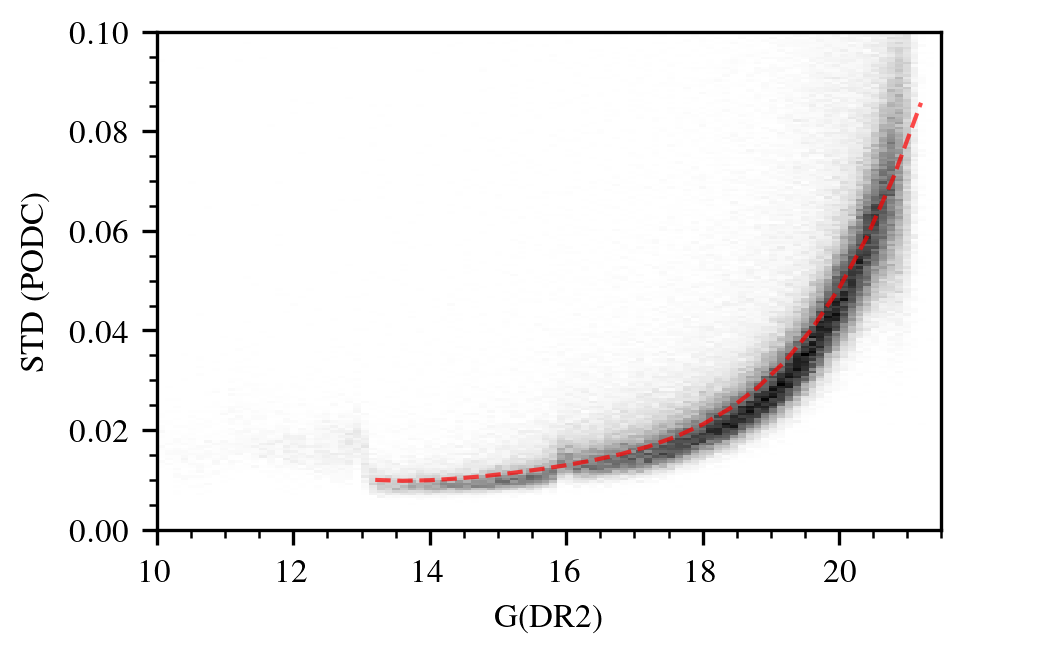}
    \caption{Standard deviation of the PODC per-transit photometry (in magnitudes), as a function of \gaia{} DR2 magnitude. We overlay a fourth-order polynomial model, fit to the median standard deviation as a function of $G_{\rm DR2}$ magnitude.}
    \label{fig:podc_rms}
\end{figure}

To test the precision of PODC, we selected a random sample of 184\,000 sources which lie in the SDSS DR7 footprint (avoiding the most crowded regions of the Galactic plane). We required the sources to have a minimum of 10 \gaia\, field-of-view transits, and we used the median of the per-CCD PODC calibrated fluxes as the per-transit CCD flux. We used the standard deviation of multiple transits for a source as a measurement of the precision of a single \gaia\, transit in the Alerts system. In Fig.~\ref{fig:podc_rms} we show that the precision reaches 1 per~cent for sources around $G=13$, falling to around 10 per~cent near the limit of the survey ($G=20$). Most alerting sources must reach $G=19$, where the median standard deviation is 0.031~mag.

We note the features (steps) at $G=13$ and at $G=16$, which are consistent with changes to the window-class (see Appendix~\ref{appx:podc} for more details). Sources which can swap window class between observations will end up with additional scatter in their photometry, because PODC does not attempt to homogenise the different calibration units. As an approximation, we fitted a model to the median standard deviation versus magnitude:

\begin{equation}
STD = 3.44 - 0.879G + 0.084G^2 - 0.0036G^3 + 0.00006G^4,
\label{eq:stdvmag}
\end{equation}

where $G$ is from DR2, and the model is valid only for the range $13<G<21$. For brighter sources, we suggest using a conservative value of 0.02 mag. Discussion on the {\it accuracy} of the PODC calibration, and comparison with \gaia{} DR2 is included in Appendix~\ref{appx:podc}.

\subsection{Transient variability in the Hertzsprung-Russell diagram}

\gsa{} has a unique feature among other transient surveys: in addition to the G-magnitude, each datapoint (transit) in a Gaia transient lightcurve has simultaneous broad-band colour information delivered by the red and blue photometers. Although this colour is essentially uncalibrated for \gsa{} (i.e. derived from the raw pixel samples), it can nevertheless be used to trace the colour evolution of transients as they evolve over time.

Unlike the \gaia\ colours released as part of DR2, at the moment of processing \gsa{} does not have available all the necessary information to accurately calibrate the BP/RP measurements (either in wavelength or in flux). For DR2, this calibration was derived from a large sample of spectrophotometric standard stars within a narrow spectral range, to accurately assess and correct for the relevant distortions \citep{evans2018}. Here we used uncalibrated colours derived by integrating the raw BP and RP spectra, uncorrected for any throughput variations or wavelength offsets that may have been present. Although this is an approximation, we show that the raw BP--RP colour provides a reasonable assessment of the transient properties.

In order to mitigate the effects of cosmic rays, charge injections and other artefacts that artificially distort the colour, we used a 3\,$\sigma$ outlier rejection in the sample values of the spectra before we computed the integrated fluxes. Although this sometimes removed real features in exceptionally strong emission line sources, it significantly improved the overall consistency of the BP--RP colour. For a higher degree of consistency, we performed a median colour correction such that it agreed with the \gaia\ DR2 colour.

\begin{figure}
\includegraphics[width=0.49\columnwidth]{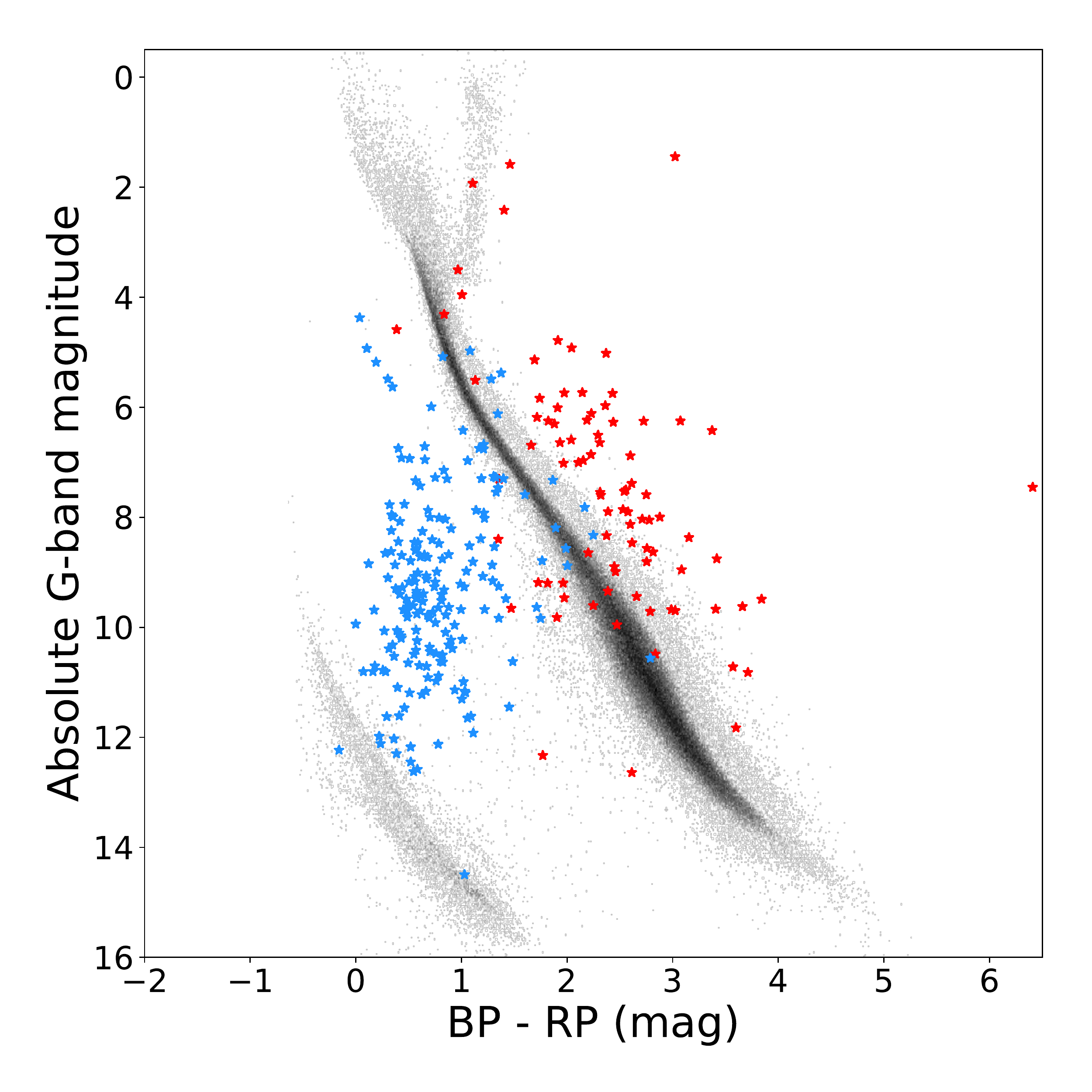}
\includegraphics[width=0.49\columnwidth]{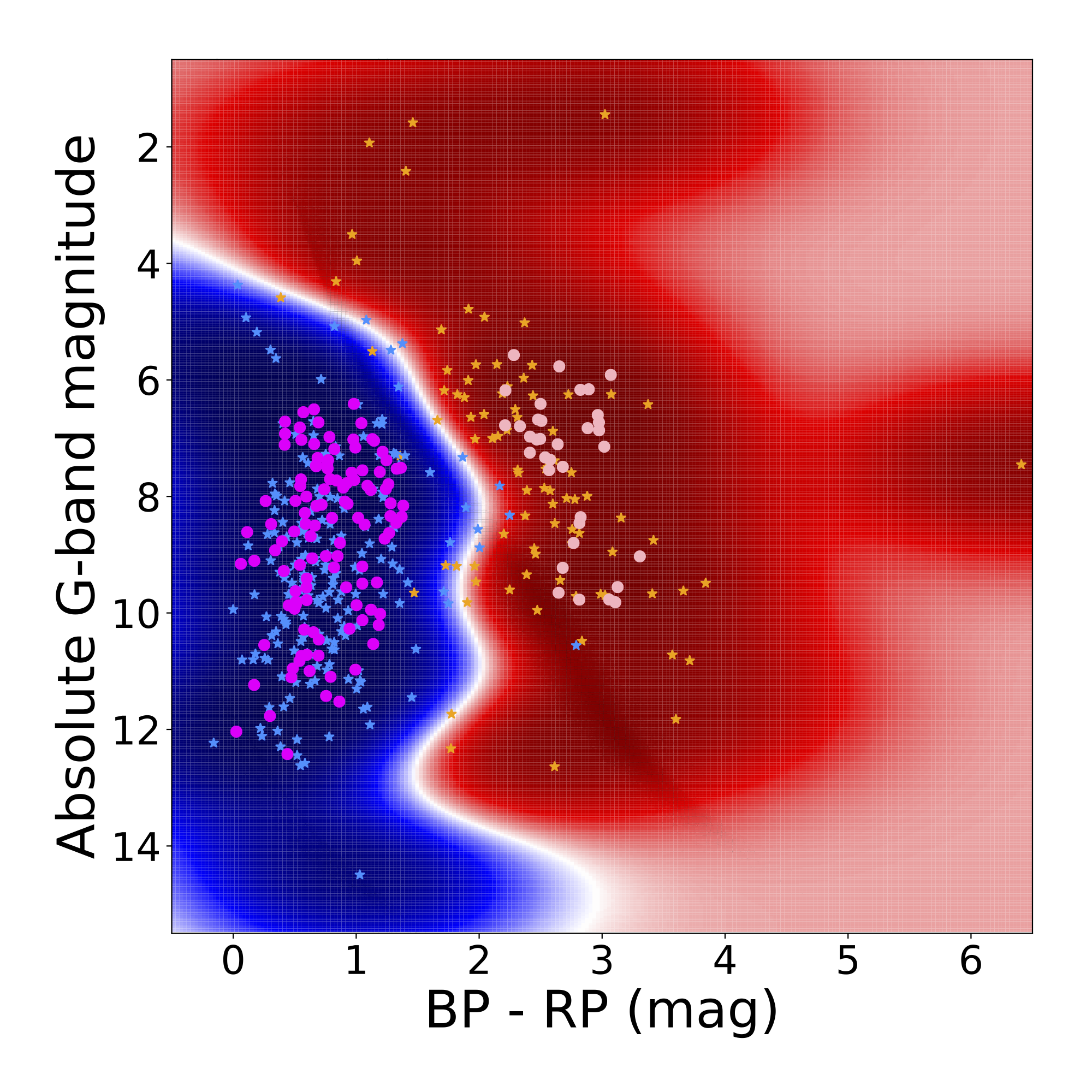}
\caption{Left: colour-magnitude diagram of \gaia\, DR2 counterparts to Gaia Alerts (stars), overlaid on a clean 100\,pc sample (grey background) to illustrate the position relative to the main and white dwarf sequences. We colour-code CVs as blue and YSOs as red. Right: probability map of a binary SVM classifier (see text). Blue regions indicate CV parameter space, red regions YSOs. The white line is the decision boundary; training samples are shown as coloured stars, while new classifications (with P$>$0.95) are shown as magenta/pink circles.}
\label{fig:hrd_gdr2colours}
\end{figure}

For those alerts that have a parallax measurement in \gaia\, DR2 (which are therefore necessarily restricted to Galactic sources), these observed quantities can be combined to trace their evolution in a colour magnitude diagram (CMD). We used the distance estimates of \citet{Bailerjones18} to calculate the absolute magnitude. Figure \ref{fig:hrd_gdr2colours} shows the sample of CVs and YSOs to illustrate the typical parameter space covered in the CMD. Here we show the average DR2 colour of each, which already illustrates that while YSOs and CVs might be discriminated by colour to zeroth order, more information (e.g. parallax) is required to provide an accurate separation of the two classes. No correction was made for reddening.

To illustrate the power of colour as well as parallax information, we created a binary classifier using a support vector machine (SVM). We used the standard radial basis function (RBF) kernel in the {\sc scikit-learn} package in Python. Probabilistic output was obtained through 5-fold cross-validation. We used the classified sources as a training set and predicted classifications for 1815 unknown alerts that have a counterpart in DR2. Because we used DR2 colours and absolute magnitudes, the results should be valid more broadly for transients with a DR2 counterpart discovered by other surveys as well. The results are visualised in Figure \ref{fig:hrd_gdr2colours}, where the colour map traces the (binary) classification probability. Blue regions indicate parameter space covered by CVs, while red regions indicate parameter space inhabited by YSOs. The white line indicates the decision boundary between the classes. We overplot a subset of newly classified sources as magenta circles to illustrate the high confidence (probability P$>$0.95) parameter space for each class. Using this simple algorithm, 
we classify 638 sources as CVs for P$>$0.95, while 202 new YSOs are classified. We include a table of these newly classified objects in the online-only material\footnote{The table of classified CVs and YSOs is only available in electronic form at the CDS via anonymous ftp to cdsarc.u-strasbg.fr (130.79.128.5) or via http://cdsweb.u-strasbg.fr/cgi-bin/qcat?J/A+A, and contains the following information. Column 1: name of the \gaia{} alert, Column 2: \gaia{} DR2 sourceId, Column 3: Ra, Column 4: Dec, Column 5: parallax, Column 6: parallax error, Column 7: G-band magnitude, Column 8: BP-RP colour, Column 9: classification}. The remainder of the 1815 alerts were not classified with high enough confidence to be included. We caution that this is a very simplistic classifier which uses only the magnitude, colour and parallax of the transients, to show where CVs and YSOs are most likely to be found. Since the colour is available for all our transients, it is a very useful parameter, but the classification will not be perfect. This classifier also only considers two types of objects, so the list may be contaminated with a small number of other objects such as flare stars, variable stars or QSOs. Future \gaia\, data releases, based on more observations, will remove the apparent parallaxes of QSOs included in GDR2.

\begin{figure}
\includegraphics[width=\columnwidth]{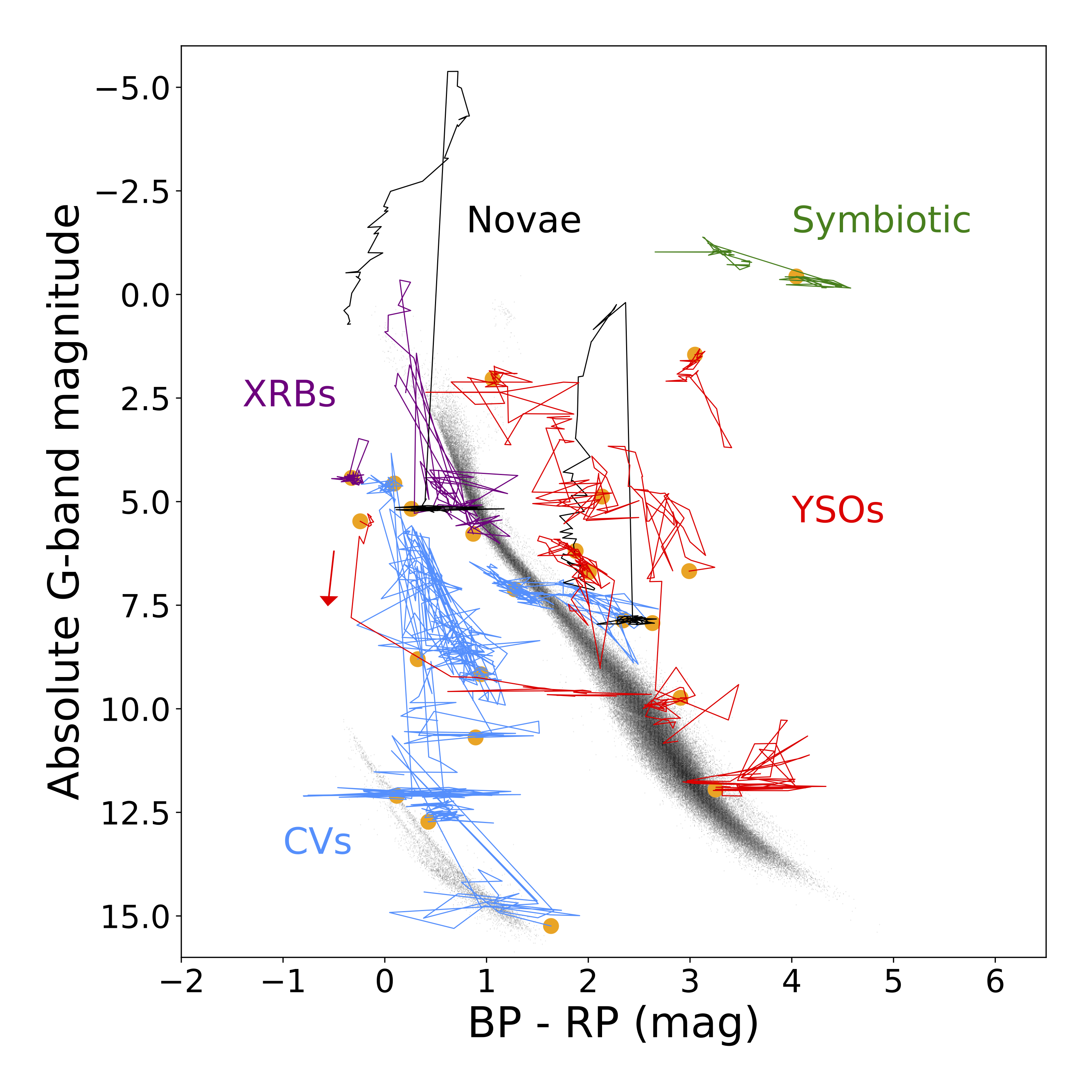}
\caption{Colour-magnitude diagram showing the colours per epoch derived from Gaia Alerts spectro-photometry, to show the evolution of a selected subsample of sources through the HRD. Orange filled circles mark the beginning of the lightcurve.}
\label{fig:hrd_class2}
\end{figure}

Another application of the alert colour information is demonstrated in Figure \ref{fig:hrd_class2}, where we show a subsample of sources and follow their evolution through the diagram as their properties vary in time. A similar figure is shown in Figure~11 of \cite{Eyer19}, which includes a large sample of periodic and non-periodic variables. We have colour coded sources by classification. Note that some sources show a large spread in BP--RP, indicating that our simple data curation may be insufficient, and a proper spectral calibration is required for more detailed analysis; fully calibrated epoch BP and RP spectra will be released in DR4. Nevertheless, we note that CVs and XRBs have bluer colours in outburst, as expected. We can also see the reddening of emission that occurs in novae between 10 and 100 days post-peak (e.g. \citealt{Hachisu14}), when the wind ejecta expands while the photosphere recedes. 

%
%
%
%

\section{Discussion}
\label{sec:discussion}

\subsection{Purity} \label{sec:purity}

\begin{figure}
\includegraphics[width=\columnwidth]{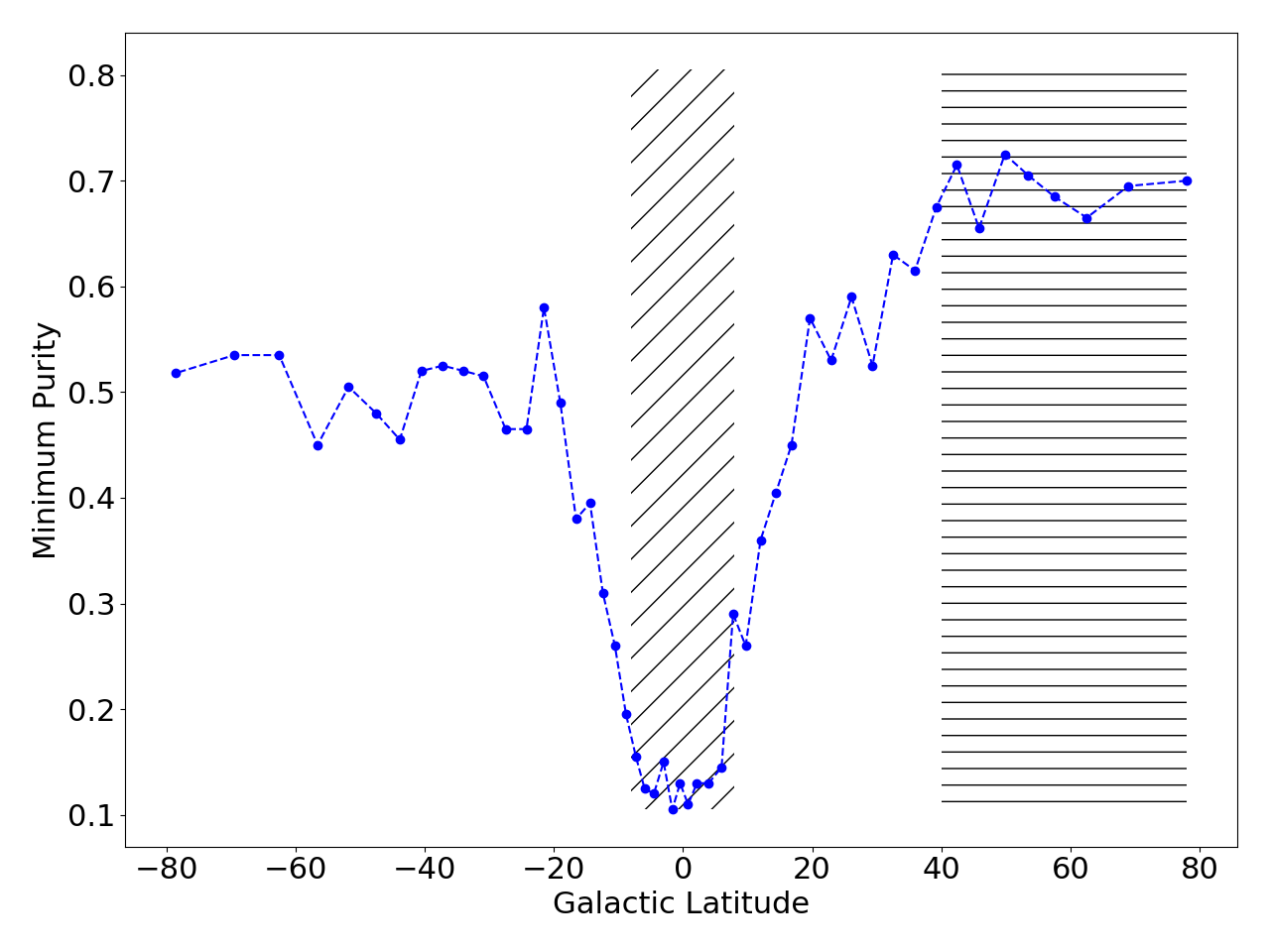}
\caption{The minimum purity of the published alert stream as a function of Galactic latitude (see text for details). The behaviour as a function of Galactic latitude may be understood by considering the coverage of other transient surveys which generally avoid the plane, are not full sky and are biased towards northern skies (ASAS-SN is the only other all sky transient survey). Additionally, very few alerts towards the Galactic plane are followed up. The hatched regions show ranges in Galactic latitudes which are used in the purity analysis (Sections~\ref{sec:purityGalLat}~and~\ref{sec:purityNewOld}). The horizontal hatched region contains 1661 published alerts and has an overall minimum purity of 0.69. The diagonal hatched region covering $\pm 8$ degrees around the Galactic plane contains 1628 published alerts and has an overall minimum purity of 0.09.}
\label{fig:purity_vGlat}
\end{figure}
\begin{figure*}
\includegraphics[width=0.5\textwidth]{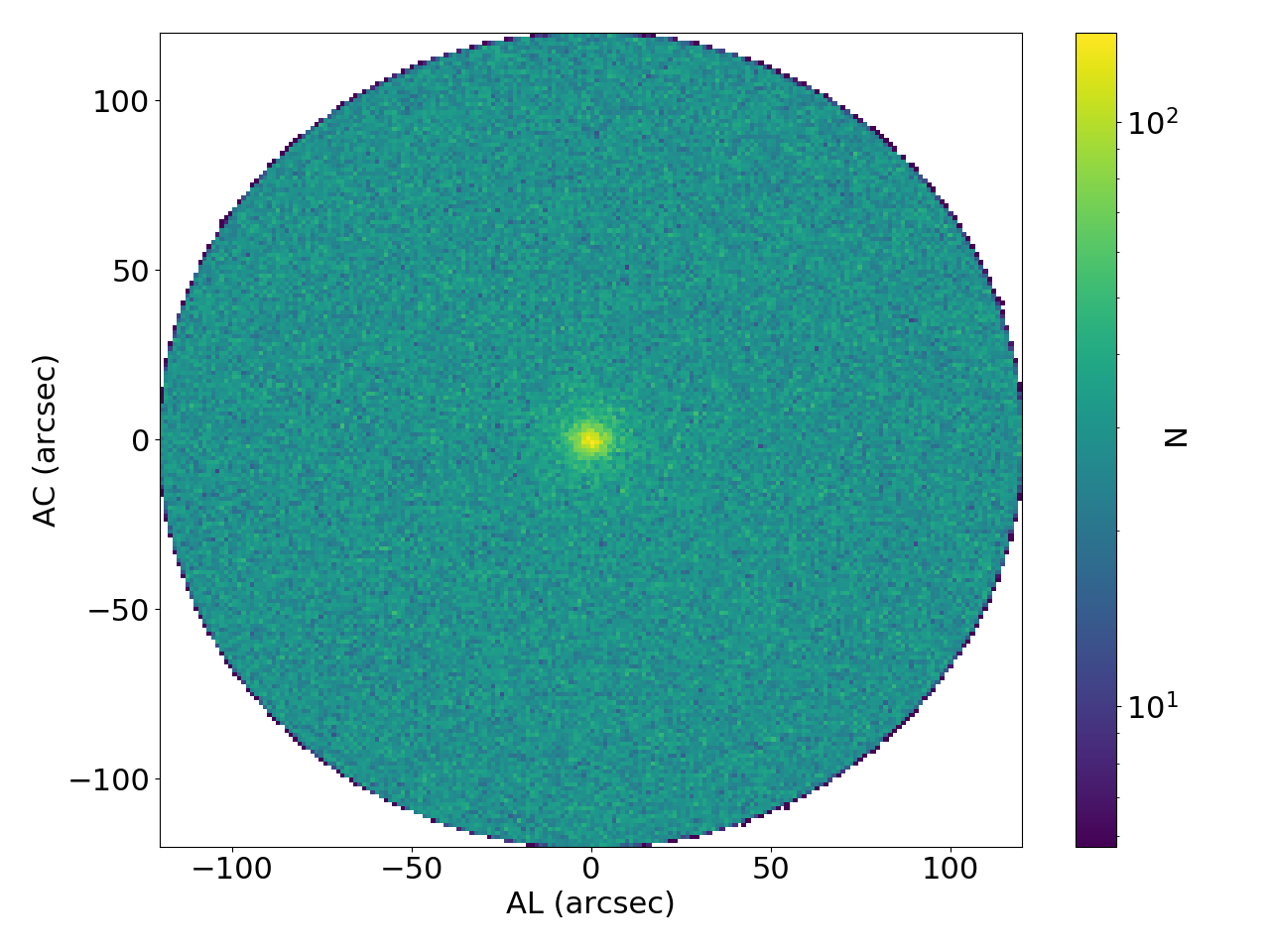}\includegraphics[width=0.5\textwidth]{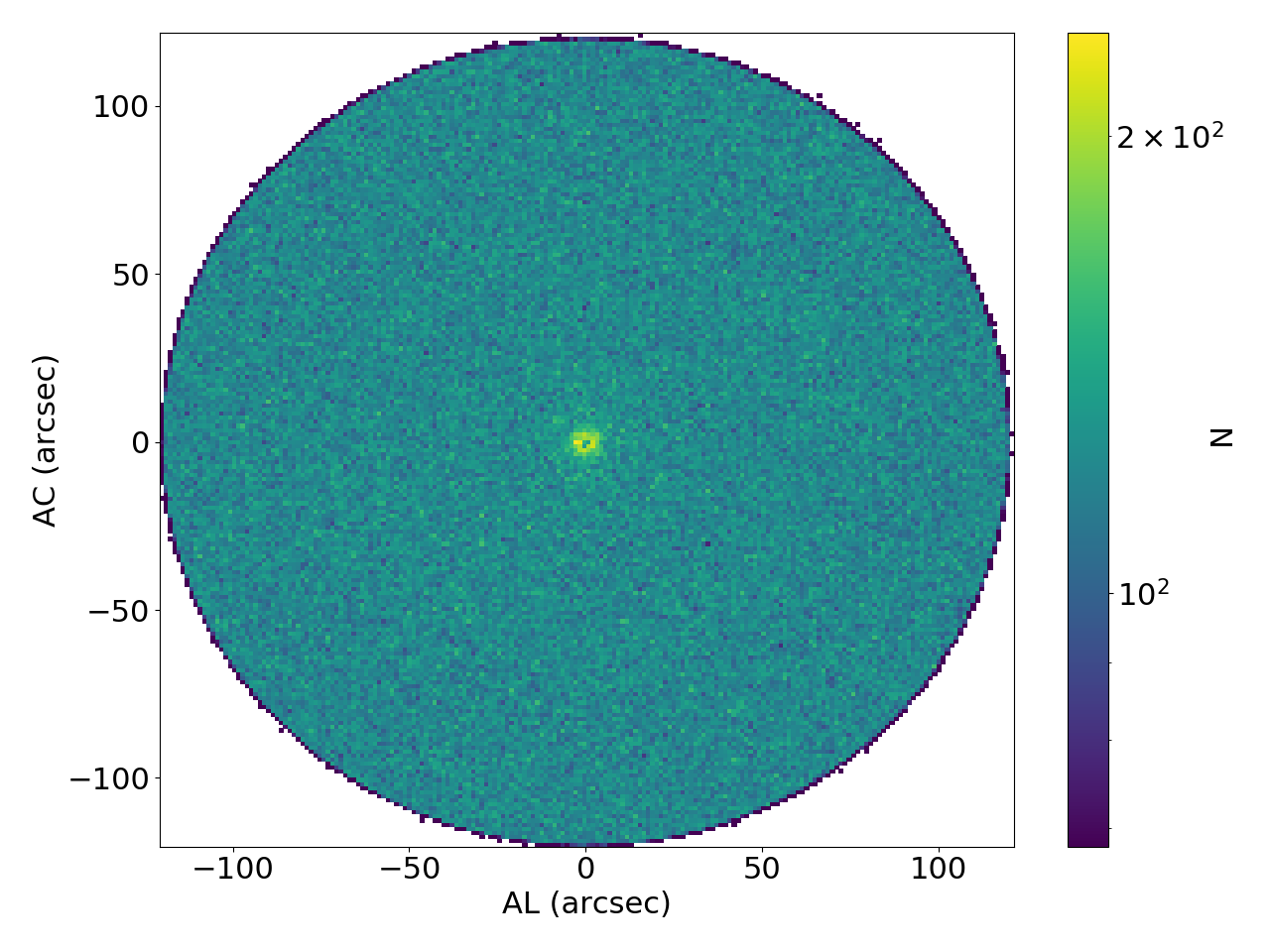}
\caption{The number of neighbouring sources within $2^{\prime}$ of each published alert in the AL and AC directions at the time of the alert, accumulated for all alerts which are confirmed by ancillary data (Left), and for the remaining alerts (Right).
Note that all sources within $0.2^{\prime\prime}$ of the alert positions are excluded as they are deemed to belong to the alerts themselves.
The distribution of sources around the alerts shows no excess in the AL and AC directions, as might be expected were there some residual contamination due to the PSF spikes of bright stars causing false detections.
There is an excess in the number of sources at very close angular separations, but this is more pronounced for the confirmed alerts than those which are not confirmed by ancillary data.}
\label{fig:confirmed}
\end{figure*}

We would like to assess the fraction of the published alerts which are astrophysically real and not due to an artefact or systematic effect in the \gaia\ data. We define this fraction as the purity of our published alerts. 
While purity is not a quantity frequently reported by surveys, it is important in the context of any population studies based on the survey as well as telescope time spent in the follow up of alerts.
As in the rest of this paper we consider those published alerts which were originally detected in or between IDT runs 1046 and 4724 (the last run of 2019, see Section~\ref{sec:introduction}).

Firstly, we examined how many of our published alerts had been observed by another transient survey, using the Transient Name Server (TNS). This could be either before or after it was discovered by \gaia. Given that a different survey should not be subject to the same artefacts, we can reliably class these alerts as astrophysically real. Additionally, we consider all alerts which have an ATEL or a spectroscopic classification to be real.
The fraction of our published alerts which are real based on this analysis is 0.44. This gives an indication of the {\em minimum} level of purity in our sample, as not all alerts were followed up (i.e. had an ATEL or were classified) and most of the other transient surveys are not full sky (with the exception of ASAS-SN).

\subsubsection{Purity: dependence on Galactic latitude}
\label{sec:purityGalLat}
 
Figure~\ref{fig:purity_vGlat} shows a strong dependence of the minimum purity level on Galactic latitude. 
This does not mean however, that our alerts are unreliable in the plane, but rather highlights the absence of coverage by other surveys, as shown in Figure~\ref{fig:survey_coverage},\footnote{To see the coverage footprints of the other surveys in TNS go to {\em https://www.wis-tns.org/stats-maps/maps}} and a low rate of follow-up.

Figure~\ref{fig:purity_vGlat} also shows a bias towards northern skies. The purity for positive Galactic latitudes appears higher than for negatives ones, as the majority of ground-based transients surveys are based at northern latitutes.
A strong dependence on the magnitude of the alert was also found, which is again unsurprising as brighter alerts are more likely to be followed up. 
It is worth noting that for $b \ge 40^{\circ}$ and $G_{\rm mag} < 17$ the fraction of our alerts confirmed by ancillary data is 0.93.

\begin{figure*}
\includegraphics[width=0.5\textwidth]{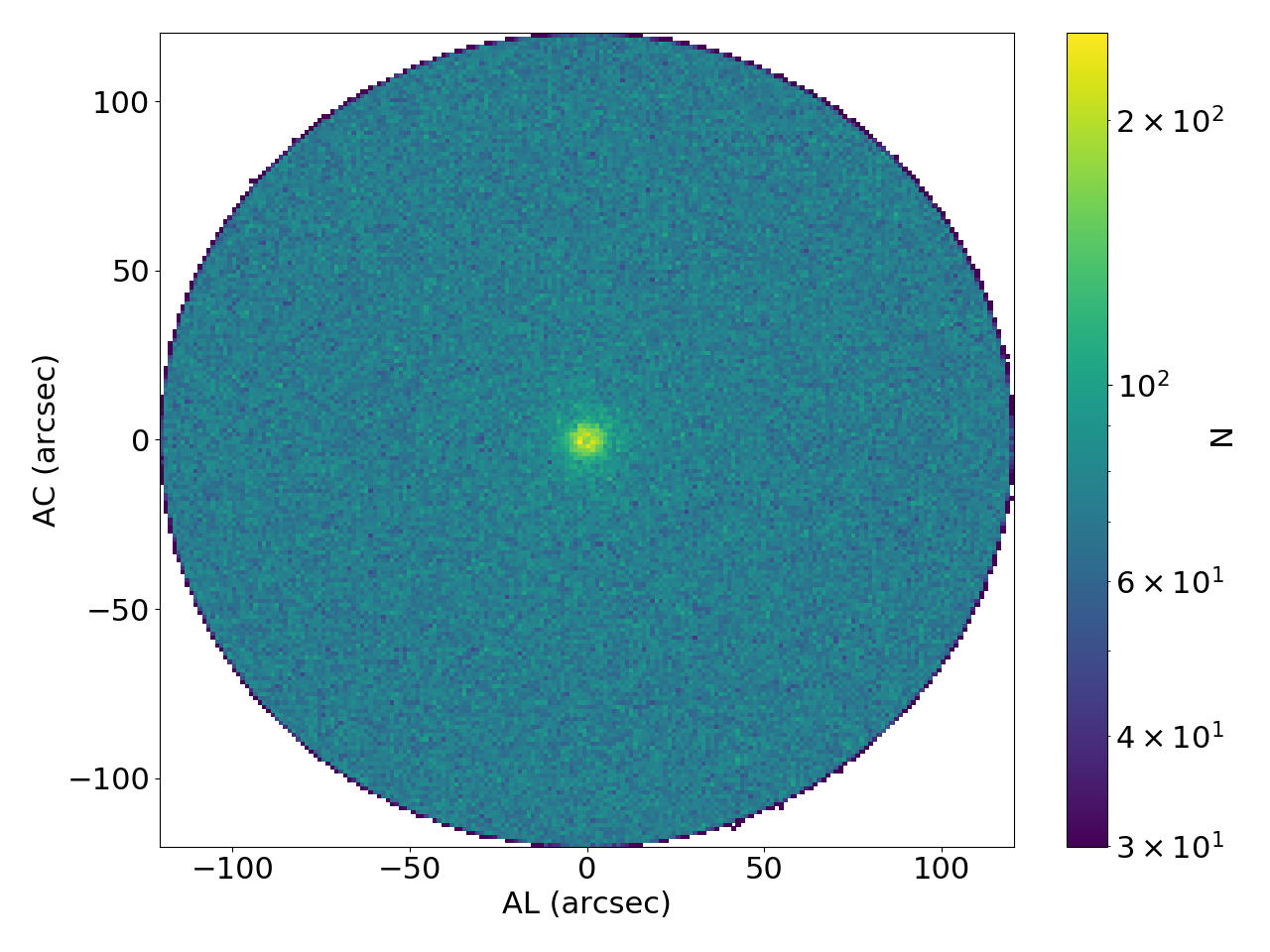}\includegraphics[width=0.5\textwidth]{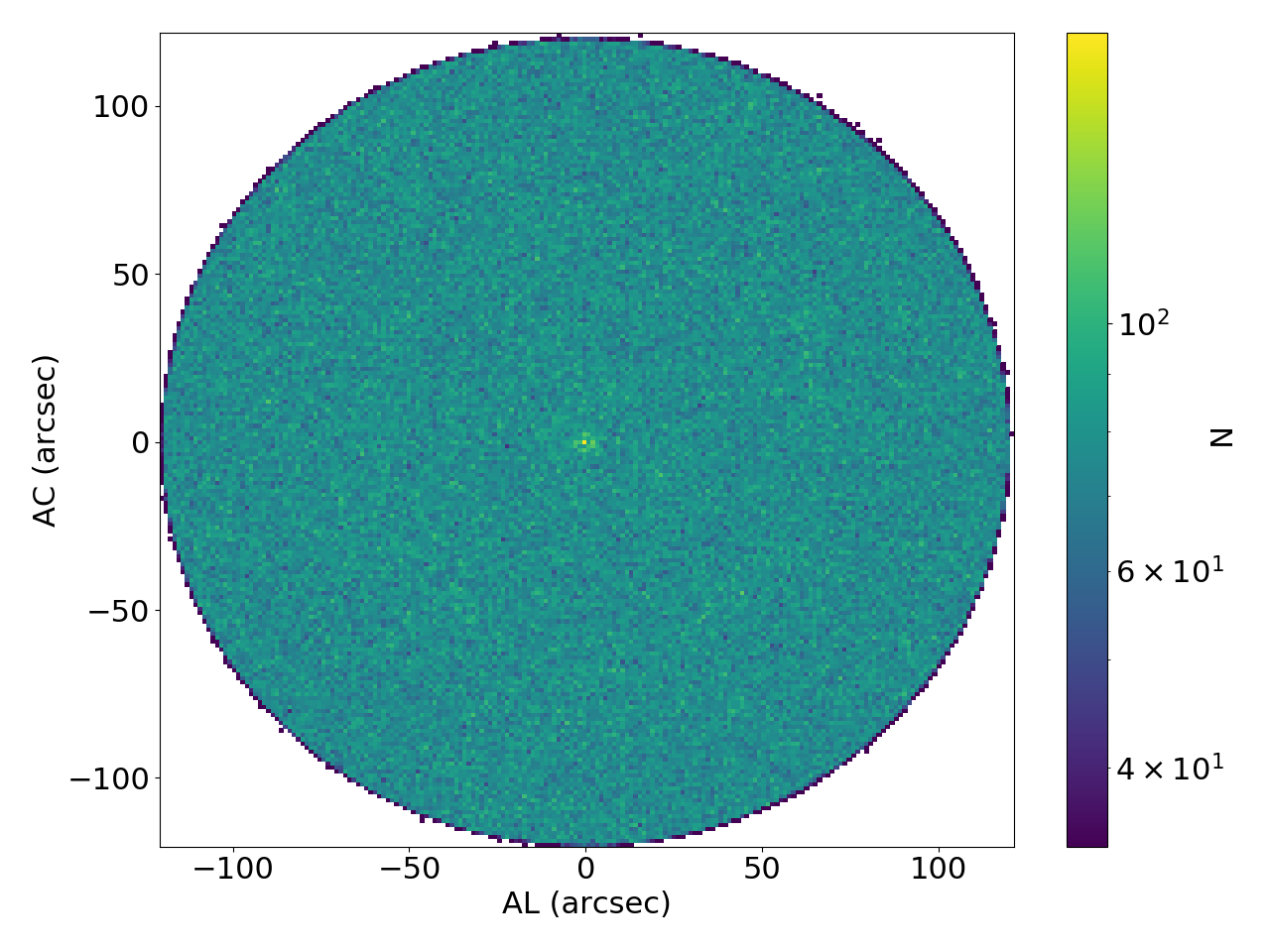}
\includegraphics[width=0.5\textwidth]{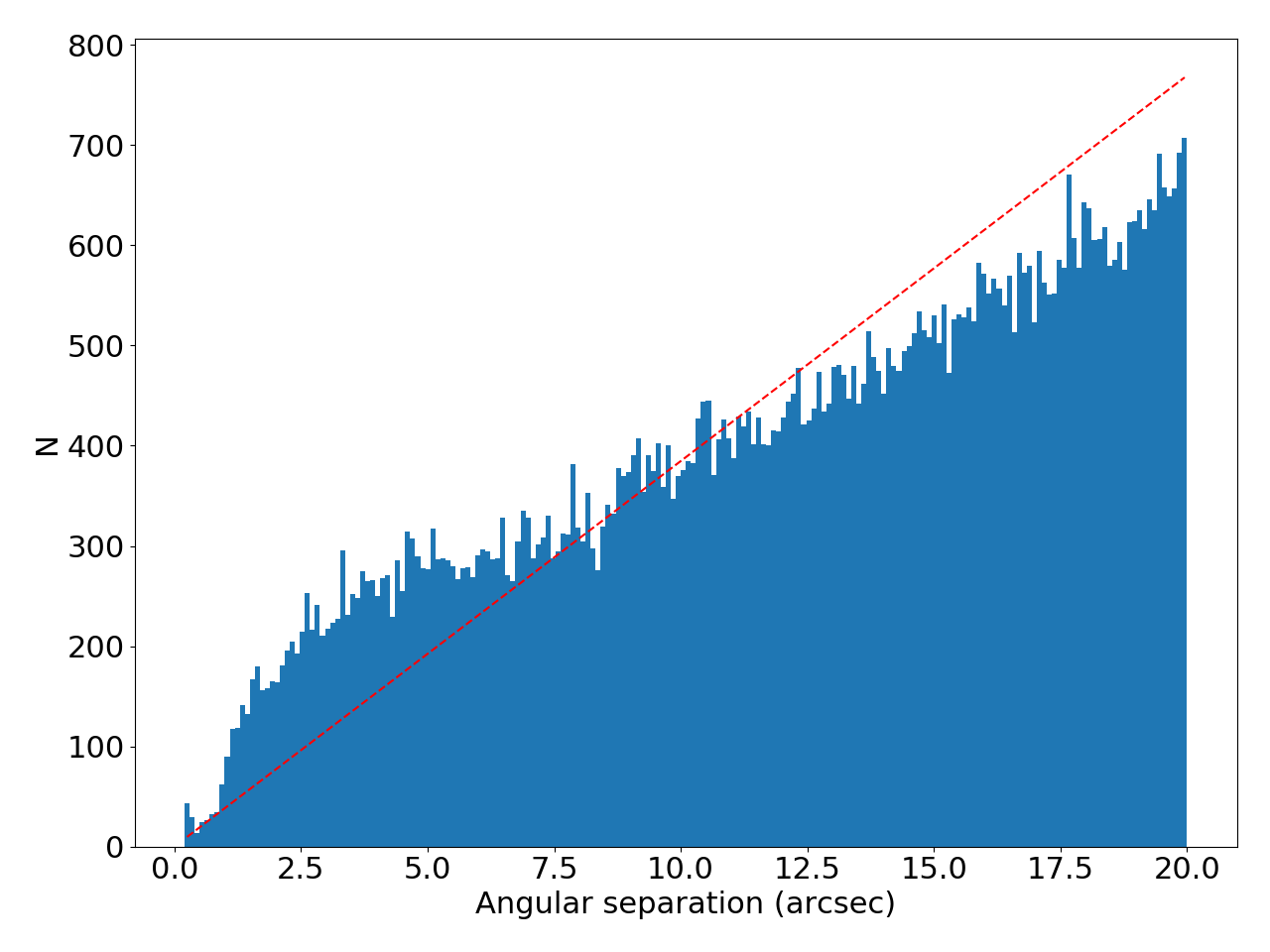}\includegraphics[width=0.5\textwidth]{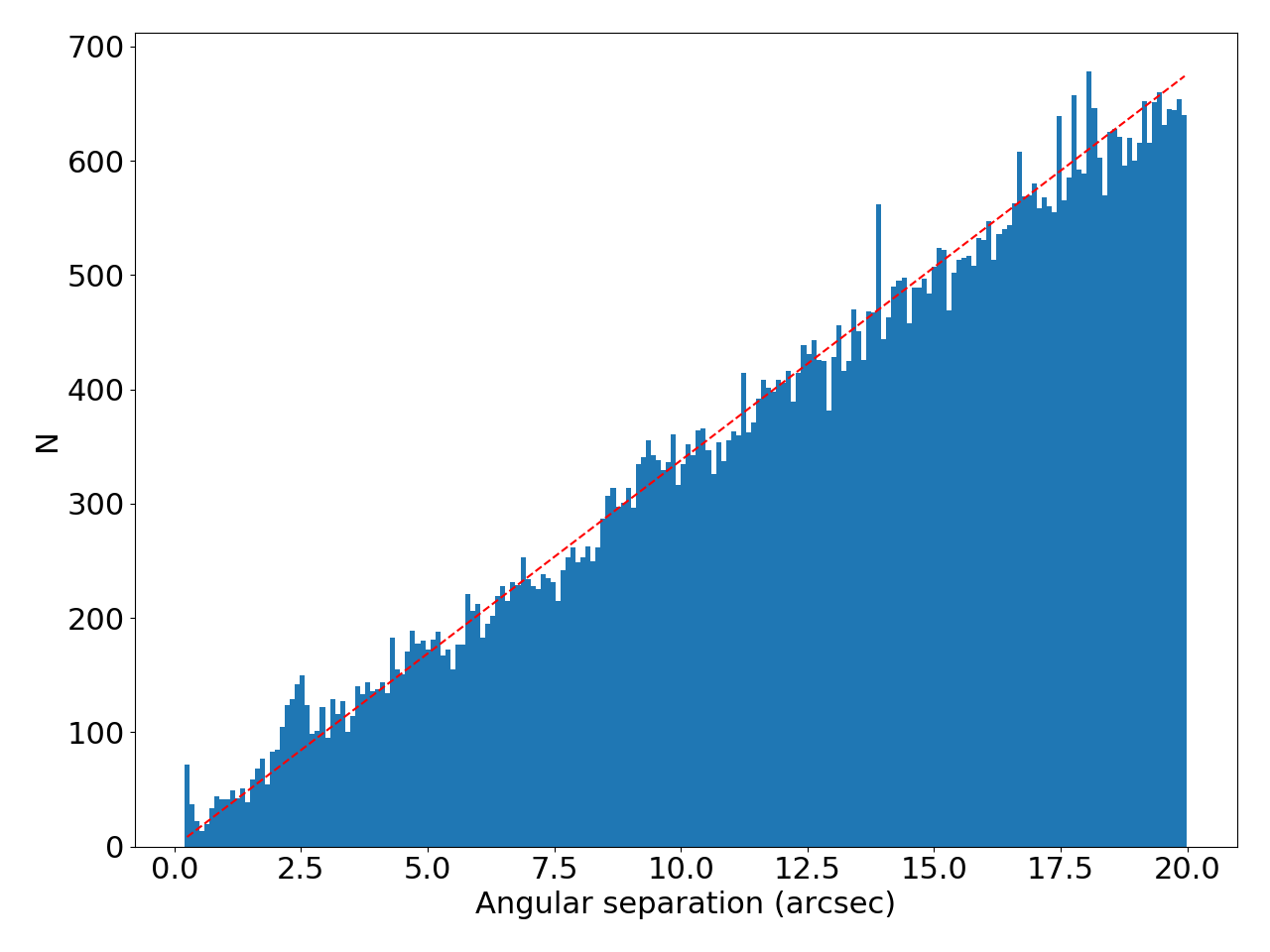}
\caption{Top: The number of neighbouring sources within $2^{\prime}$ of each published alert in the AL and AC directions at the time of the alert, accumulated for all \newsource\, alerts (Left), and for all \oldsource\, alerts (Right).
Again all sources within $0.2^{\prime\prime}$ of the alert positions are excluded as they are deemed to belong to the alerts themselves.
Bottom: The number of neighbouring sources as a function of angular separation out to $20^{\prime\prime}$ around the published alert, accumulated for all \newsource\, alerts (Left), and for all \oldsource\ alerts (Right). The red dashed line is the expectation for the number of sources based on the assumption of the same number of sources per unit area.
Here we can see that the excess in the number of sources at very close angular separations, is much more pronounced for \newsource\, alerts than for \oldsource\, alerts. The range of angular separations in which there is an excess in the number of neighbouring source is more clearly seen in the bottom panels. 
A small excess is still visible around $2.5^{\prime\prime}$ for \oldsource\, alerts.}
\label{fig:angsep_excess}
\end{figure*}

\subsubsection{Purity: investigation of alert environment}

As described in Section~\ref{sec:spurious} the majority of the causes of spurious alerts were found to be due to environmental effects.
Therefore, we chose to study the environment in the vicinity of all the published alerts out to a larger angular separation distance than is possible (due to CPU and timely operation constraints) in our normal processing. The results of our analysis, aimed to detect possible differences between alerts with and without a confirmation in ancillary data, is shown in Figure~\ref{fig:confirmed}. Here we accumulated the relative locations in the AL and AC directions at the time of the alert, of any neighbouring sources in the vicinity out to an angular separation of $2^{\prime}$.
In the absence of any environmental effects we should expect a uniform distribution of sources around the location of the alerts. However, this is not what is seen in Figure~\ref{fig:confirmed} where we see an excess in the number of sources at narrow angular separations from the alert. While also visible for the unconfirmed alerts, this excess is more pronounced for the confirmed alerts meaning this environmental effect must be due to the nature of the alerts themselves rather than a spurious detection caused by an environmental effect.

\subsubsection{Purity: \newsource\, versus \oldsource\, alerts}
\label{sec:purityNewOld}

In an effort to understand this, these plots were regenerated, but this time the alerts were divided between \newsource\, and \oldsource\, alerts. These are shown in Figure~\ref{fig:angsep_excess}, where we see that this excess is barely visible for \oldsource\, alerts while being clearly visible for \newsource\, alerts. In addition, in Figure~\ref{fig:angsep_excess}, we plot histograms of the number of sources as a function of angular separation out to $20^{\prime\prime}$ to more easily see the location of the excess.
As the majority of the \newsource\, alerts are SNe, the reason for the excess becomes clear; these are sources associated with the host galaxy.
The angular extent of many galaxies is such that the on-board detection may record multiple detections at various regions of brighter emission in the galaxy. It is these sources which are being found by the environmental search, and that result in the excess at narrow angular separations. 
As evidenced by Figure~\ref{fig:angsep_excess}, \oldsource\, alerts are not completely immune to this effect as SNe are occasionally detected via the \oldsource\, route as a brightening of a previously observed source associated with a galaxy. 
This also explains the difference in the size of this effect between the confirmed and unconfirmed alerts in Figure~\ref{fig:confirmed} as more SNe are followed-up and hence classified than other classes of transients leading to the more pronounced excess for the confirmed alerts.

Importantly, this demonstrates the effectiveness of this technique, of searching in the vicinity of all the published alerts and accumulating their neighbouring sources as a function of their AL and AC positions at the time of the alert. It also demonstrates that there are no other visible environment effects.
As an additional check we reproduced the plots of accumulated sources in the AL and AC directions for the alerts in the horizontal hatched and diagonal hatched regions in Figure~\ref{fig:purity_vGlat}, corresponding to high ($b \ge 40^{\circ}$) and low ($|b|< 8^{\circ}$) Galactic latitude regions. We find a completely uniform distribution of sources at low Galactic latitudes and that the excess is caused by the host galaxies at high Galactic latitudes.
Finally, we split the alerts based on their magnitude at the time of detection, and see essentially the same behaviour for the bright and faint alerts.

\subsubsection{Purity: summary}

It is now possible to conclude that we find absolutely no evidence for any remaining environmentally induced spurious alerts in our sample of published alerts. 
There is no evidence that the published alerts in the Galactic plane are any less reliable than those at higher Galactic latitudes, nor that fainter alerts are any less reliable than brighter ones.
We conclude that the overall purity of our published alerts is comparable to the subset where $b \ge 40^{\circ}$ and $G_{\rm mag} < 17$, and the fraction of our alerts confirmed by ancillary data is 0.93.

\subsection{Completeness}
\label{sec:tnscomp}

Completeness of a transient survey is a simple concept, but is hard to measure in practice. To be able to measure event rates, it is important to understand what fraction of transients we miss. We have demonstrated that \gsa{} is a transient survey with high purity, however we may expect this to come at the price of completeness. 

To quantify this, we have considered a sample of reasonably long lived, bright and well understood transients which \gaia{} has had a good chance of detecting. Of course, we have imperfect knowledge of what the Universe really looks like, and comparisons with external transient surveys can only be as complete as the reference material. We also comment that this approach says little about \gaia's completeness to short-lived transient events such as outbursts from Cataclysmic Variables, or flares from M dwarfs. These more complex selection functions are not considered in this paper.

For our completeness study we use TNS as a starting point, which also records spectroscopic classifications for significant numbers of SNe. From 11 July 2016 until 31 December 2019, a period through which \gsa{} has been operating in a stable mode, there were a total of 5367 classified SNe reported to TNS across all magnitudes. We restrict ourselves to a subset of 2826 SNe to which \gsa{} should be sensitive, with a reported magnitude of $m=19$ or brighter, noting that these come from a large variety of independent transient surveys, with observations made in different photometric systems.

From this set, we note that 1314 were observed and reported by \gsa{} (379 were reported first by \gsa, 39 per cent). A small number of \gsa{} detections (six) came through our salvaging process (see Section~\ref{sec:salvage}), the rest were entirely independent detections. It may be that \gsa{} would have detected them following another scan, however in this analysis we consider them to be non-detections by \gaia{}, to be as conservative as possible. Our overall external completeness $C_E = 1308 / 2826 = 0.46$.

\subsubsection{Completeness: scanning law}

\begin{figure}
\includegraphics[width=\columnwidth]{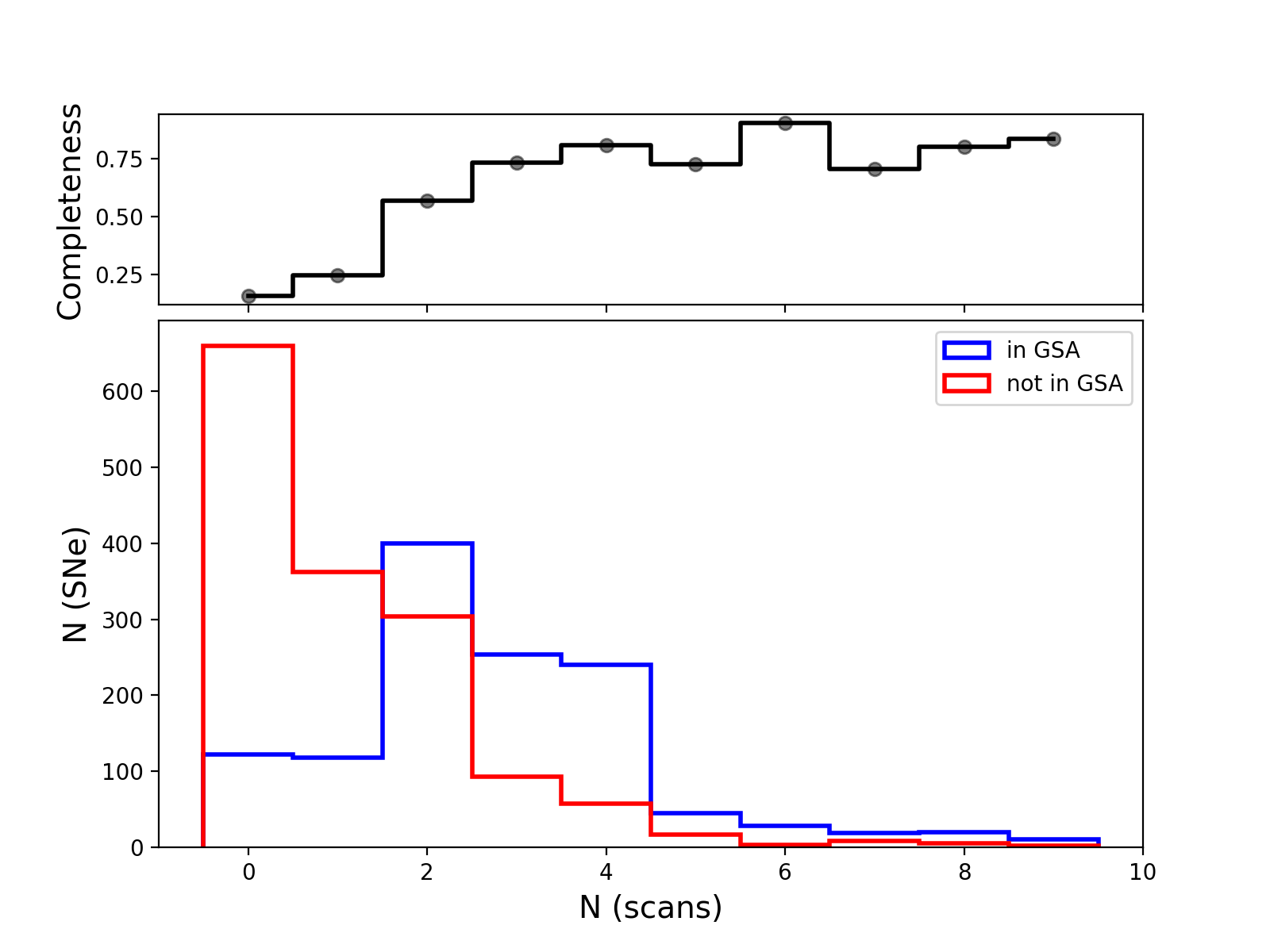}
\caption{Main panel: histograms of the numbers of SNe reported to TNS between 11 July 2016 and 31 December 2019, as a function of the number of times Gaia scanned the location of the event from the date of detection of the event until 40 days after. The histograms are divided into two samples: {\bf blue} independently detected and published by \gsa, {\bf red} not detected and published by \gsa. Upper panel: fraction of the total number detected by \gsa. Note that some of the SNe were first detected by \gaia\, more than 40 days after the event was reported to TNS, thus the N(scans)=0,1 bins are occupied.}
\label{fig:completeness_scans}
\end{figure}

We expect that the largest contributor to the \gsa{} missing events will be the requirement to have two detections from distinct fields-of-view, and separated by less than 40 days. Figure ~\ref{fig:completeness_scans} shows the distribution in the number of scans by Gaia of the selected 2826 SNe within 40 days of the event notification date, for both the events detected and missed by \gsa. This figure supports our expectations, and shows that the internal completeness $C_I=0.57$ where we have exactly two scans. The median completeness for N(scans) $>2$ is 0.8, implying that the \gaia\, scanning law, and the need to minimise the false alarm rate, dominates the completeness of \gsa. For all subsequent analysis, we only include SNe for which there are at least two \gaia\, scans within 40 days of the trigger. A total of 1073 SNe were detected in this subset, and 491 were not, thus our overall internal completeness $C_I=0.69$ on average. 

\subsubsection{Completeness: magnitude}

\begin{figure}
\includegraphics[width=\columnwidth]{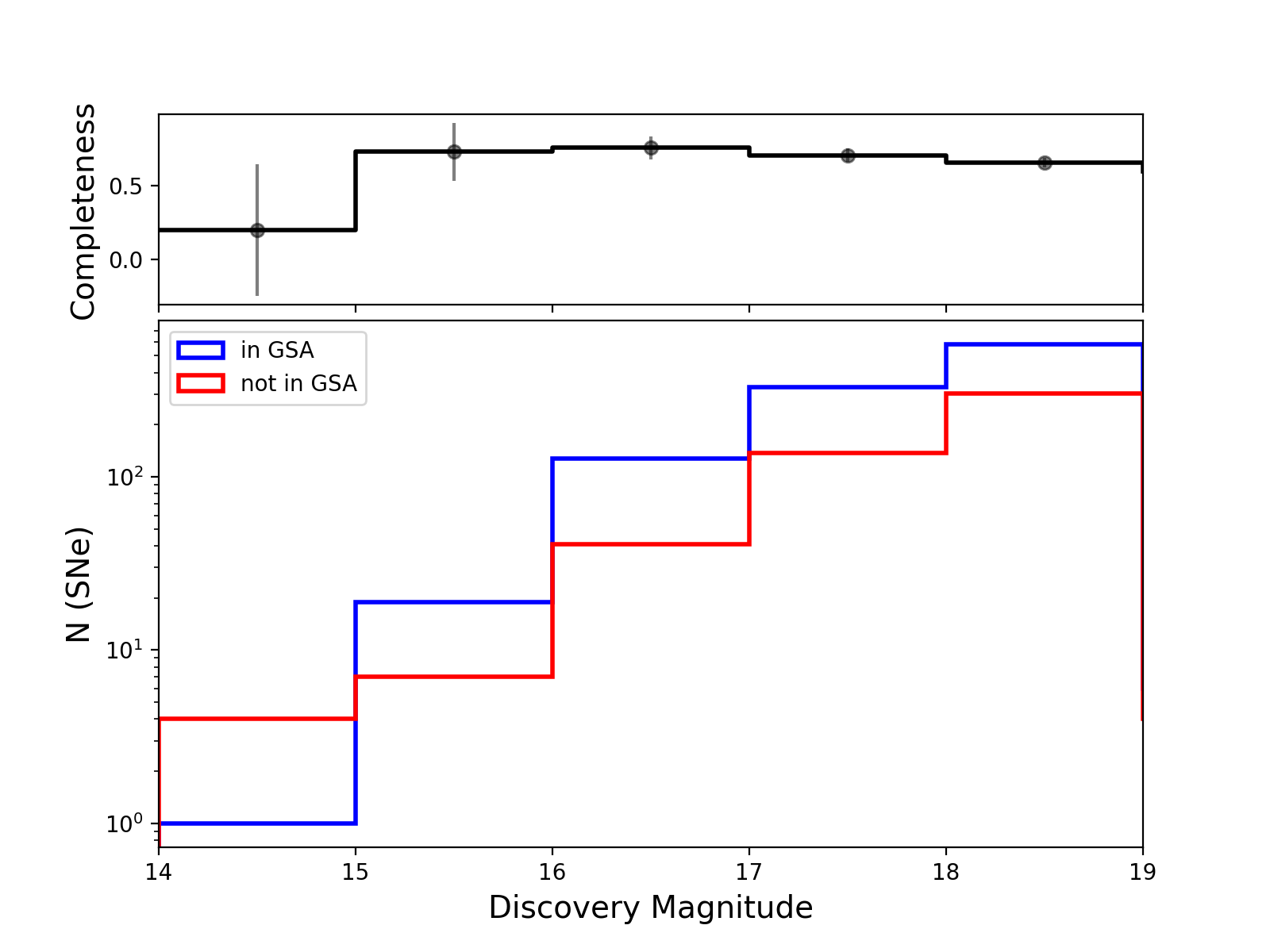}
\caption{Completeness as a function of magnitude for SNe with 2 or more \gaia\, scans. Main panel: histograms are divided into two samples: {\bf blue} independently detected and published by \gsa, {\bf red} not detected and published by \gsa. Upper panel: fraction of the total number detected by \gsa. }
\label{fig:completeness_mag}
\end{figure}

Figure~\ref{fig:completeness_mag} shows the dependence of \gsa{} detection on SN magnitude. The apparent incompleteness for bright SNe is dominated by small number statistics. Moving towards fainter magnitudes, there does seem to be a roll-off in the completeness which falls from 0.75 ($\pm 0.08$, $G=$16--17) at peak, to 0.60 ($\pm 0.03$, $G=$18--19). This may be in part due to the distinct filters used by the surveys, but also because the SNe, in some cases, may already be declining at the time of announcement.

\subsubsection{Completeness: nuclear events}

\begin{figure}
\includegraphics[width=\columnwidth]{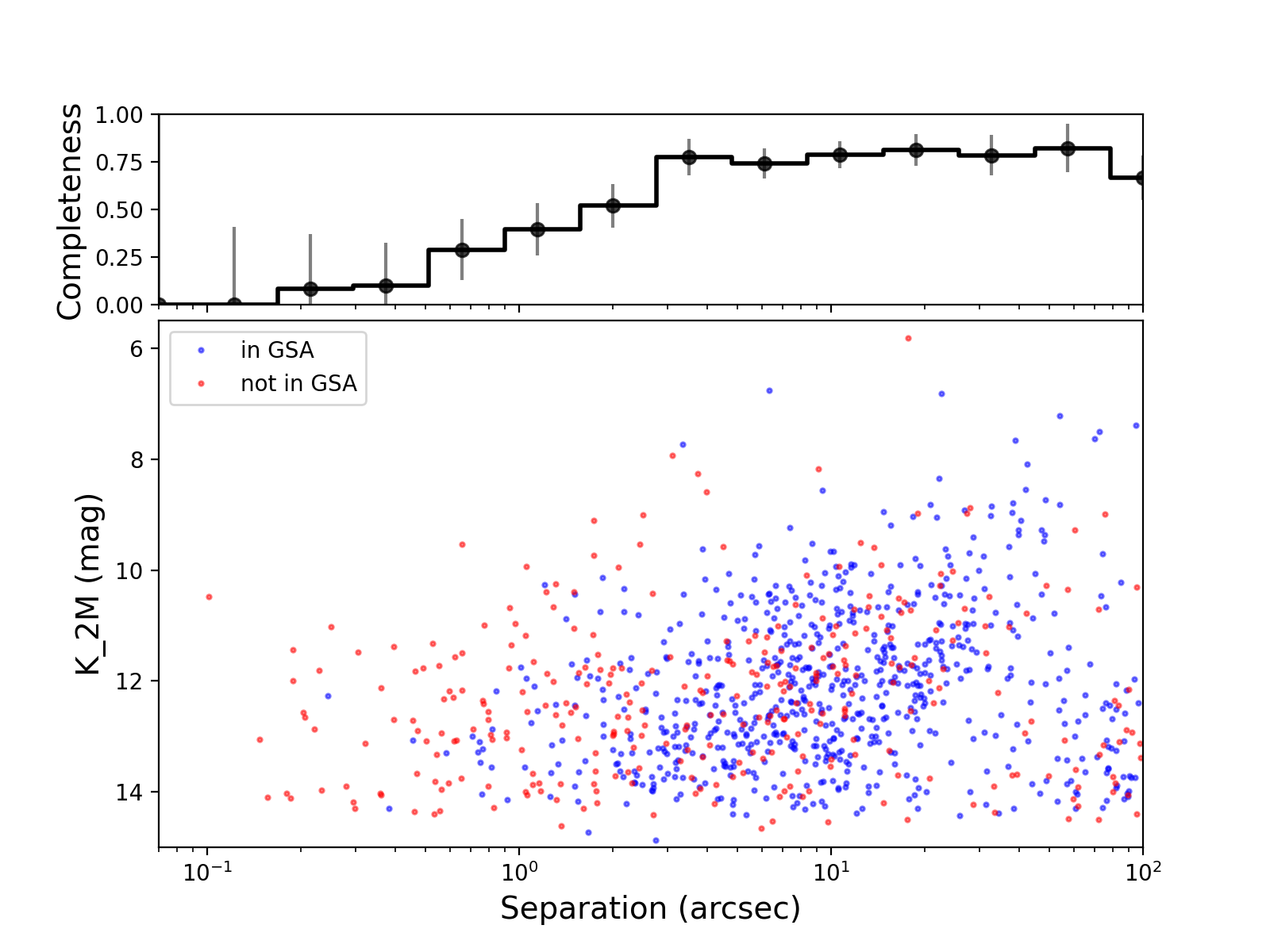}
\caption{Main panel: 2MASS `host' galaxy magnitude (K-band) versus distance to SN (arcseconds) for those events published by \gsa{} (in blue), and those missed (in red). Upper panel: completeness as a function of separation. We also note the likely incorrect host identification for separations greater than 30$^{\prime\prime}$.}
\label{fig:completeness_galsep}
\end{figure}

Finally, we explored completeness versus separation from the host galaxy, and this is shown in Figure~\ref{fig:completeness_galsep}. The sample of `host' galaxies was created by cross-matching the coordinates of the SN sample (again only those with $\geq 2$ scans are considered) against the 2MASS extended source list. There is a cluster of datapoints at separations larger than $\sim30^{\prime\prime}$ at the bottom-right of Figure~\ref{fig:completeness_galsep}. These are possibly mismatches and suggestive of incompleteness in the galaxy sample. There is also a very clear deficit of \gsa{} SN discoveries within 3$^{\prime\prime}$ of the centre of galaxies (between 3$^{\prime\prime}$ and 80$^{\prime\prime}$ the average completeness is 0.79). A similar deficit has previously been ascribed (at least in part) to source confusion in the cross-match phase of IDT for transients in the centres of galaxies. \cite{Kostrzewa2018} found that around 45 per cent of nuclear transients were missed by \gsa{} for this reason. Our completeness for SNe appears to be even lower than this in the nuclear region (though note the significant error bars). Our result also contrasts with simulations (\citealt{blagorodnova16}), which predicted that 90 per cent of SNe would be resolved from their host galaxies by \gaia{}, and detected as \newsources\ for separations larger than 0.3$^{\prime\prime}$, provided that the magnitude of the SN is comparable to the galaxy's bulge. 

It is also worth comparing these results to the discussion on purity in Section~\ref{sec:purity}. Particularly in the bottom left-hand panel of Figure~\ref{fig:angsep_excess} we can see a significant excess of NewSources associated with neighbours. This is presumably dominated by SNe associated with galaxies, although there is likely a contribution from Cataclysmic Variables in crowded environments (the Galactic plane). This excess falls sharply at close separations, around 1$^{\prime\prime}$, indicative of a reduction in the sensitivity of the \newsource\ detector to complex environments.

%
%
%
%

\section{Summary}
\label{sec:summary}

In this paper we have described the \gaia{} Photometric Science Alerts system covering the first 5 years of operations. Our ingestion and processing system handles upwards of 60 million transits per day, searching for new sources, and sources which change significantly in brightness (both upwards and downwards). The flux-change detectors work with the lightcurves, and thus are sensitive not only to sudden changes in brightness, but also to slowly varying sources.

A series of (mostly environmental) filters is applied to reduce the detection rate from a few thousand provisional candidates to a shortlist of several tens of events per day. These checks also identify likely long-period variables (such as Miras) and known Solar System objects. Surviving candidates are subjected to human scrutiny before being published via our Alerts Website, TNS discovery reports and VOEvents.

When an alert is published, all available internally calibrated \gaia{} $G$ band photometry, and uncalibrated BP/RP spectra, of that object becomes public. This includes subsequent measurements of the same object, with the data added to the alert webpages when processed.

We currently publish alerts at a rate of approximately 12 events day$^{-1}$, and almost 25 per cent of them are ultimately classified. The published classifications are dominated by SNe, but we show that this is biased by the extensive supernova follow-up campaigns. The bulk of our unclassified alerts reside in the Galactic plane, and are therefore likely to be Galactic in origin. As an experiment, we built a simple probabilistic alert classifier using uncalibrated BP--RP colour and \gaia{} DR2 parallax (where available), leading to the identification of 638 new candidate CVs and 202 new candidates YSOs. We also show that per-transit data ($G$-band and uncalibrated BP and RP colours) for alerting sources with available DR2 parallaxes can be used to trace the evolution of the transient's position in the colour-magnitude diagram, revealing its nature without the need for spectroscopic confirmation.

We investigated the astrometry of \gsa{} and showed that the accuracy of individual alert detections is 55 milliarcseconds when compared to \gaia{} DR2, and is independent of source magnitude. The photometry of our alerts has a precision of 1 per cent for sources around $G=13$, falling to around 10 per cent at $G=20.7$. Most alerting sources must reach $G=19$, where the median standard deviation is 0.031~mag.

\gsa{} suffers very low levels of contamination from false positives. We showed that the minimum purity of the survey in uncrowded regions for sources with $G<17$ is 93 per cent. Ancillary data is harder to find in the Galactic plane, and for fainter sources, but we find no evidence of additional sources of contamination at faint magnitudes, or in crowded regions.

We also investigated the completeness of the survey, by comparing \gaia{} discoveries to classified supernovae published via TNS (across all sub-types). We measure an overall completeness of 46 per cent, and find that the \gaia{} scanning law, combined with our requirement for two \gaia{} detections, dominates our ability to detect a supernova. Our completeness for supernovae with two or more scans is 79 per cent, unless within 3 arcseconds of the nucleus of the host galaxy, where it drops significantly.

Finally, we note that a total of 2612 alerts spanning observations taken between 25 July 2014 and 28 May 2017 will be included in \gaia\, DR3 in a supplementary table.


\begin{acknowledgements}

We thank the anonymous referee for comments and suggestions that improved this article.

This work has made use of data from the European Space Agency (ESA) mission \gaia\ (\url{https://www.cosmos.esa.int/gaia}), processed by the \gaia\ Data Processing and Analysis Consortium (DPAC,
\url{https://www.cosmos.esa.int/web/gaia/dpac/consortium}). Funding for the DPAC has been provided by national institutions, in particular the institutions participating in the \gaia\ Multilateral Agreement. Further details of funding authorities and individuals contributing to the success of the mission is shown at \url{https://gea.esac.esa.int/archive/documentation/GEDR3/Miscellaneous/sec_acknowl/}.

We thank the United Kingdom Particle Physics and Astronomy Research Council (PPARC), the United Kingdom Science and Technology Facilities Council (STFC), and the United Kingdom Space Agency (UKSA) through the following grants to the University of Bristol, the University of Cambridge, the University of Edinburgh, the University of Leicester, the Mullard Space Sciences Laboratory of University College London, and the United Kingdom Rutherford Appleton Laboratory (RAL): PP/D006511/1, PP/D006546/1, PP/D006570/1, ST/I000852/1, ST/J005045/1, ST/K00056X/1, ST/K000209/1, ST/K000756/1, ST/L006561/1, ST/N000595/1, ST/S000623/1, ST/N000641/1, ST/N000978/1, ST/N001117/1, ST/S000089/1, ST/S000976/1, ST/S001123/1, ST/S001948/1, ST/S002103/1, and ST/V000969/1.

This paper made use of the Whole Sky Database (WSDB) created by Sergey Koposov and maintained at the Institute of Astronomy, Cambridge with financial support from the Science and Technology Facilities Council (STFC) and the European Research Council (ERC).

We thank the William Herschel and Isaac Newton Telescopes on the Roque de los Muchachos Observatory, La Palma, Spain, as well as the Optical Infrared Coordination Network for Astronomy (OPTICON) for their support of this project through telescope time, especially during the commissioning and verification phases.

We thank the Copernico 1.82m telescope (Mt. Ekar, Asiago Italy) operated by INAF Padova for supporting the project through telescope time (under the Large Programme Tomasella-SNe) during the verification phases.

We acknowledge observations taken as part of the PESSTO project collected at the European Organisation for Astronomical Research in the Southern Hemisphere under ESO programme 199.D-0143.

Authors at the ICCUB were supported by the Spanish Ministry of Science, Innovation and University (MICIU/FEDER, UE) through grant RTI2018-095076-B-C21, and the Institute of Cosmos Sciences University of Barcelona (ICCUB, Unidad de Excelencia ’Mar\'{\i}a de Maeztu’) through grant CEX2019-000918-M.

This work is supported by Polish NCN grants: Daina No. 2017/27/L/ST9/03221, Harmonia No. 2018/30/M/ST9/00311, Preludium No. 2017/25/N/ST9/01253 and MNiSW grant DIR/WK/2018/12 as well as the European Commission's Horizon2020 OPTICON grant No. 730890. The Authors would like to thank the Warsaw University OGLE project for their continuous support in this work. 

AB acknowledges financial support from the Netherlands Research School for Astronomy (NOVA).

AG acknowledges the financial support from the Slovenian Research Agency (grants P1-0031, I0-0033, J1-8136, J1-2460).

AH was funded in part by the Leverhulme Trust through grant RPG-2012-541 and by the European Research Council grant 320360.

AP acknowledges support from the NCN grant No.~2016/21/B/ST9/01126.

CM acknowledges support from Jim and Hiroko Sherwin.

DAK acknowledges support from the Spanish research projects AYA 2014-58381-P, AYA2017-89384-P, from Juan de la Cierva Incorporaci\'on fellowship IJCI-2015-26153, and from Spanish National Research Project RTI2018-098104-J-I00 (GRBPhot).

EB and STH are funded by the Science and Technology Facilities Council grant ST/S000623/1. TW was funded in part by European Research Council grant 320360 and by European Commission grant 730980.

GC acknowledges the Agenzia Spaziale Italiana (ASI) for its continuing support through contract 2018-24-HH.0 to the Italian Istituto Nazionale di Astrofisica (INAF).

GD acknowledges the observing grant support from the Institute of Astronomy and Rozhen NAO BAS through the bilateral joint research project "Gaia Celestial Reference Frame (CRF) and fast variable astronomical objects" (during 2020-2022, leader is G.Damljanovic), and support by the Ministry of Education, Science and Technological Development of the Republic of Serbia (contract No 451-03-68/2020-14/200002).

G. Marton acknowledges support from the EC Horizon 2020 project OPTICON (730890) and the ESA PRODEX contract nr. 4000129910.

MF is supported by a Royal Society - Science Foundation Ireland University Research Fellowship

NB acknowledges support from the research programme VENI, with project number 016.192.277, which is (partly) financed by the Netherlands Organisation for Scientific Research (NWO).

NI is partially supported by Polish NCN DAINA grant No. 2017/27/L/ST9/03221.

PAW acknowledges research funding from the South African National Research Foundation.

RWW was funded by the Science and Technology Facilities Council grant ST/P000541/1.

V.A.R.M.R.\ acknowledges financial support from Radboud Excellence Initiative, the Funda\c{c}\~{a}o para a Ci\^encia e a Tecnologia (FCT) in the form of an exploratory project of reference IF/00498/2015/CP1302/CT0001, FCT and the Minist\'erio da Ci\^encia, Tecnologia e Ensino Superior (MCTES) through national funds and when applicable co-funded EU funds under the project UIDB/EEA/50008/2020, and supported by Enabling Green E-science for the Square Kilometre Array Research Infrastructure (ENGAGE-SKA), POCI-01-0145-FEDER-022217, and PHOBOS, POCI-01-0145-FEDER-029932, funded by Programa Operacional Competitividade e Internacionaliza\c{c}\~ao (COMPETE 2020) and FCT, Portugal.

ZKR acknowledges funding from the Netherlands Research School for Astronomy (NOVA).

ZN acknowledges support from the ESA PRODEX contract nr. 4000129910.

\end{acknowledgements}


\bibliographystyle{aa} 
\bibliography{gsabib} 


\begin{appendix} 

\section{Cyclic processing and catalogue changes}
\label{appx:idu}
Cyclic processing is a reprocessing of all the \gaia\, data that occurs between IDT processing and the main data releases. It includes some of the same kinds of processing as IDT (but there are also many additional activities), but at greater leisure and with more computationally-expensive algorithms. In particular, cyclic processing repeats and revises the mapping of transits to sources, using an improved spatial cross-match with improved astrometry (see \citealt{torra20}). Cyclic processing happens long after \gsa{} transients have been raised and published from a given set of observations. The published alerts are not revised to reflect changes to flux measurement.

Of necessity, alerts processing uses IDT's working source-catalogue to build the lightcurves. Periodically (on exactly two occasions up to the end of 2020) IDT changes its working catalogue to the most-recent cyclic catalogue. To accommodate this, the alerts database has to be updated to the new catalogue, such that old and new observations may still be combined into lightcurves. 

There is never a one-to-one mapping between the old and new working catalogues: cyclic processing uses an improved astrometric solution, which can result in splits and mergers of IDT's sources to best fit a new clustering analysis of the sky (see \citealt{torra20}). And importantly for \gsa, a magnitude criterion was employed in the clustering algorithm to help disentangle valid and spurious detections into different clusters. This means that transits forming a lightcurve of a published alert could end up assigned to different sources in the new catalogue, especially for a highly variable (e.g. transient) source, where the bright and faint parts of a lightcurve may appear in the catalogue as distinct sources. An additional algorithm was later added to the cyclic processing in an attempt to improve matching for these variable sources.

When IDT's working catalogue is updated, then we map the transits of published alerts to their new sourceIds (if they have changed). Where the update results in multiple sourceIds associated with a single alert, the alert lightcurve is visually inspected including all positionally coincident transits, and if necessary additional sourceIds are assigned to the published alert (these are called {\it mixed-in sources}). In this event the published lightcurve will be updated with future transits belonging to any of the sources assigned to the published alert.

\section{Photometric One Day Calibration}
\label{appx:podc}

\begin{figure*}
\includegraphics[width=0.90\textwidth]{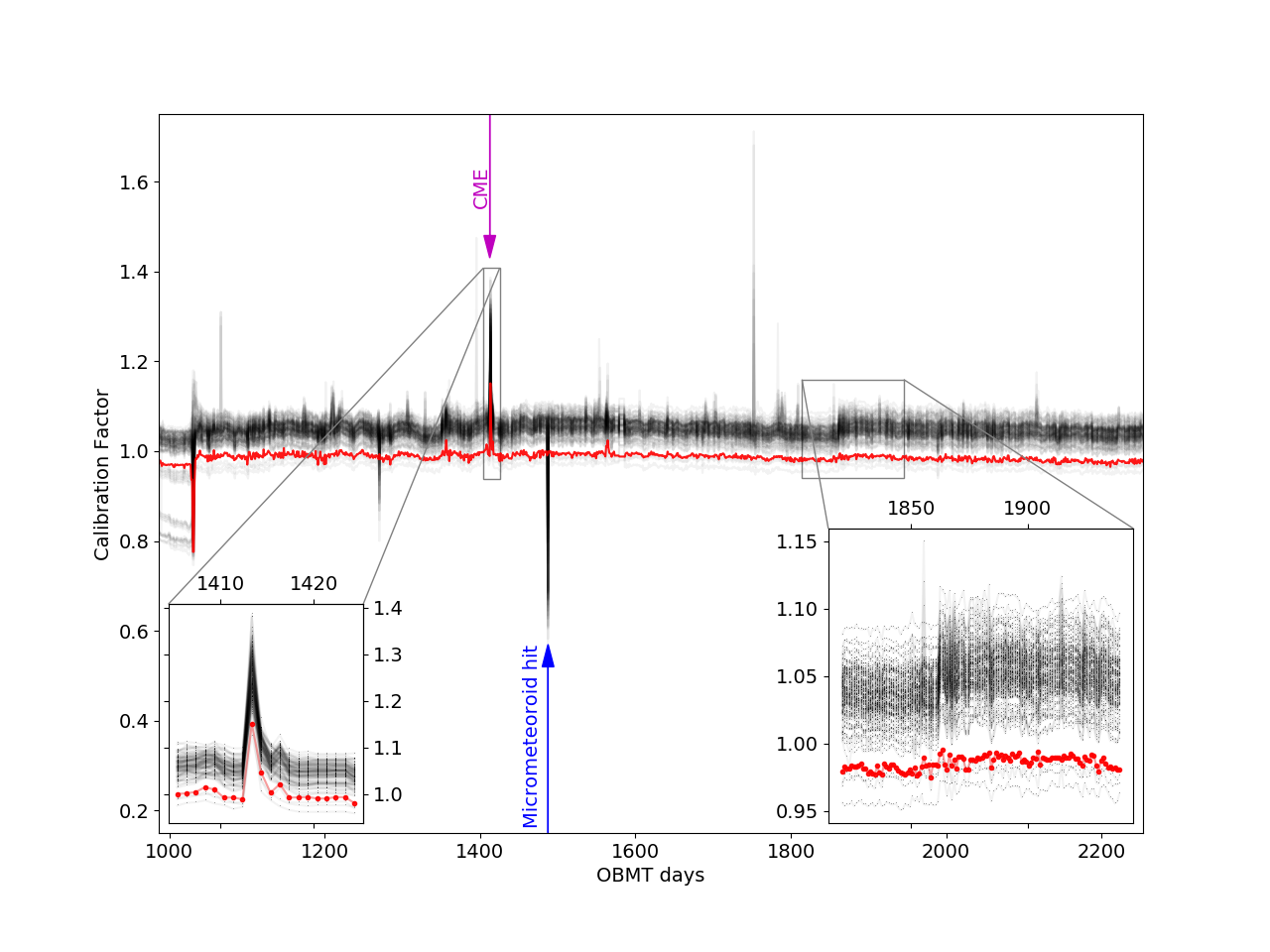}
\caption{Calibration factors returned by PODC for the 110 calibration units corresponding to the ungated windows for the faintest sources (fainter than $\rm{G_{mag}} \sim16$). The calibration factors for the calibration unit corresponding to AF8, row 2 in the following field-of-view (FFoV) is shown in red. The time interval covers the range of IDT runs from 1046 to 4724 used in this paper, and is expressed in days in on-board mission time (OBMT).
Excursions in the daily measured calibration factor can and do occur, these can result from large numbers of cosmic rays/charged particle events as well as micrometeroid hits.
Insert lower left: (indicated by magenta arrow) excursion which occurred roughly 2 days after the CME X9.3 of September 2017.
Insert lower right: close-up of the more typical stable behaviour of the calibration factor derived by PODC. We note the jump around OBMT 1861.9 is a refocusing of both FOVs.
}
\label{fig:calibrationFactor}
\end{figure*}

In \gsa{} we store raw IDT fluxes in the database, and calibrate all transit photometry on-the-fly via a database function. The parameters used by the function are derived from the Photometric One Day Calibration (PODC). PODC is generated on an approximately monthly timescale by DPCI, but with a one-day cadence. Thus the \gsa{} photometric calibration is always out-of-date, and relies on the generally stable and slowly varying throughput of the \gaia\ instruments. The PODC calibration is a simplified and non-iterative version of the calibration applied by DPCI for the production of the main \gaia{} data releases. We also note that PODC does not include a {\em link} calibration (\cite{evans2018}) to bring the distinct \gaia{} instrument configurations (gates and window classes) into agreement. Finally, although the PODC calibration includes colour terms, we decided to neglect these in implementation. This is because there are significant numbers of transits of alerting sources where we are missing BP and/or RP photometry (sometimes these data are delivered later). In Figure~\ref{fig:calibrationFactor} we show the time evolution of the PODC derived calibration factors for the 110 calibration units corresponding to the ungated windows for the faintest sources (fainter than $\rm{G_{mag}} \sim16$).

We compared the PODC calibration to the iterative \gaia\, DR2 $G$-band calibration \citep{evans2018,riello2018} for a set of 184\,000 sources, which have been randomly selected to lie in the SDSS DR7 footprint (this avoids the most crowded regions of the Galactic plane). We required the sources to have a minimum of 10 \gaia\, field-of-view transits, and we use the median of the per-CCD PODC calibrated fluxes as representative of the per-transit CCD flux. Because PODC was initialised twice, with two distinct magnitude zeropoints, we apply these internal zeropoints to transform from fluxes to magnitudes. For the DR2 fluxes we use the revised photometric zeropoints\footnote{\url{https://www.cosmos.esa.int/web/gaia/iow_20180316}} published in March 2018.

\begin{figure}
    \centering
    \includegraphics[width=\columnwidth]{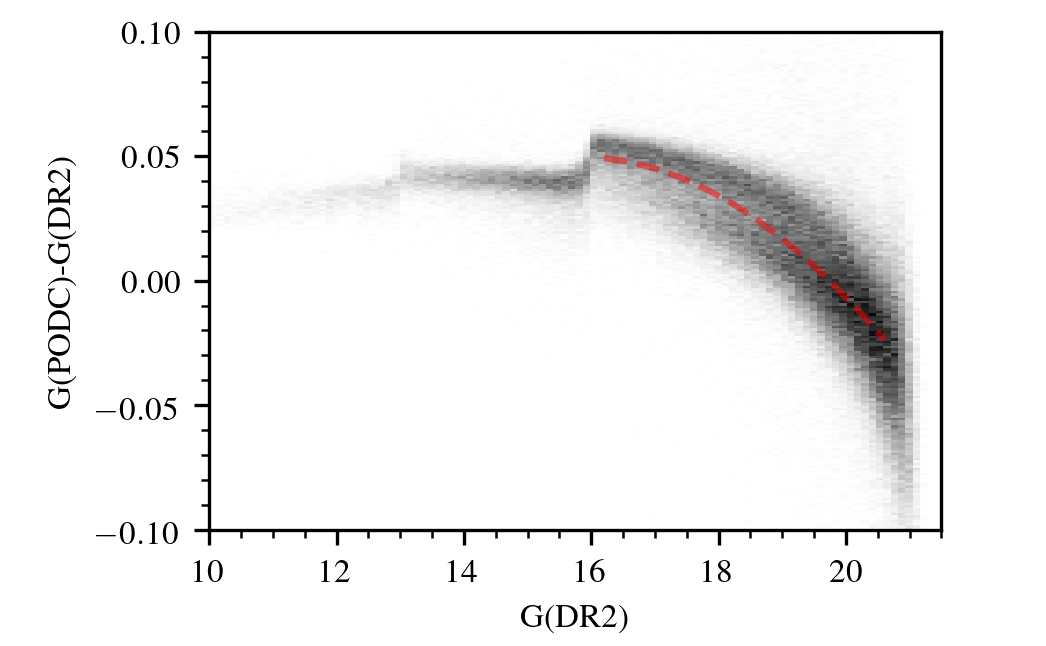}
    \caption{Offset between PODC and DR2 calibrated sources as a function of DR2 $G$ magnitude. Sources must have a minimum of 10 \gaia\, transits for inclusion, and were selected to have overlaps with the SDSS footprint, hence avoiding regions of low Galactic latitude and high extinction. A simple quadratic model (shown as a dashed red line) shows the residual magnitude dependence for sources fainter than $G=16$.}
    \label{fig:gvp_mag}
\end{figure}

Some striking features can be seen in the difference between the PODC and the DR2 photometry (see Fig~\ref{fig:gvp_mag}). The first is that there are two discontinuities (at the few per\,cent level) at $G$(DR2)$\sim$13 and $G$(DR2)$\sim$16. These both correspond to changes in the on-board window class allocated to a source, thus the size and binning of the readout window \citep{debruijne2015}. At $G=13$ the window changes from 2D (for brighter sources) to 1D (window Class=1), while at $G=16$ the 1D window changes size from 18 to 12 pixels Along Scan (window Class=2). These steps are best explained by the lack of a {\it link} calibration in PODC (see earlier).

The second feature visible in Fig.~\ref{fig:gvp_mag} is a clear non-linear trend between offset and magnitude for sources fainter than $G=16$. The model overlaid on the figure is a fit to the median offset between PODC and DR2 in discrete magnitude bins, ranging from $+0.05$ at $G=16.2$ to $-0.025$ at $G=20.4$. We use a quadratic model fitted to the magnitude term:

\begin{equation}
    G_{\rm PODC}-G_{\rm DR2} = -0.69 + 0.094 G_{\rm DR2} - 0.003 G_{\rm DR2}^2
    \label{eq:a1}
\end{equation}

shown as a dashed red line in the same Figure. It is not clear where this apparent non-linearity arises, but we note that \gaia{} DR2 takes Intermediate Data Update fluxes as input, which have improved image parameter determination upon that implemented in IDT, as well as a better PSF/LSF model, better background treatment and other improvements (see \citealt{lindegren18}).

\begin{figure}
    \centering
    \includegraphics[width=\columnwidth]{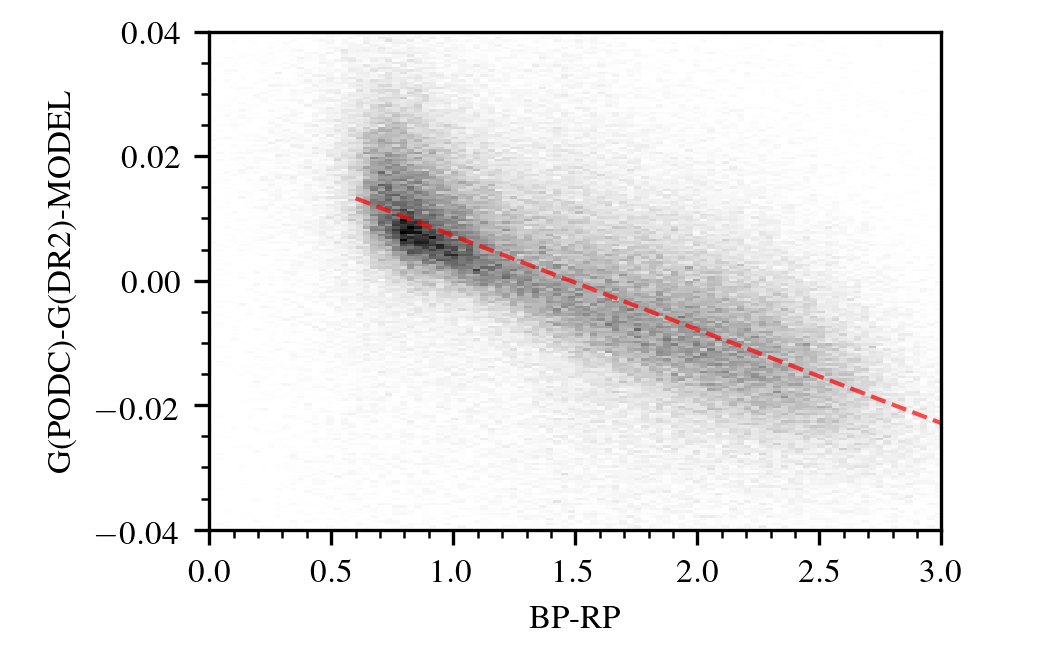}
    \caption{Difference between data and quadratic magnitude model (see Fig~\ref{fig:gvp_mag}) as a function of \gaia\, BP--RP colour.}
    \label{fig:gvp_bprp}
\end{figure}

The third feature of note in Fig.~\ref{fig:gvp_mag} is the large scatter in $G_{PODC}-G_{DR2}$ at any $G$(DR2), and the sequence actually looks bimodal beyond $G$(DR2)$\sim17$. In order to test the hypothesis that this is a colour effect (because we excluded the colour-term in our implementation of PODC), we show the residuals from the model for sources fainter than $G=16$ in Fig.~\ref{fig:gvp_bprp}. Indeed, there is a significant linear trend with BP--RP which explains most of the observed scatter. For completeness, the model fit to the data shown in the figure leads to an updated version of Equation~\ref{eq:a1}, such that:

\begin{equation}
    G_{\rm PODC}-G_{\rm DR2} = -0.67 + 0.094 G_{\rm DR2} - 0.003 G_{\rm DR2}^2 - 0.015 (BP-RP)
\end{equation}

\newpage
\section{Computing arrangements}
\label{appx:comp}
The alerts cluster is co-located with, but distinct from, DPCI. 
Our cluster is built around a large, PostgreSQL Relational Database Management System (RDBMS). A dedicated, 32-core server holds the primary copies of the databases and an identical machine holds a secondary, read-only copy of each database, kept current by synchronous replication. These replica databases were originally installed as protection against data loss, a role that has never been exercised. They now serve to increase performance by spreading the query load across two servers. Each database server has 176 TB of bulk-data space, arranged as a RAID 6. The alerts computers are interconnected, and connected to DPCI, by an Infiniband network.

The alerts pipeline is divided into: a batch application for ingestion of data, and primary data-reduction; a batch application for filtering of candidate alerts; a web application for human assessment of candidates surviving automatic filtering; and further web-applications for publishing selected alerts. 
The data reduction application, which handles the greatest volume of data in each run, is assigned its own 64-core server. An identical server is held in readiness to take over should the first one fail. In practice, this reserve server is used for testing, for exploration of new algorithms, and occasionally for large-scale rearrangements of the main database. Each data-reduction server has 22 TB of data space to receive new data from DPCI. This holds roughly six months of data and allows reprocessing to correct errors or improve results over this span of the mission.

The other applications run in virtual machines hosted on a pair of smaller servers. There is also a disc-server for back-ups of the main databases, with a capacity of 284 TB.

The main data-reduction application was written in Java, for performance and to conform to DPAC standards. The web applications were written in Python, using the Django framework, for ease of development. The batch-filtering application was also written in Python.

The most notable feature of the computing design is that the bulk of the data is stored {\it in the PostgreSQL database}. Compare this with the more common arrangement where only metadata are stored in the RDBMS, while the bulk is stored in flat files. Ingesting both the full input data-set, and the results of processing, reduces significantly the amount of code needed to simply move data, and allows extreme flexibility in writing the applications that consume, analyse and display the results. However, it is challenging to obtain sufficient throughput from the relational architecture. Many design choices in the database and applications are forced by the need to reduce a day’s data in six hours or less.

At the time of writing, the PostgreSQL system occupies 110 TB of its storage (with some unknown and variable fraction of this reclaimable for new data), almost all of which space is taken up in ingesting and recording the transit data. This is exceptionally large for a database that is extended daily and in which roughly 5\% of the contents must be scanned for daily processing. 

The data-set is far too large to fit comfortably into a single table for each record type, and is therefore partitioned by position of the sources into level-5 ($n_{side}$ = 32) HEALpixels; there are 12,288 tables for each record type that deals with individual transits or sources. This partitioning is much coarser than the level-12 mesh used elsewhere by DPAC to group Gaia sources into catalogues. The level-5 mesh was chosen because it is the closest match to the width of the Gaia field of view. Empirically, it is most efficient to query historic transit-data for a whole HEALpixel, selecting transits only for those sources that have received new data in current run, than to make one query per source with new data. The level-5 mesh minimises simultaneously the number of these expensive queries and the number of rows touched by a query that relate to sources outside the current scan.

The queries to ingest and to raise data cannot be straightforward in such a highly-stressed system; they have to be highly tuned for performance. This precludes the use of generic, object-relational-mapping libraries to generate queries. Instead, the application code forms its queries directly.
In order to achieve performance, much parallelism is needed, and the number of threads accessing the data has to be chosen carefully. We knew from pre-mission tests that the application was not the bottleneck and could be given sufficient threads of execution to saturate the database engine. Later experiments showed that the database performance, for this application, is limited by CPU power on the database server, not by I/O capacity; the complexity of the queries dominates. Therefore, the ideal arrangement is to have one application thread accessing the database per CPU core on the database server: 32 such threads in our system. The data-reduction servers have 64 CPU cores and to use these we run two second-level threads per HEALpixel to consume the raised data from the memory of the data-reduction server.
We feel that we have reached a practical limit in the size of database that can be hosted on a single RDBMS-server. Any major expansion (beyond routine addition of daily data for the remaining mission) or rearrangement of the database would drive us to a multi-server solution with partitioning across computers. This would be massively more expensive and we are happy that we have managed to do the work with single servers. We note that while it is possible to tune the daily data-flow for acceptable performance, whole-database operations are uncomfortably slow. This includes data-model changes that add columns; weekly back-ups of the database; and especially changes to the IDT working-catalogue that force us to update the transit-source mapping for the whole data-set. The latter operation requires down-time of at least one week. When using PostgreSQL at this scale, any routine operation becomes a significant campaign that requires careful planning.

\newpage
\section{Abbreviations used in this paper}
\label{appx:abbreviations}

\begin{tabular}{lcl}
 AC & -- & across-scan direction\\
 ADS & -- & Astrophysics Data System\\ 
 AF & -- & Astrometric Field\\
 AGIS &--& Astrometric Global Iterative Solution\\
 AL & -- & along-scan direction\\
 ASAS & -- & All-Sky Automated Survey\\
 ASAS-SN & -- & The ASAS Supernova survey \\
 ATEL &--& Astronomer's Telegram \\
 BP/RP & -- & Blue Photometer/Red Photometer \\ 
 CBAT & -- & Central Bureau for Astronomical \\
      &    & Telegrams \\
 CCD & -- & charge-coupled device (detector)\\ 
 CME & -- & Coronal Mass Ejection\\
 CPU & -- & Central Processing Unit\\
 csv & -- & comma separated values (file type)\\
 CV  & -- & Cataclysmic Variable\\
 Dec & -- & Declination\\
 DPAC & -- & Data Processing and Analysis Consortium \\ 
 DPCI & -- & Data Processing Centre in Cambridge\\
 DR2 & -- & \gaia's Data Release 2, 25 April 2018\\
 DR3 & -- & \gaia's Data Release 3, expected 2022\\
 ESA & -- & European Space Agency \\
 ESO & -- & European Southern Observatory\\
 ESOC & -- &  European Space Operations Centre\\
      &    & (Darmstadt, Germany)\\ 
 ETL & -- & Extract Transform Load\\
 FOV & -- & Field of View\\
 $G$ & -- & \gaia's `white light' photometric band \\
 GDR2 & -- & \gaia\, Data Release 2 \\
 GoF & -- &  Goodness-of-Fit\\
 GSA & -- & \gaia\, Science Alerts \\ 
 HEALpix & -- & Hierarchical Equal Area\\ 
         &    & isoLatitude Pixelisation\\  
 HRD & -- & Hertzsprung-Russell diagram\\
 IAU & -- & International Astronomical Union\\
 IDT & -- & Initial Data Treatment\\
 IGSL & -- & Input \gaia\ Source List\\
 IoA & -- & Institute of Astronomy, Cambridge\\ 
 IPAC & -- & Infrared Processing and Analysis \\
      &    & Center (NASA)\\
 iPTF & -- & intermediate Palomar Transient Factory \\      
 LSF & -- & Line Spread Function\\ 
 MAD & -- & Median Absolute Deviation\\ 
 MASTER & -- & Mobile Astronomical System of \\
        &    & Telescope-Robots   \\
 MOC & -- & Mission Operations Centre (Darmstadt,\\ 
      &    & Germany)\\ 
 NASA & -- & National Aeronautics and Space\\
      &    & Administration\\
 NED  & -- & NASA/IPAC Extragalactic Database\\
 NUTS & -- & Nordic Optical Telescope Unbiased\\
      &    & Transient Survey \\
 OBMT & -- & On Board Mission Time\\
 OGA & -- & On-ground Attitude \\
 OGLE & -- & Optical Gravitational Lensing Experiment\\
 PanSTARRS & -- & Panoramic Survey Telescope and Rapid\\
           &    & Response System \\
 PESSTO & -- & Public ESO Spectroscopic Survey of\\
        &    & Transient Objects \\ 
 PODC & -- & Photometric One-day Calibration\\
 PPE & -- & Prompt Particle Event\\
 PSF & -- & Point Spread Function\\
 \end{tabular}
\newpage
\begin{tabular}{lcl}
 QSO & -- & Quasi Stellar Object\\
 RA & -- & Right Ascension\\  
 RAID & -- & Redundant Array of Independent Discs\\
 RBF & -- & Radial Basis Function\\
 RDBMS & -- & Relational Database Management System\\
 RMS & -- & root mean square\\
 RSS & -- & Really Simple Syndication\\
 RVS & -- & Radial Velocity Spectrograph (\gaia)\\
 SDSS & -- & Sloan Digital Sky Survey\\
 SEDM & -- & Spectral Energy Distribution Machine\\
      & -- & (spectrograph)\\
 SLSN & -- & Superluminous Supernova\\
 SM & -- & Sky mapper (\gaia)\\
 SN  & -- & Supernova\\
 SOC & -- & Science Operations Centre (Madrid, Spain)\\ 
 SSO & -- & Solar System object\\
 SVM & -- & Support Vector Machine\\
 TCB & -- & Barycentric Coordinate Time\\
 TNS & -- & Transient Name Server\\
 URL & -- & Uniform Resource Locator (web address)\\
 UTC & -- & Coordinated Universal Time \\
 VOEvent & -- & the Virtual Observatory Event language\\
 XP & -- & shorthand for BP/RP\\
 XRB & -- & X-ray binary star\\
 VSX & -- & International Variable Star Index database\\
 YSO & -- & Young Stellar Object\\
 ZTF & -- & Zwicky Transient Facility
\end{tabular}

\end{appendix}

\end{document}